\documentclass[12pt, letterpaper]{article}

\usepackage{dsfont}
\usepackage{xcolor,colortbl}
\usepackage{tabularx}
\usepackage{svg}
\usepackage{caption}
\usepackage{titlesec}

\usepackage{makecell}
\usepackage{adjustbox}
\usepackage[flushleft]{threeparttable}
\usepackage{booktabs}
\usepackage{helvet}
\usepackage{times}
\usepackage{placeins}
\usepackage{multirow}
\usepackage{subcaption}
\usepackage{arydshln}
\usepackage{longtable}
\usepackage{float}
\usepackage{multicol}
\usepackage{supertabular}
\usepackage{booktabs}
\usepackage[framemethod=TikZ]{mdframed}
\usepackage{fullpage}
\usepackage[switch]{lineno}
\usepackage{amssymb}
\usepackage{minitoc}
\usepackage{amsmath}
\usepackage{rotating}
\usepackage{array}
\usepackage{mathtools}
\usepackage[ruled]{algorithm2e}
\usepackage{algorithmic}
\usepackage{bm}
\usepackage{comment}
\usepackage{enumitem}
\usepackage{graphics}
\usepackage{graphicx}
\usepackage{latexsym}
\usepackage{mathrsfs}
\usepackage{morefloats}

\usepackage{nicefrac}
\usepackage{authblk}
\usepackage[english]{babel}
\usepackage{blindtext}
\usepackage{siunitx}
\usepackage{graphicx}
\usepackage{url}
\usepackage{hyperref}
\usepackage{xr-hyper}
\usepackage{dcolumn}
\usepackage{todonotes}
\makeatletter
\newcommand*{\addFileDependency}[1]{
  \typeout{(#1)}
  \@addtofilelist{#1}
  \IfFileExists{#1}{}{\typeout{No file #1.}}
}
\makeatother

\graphicspath{{main_figures/}{../main_figures/}}

\newcolumntype{d}[1]{D{.}{.}{#1}}

\title{Characterizing the effect of retractions on publishing careers}

\author[1,2]{Shahan Ali Memon}
\author[1*]{Kinga Makovi}
\author[1*]{Bedoor AlShebli}

\affil[1]{\normalsize 
Social Science Division, New York University Abu Dhabi}

\affil[2]{\normalsize 
Information School, University of Washington}

\affil[*]{\small Joint corresponding authors. E-mails:\ bedoor@nyu.edu, km2537@nyu.edu}

\date{}

\begin{document} 
\maketitle

\begin{abstract}
\noindent Retracting academic papers is a fundamental tool of quality control, but it may have far-reaching consequences for retracted authors and their careers. Previous studies have highlighted the adverse effects of retractions on citation counts and coauthors' citations; however, the broader impacts beyond these have not been fully explored. We address this gap leveraging Retraction Watch, the most extensive data set on retractions and link it to Microsoft Academic Graph and Altmetric. Retracted authors, particularly those with less experience, often leave scientific publishing in the aftermath of retraction, especially if their retractions attract widespread attention. However, retracted authors who remain active in publishing maintain and establish more collaborations compared to their similar non-retracted counterparts. Nevertheless, retracted authors generally retain less senior and less productive coauthors, but gain more impactful coauthors post-retraction. Our findings suggest that retractions may impose a disproportionate impact on early-career authors.

\end{abstract}

\section*{Introduction}

Reputation is a crucial factor in building status, particularly when quality is uncertain or unobservable \cite{Lynn_etal_2009}, and when it is produced through highly technical and complex processes. This characterizes creative fields, medicine and science alike. Therefore, when a scientist's reputation is challenged, the consequences can be severe \cite{Azoulay_etal_2017, Misttry_etal_2019, Jin_etal_2019, Hussinger_Pellens_2019}, with long-lasting effects on their career outcomes. The credibility of a scientist, a crucial currency of their reputation, is established over the course of their career based on the quality of their publications \cite{petersen2014reputation} among other factors. Therefore, when the quality of one's work is called into question, the stakes are high, and the consequences can be more significant than the outcome of a single project. While positive signals, such as citations and grants, have been linked to reputation-building \cite{Aksnes_Rip_2009, petersen2014reputation}, less is understood about the relationship between a scientist's challenged reputation and their career progression in future collaborations. Hence, further research is needed to fully comprehend the impact of such challenges on scientific collaborations and career trajectories. Retractions of scientific papers give us a window through which to study this question.

When the integrity of a scientific paper is disputed, editors and authors may choose to remove the work from the canon either together or in isolation. While the article may still be accessible, it will be accompanied by a retraction notice that explains the reason(s) behind its removal, such as misconduct, plagiarism, mistake, or other considerations. This creates a clear and visible signal associated with the authors of the paper that the quality of their work has come under scrutiny. Retractions have been used for this purpose since 1756 \cite{Vuong_etal_2020}, and contemporary journals have formal procedures to execute when the authors or readers highlight problematic content.

Prior work investigates the impact of retractions on scientific careers, examining productivity \cite{Misttry_etal_2019}, citations for retracted papers \cite{Jin_etal_2019}, citations for papers published prior to retraction \cite{Azoulay_etal_2017}, and post-retraction citations of pre-retraction collaborators \cite{Hussinger_Pellens_2019}, generally finding negative effects. However, some work also demonstrates that these effects are heterogeneous and might vary based on the reason for retraction, and/or the prominence of the retracted author \cite{Azoulay_etal_2017, Bishop_2018, Jin_etal_2019}, for instance, measured by author order. While recent research is breaking new ground on the quantitative analysis of the impact of retractions, it often focuses on specific fields, or compares different retracted authors to one another (e.g., those who have experienced a single versus multiple retractions). This existing work used various strategies to construct a comparison group for retracted authors, some examining others publishing in the same journal at the same time, while other examining coauthors of retracted scientists on non-retracted papers, or on retracted papers without assigning them blame. None use a more comprehensive strategy to create matched pairs on the basis of a series of author-level characteristics, and with few exceptions, they do not examine the impact of retraction on post-retraction collaboration networks \cite{Mongeon_Lariviere_2016, Sharma_Mukherjee_2024}. Therefore, a comprehensive analysis of retractions across fields and over time has yet to be undertaken where retracted authors are compared to otherwise similar non-retracted authors. Beyond documenting the impact of retraction on careers, it is essential to examine the mechanisms that might bring about these effects. Therefore, we focus both on the continuity of post-retraction careers, as well as the development of the collaboration network of retracted authors that is needed to succeed in publishing careers \cite{lee2005impact, he2009research, wuchty2007increasing}. 

Retractions can attract significant attention, particularly when they expose egregious misconduct. Such instances not only question the authors' reputation, but also undermine the public's trust in science, scientific findings, and institutions of science, such as universities, internal review boards, journals, and the peer-review process. Some retractions, therefore, cast a long shadow that extends far beyond the scrutinized work. For instance, the retraction of a study on political persuasion and gay marriage in \textit{Science} in 2015 \cite{LaCour_etal_2014}, which was likely based on fabricated data, led to questions about the impartiality of reviewers~\cite{konnikovaGayMarriage}. Similarly, when a paper in \textit{Nature} was retracted due to falsified images \cite{Obokata_etal_2014}, criticism went beyond concerning the conduct of the first author, and extended to the male-dominated Japanese academy and its culture of fierce pressure and competition~\cite{mcneillAcademicScandal}. In both cases, the first authors left scientific publishing careers after receiving extreme levels of attention (Altmetric scores above 1000). However, how systematic the impact of this attention is, is yet to be fully understood. This question, of course, is tied closely to how retracted scientists might rebuild their collaboration network, as future collaborators may or may not learn about past events, depending on the level of attention they received.

The value in social relationships, and the theory that resources encapsulated in them may be leveraged is longstanding \cite{Coleman_1988}. The assumption, specifically that larger collaboration networks are beneficial, is rooted in prior work that documents the benefits of research collaborations, and that of larger collaboration networks. Qualitative evidence suggests that researchers collaborate for both instrumental and strategic reasons, such as access to specialized expertise, equipment and other resources, visibility for professional advancement, and enhanced research productivity, as well as emotional reasons, since many regard collaborative work as energizing and fun \cite{Beaver_2001}. These self-reports are reflected in empirical evidence, such as the association between the size of collaboration networks and citations \cite{Bosquet_Combes_2013}, in addition to future productivity \cite{Lee_Bozeman_2005, Ductor_etal_2014}. Importantly, Ductor and colleagues suggest that the quality of one's coauthor network signals important information about researchers' quality, and that these signals are crucial to assess one's research potential, especially at the beginning of the career \cite{Ductor_etal_2014}. Additionally, prior work reveals that coauthor networks show higher levels of triadic closure than expected by chance, that is, authors of scientific papers tend to work with former coauthors of their coauthors in the future \cite{Shi_etal_2015, Kim_Diesner_2017}. Such regularity is based on similarity, but also on strategic considerations, where a scientist brokers relationships among their unconnected coauthors, thereby communicating information about the qualities of those they connect that are challenging to observe otherwise \cite{Zhelyazkov_2018}, such as their skill or integrity in the context of scientific publishing. Extending these arguments to authors who experience a retraction, the collaboration networks they maintain, or build could be crucial to recover from a negative signal about the quality of their work, and some processes, such as triadic closure, might help them in particular to do so.

Drawing on retractions as a (potentially stigmatizing) signal that challenges authors' reputations, we offer three key empirical observations. First, we find that the extent of attention received by a retraction is positively associated with the likelihood of retracted authors leaving publishing careers. That is, the more public the retraction, the more profound its consequences appear to be for authors' careers. This finding is especially significant since most attention received by papers extends beyond the content of the science, and involves discussions of great societal importance about the context within which scientific findings are produced. Second, perhaps counterintuitively, we demonstrate that conditional on staying in scientific publishing, retracted authors retain and gain more collaborators compared to otherwise similar authors without retractions. Third, while these larger collaboration networks may benefit retracted authors, retracted authors generally build qualitatively different, and weaker networks compared to their similar counterparts in terms of their collaborators' seniority, and productivity post-retraction. 

\section*{Results}

The consequences of retractions on authors' careers can be severe, sometimes resulting in them leaving scientific publishing entirely. Analyzing the timing of an author's departure from publishing following a retraction can yield valuable insights. To facilitate this analysis, we utilize two main data sets: Retraction Watch (RW) \cite{retractionwatch}, the most extensive publicly available database of retracted papers with over 26,000 publications in around 5,800 venues, and Microsoft Academic Graph (MAG) \cite{sinha2015overview, wang2019review}, which provides comprehensive records and citation networks for over 263 million scientific publications and collaboration networks for over 271 million authors. From RW, we exclude bulk retractions (e.g., when all papers are retracted from a conference proceeding as a result of questionable peer-review \cite{McCook_2018}), as well as authors with multiple retractions to center on a singular event in our analysis. Furthermore, we focus on papers retracted between 1990 and 2015 to allow us to examine post-retraction outcomes. After filtering RW, we merge it with MAG, identifying over 4,578 retracted papers involving 14,579 authors (the ``filtered'' sample). Linking these two data sets allows us to characterize these retracted authors, and describe their pre-retraction and post-retraction careers. Full details of our pre-processing steps, further justifications for exclusion criteria, and merging of datasets can be found in the Supplementary Note 1 and the ``Merging RW and MAG'' section in Materials and Methods.

In order to study the relationship between when an author leaves scientific publishing and having faced a retraction, it is necessary to define ``leaving publishing'' or ``attrition.'' Our data, however, are right-censored, meaning that especially for authors who started publishing recently, we may not observe their entire careers. This makes it difficult to accurately determine their true attrition year or whether they have indeed left scientific publishing. For those whose attrition year can be calculated, we define it by first identifying either the last observed publication year or the start of the first prolonged gap in an author's career, recognizing that such gaps indicate a significant interruption. For those whose attrition year cannot be determined due to the right-censored nature of the data, we assume that they are still active and highlight how this is handled in each analysis.

To identify attrition, we analyze the distribution of the longest gaps in authors' publishing careers in STEM fields, which make up the majority of our data in RW, across the entire MAG dataset. For each cohort, the length of this gap is selected to be the 95\textsuperscript{th} percentile of all gaps, indicating that 95\% of authors have maximum gaps this long or shorter (see Supplementary Figure 1). The variability in gap lengths across cohorts can be attributed to two main factors: first, MAG relies on digitized publication records, which may result in missing publications, particularly for earlier years; second, over time, the frequency of scientific publishing has increased, thereby reducing the typical gap size. Thus, we determine an author's attrition year either by identifying the onset of such a gap in their career or, if no such gap exists, by their final year of publishing activity. Authors who have not experienced a gap relevant for their cohort by 2020, when our observation window ends, are presumed to be active.

Next, we categorize retracted authors into three groups based on the relationship between their retraction year and their departure from scientific publishing: those who left publishing careers (i) around the time of retraction (years 0 and -1; blue in Figure 1), (ii) after retraction (years 1 onward, pink in Figure 1) or continue to have ongoing careers, and (iii) before retraction (years -2 and earlier, grey in Figure 1a). A notable trend is revealed, showing that approximately 45.9\% of authors who have left their publishing careers do so around the time of retraction (Figure 1a, note that authors with ongoing publishing careers in 2020 are not included in this panel). Specifically, 29\% leave in the year of retraction (year 0), 16.9\% depart shortly before (year -1). In addition to exploring this aggregate pattern, we further investigate the probability of authors remaining in scientific publishing across different academic ages (Supplementary Figure 2), replicating Figure 1a but disaggregated by age. Our analysis reveals that early-career authors, specifically those whose retraction falls within 0-3 years from their first publication, are much more likely to leave publishing when experiencing a retraction. Furthermore, we explore these patterns by affiliation rank, author order, and retraction reason  (Supplementary Figures 3-5). We find, descriptively, that authors whose papers were retracted due to misconduct and plagiarism are more likely to leave in the retraction window compared to those retracted for a mistake.

Further descriptive statistics based on the sample of 12,742 retracted authors who either left publishing during the retraction window or later, or maintained ongoing careers in 2020 (i.e., excluding those in grey in Figure 1a who left well before retraction occurred) reveal that an overwhelming majority belong to STEM fields such as Biology, Medicine, Chemistry, Physics, with less than 1\% originating from non-STEM fields (Figure 1b). Additionally, the share of women in the group who leave around the time of retraction is greater than the share of women in the group who leave after or have continued careers (30\% vs. 25.5\%, respectively; \(\chi^2(1, N = 12742) = 20.07, p < 0.001\)) when comparing these descriptively. Moreover, our analysis indicates that authors who leave academic publishing around the time of retraction are significantly less experienced---as measured by academic age (Welch's $t(10465.17)=-82.459$, $p<0.001$), number of papers (Welch's $t(11153.30)=-57.529$, $p<0.001$), number of citations (Welch's $t(10713.92)=-37.005$, $p<0.001$), and number of collaborators (Welch's $t(10724.73)=-30.594$, $p<0.001$)---in comparison to authors who stay.

\begin{figure}[H]
\centering
\includegraphics[scale=0.6]{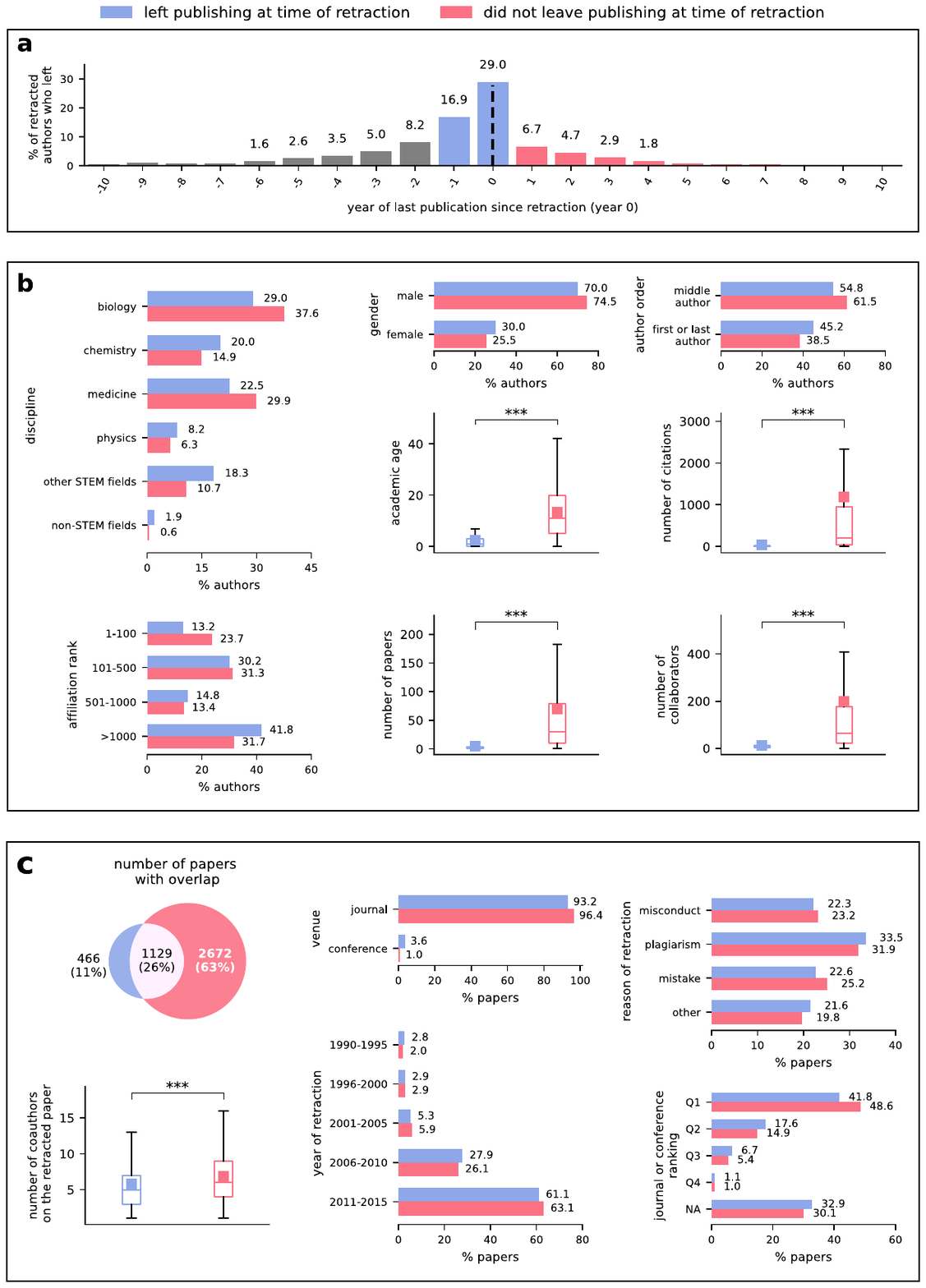}
\caption{\textbf{Characteristics of retracted authors.} \textbf{(a)} shows percentage of retracted authors who left (blue) versus those who did not (red). \textbf{(b)} shows comparisons across different author-level characteristics among retracted authors who left scientific publishing at the time of retraction, and those who have not with $N=12,742$. \textbf{(c)} similar to (b) but for paper-level characteristics with $N=4,267$. The boxes extend from the lower to upper quartile values of the data, with a line at the median; whiskers extend to the minimum and maximum values within 1.5 times the interquartile range (IQR) of the lower and upper quartiles, respectively; means are shown as additional square markers. Normality and homogeneity of variances were tested and not met; thus, $p$-values were calculated using a non-parametric two-sided Mann-Whitney U test. Outliers are removed from box plots for presentation purposes. *** $p<.001$}
\label{mainfig:attrition_exploration}
\end{figure}

Looking at the characteristics of the 4,267 retracted papers authored by the sample above, we create the following two groups: papers with authors who left around the time of retraction (blue in Figure 1c), and papers with authors who left later or had ongoing publishing careers in 2020 (pink in Figure 1c). Note that these two groups have an overlap: papers that had authors from both groups, specifically, 26\% of retracted papers are this sample. Looking at the distributions of these two groups of papers, we find that the majority of them were published in journals rather than conferences and were retracted after 2010 (Figure~\ref{mainfig:attrition_exploration}c). Most of these papers fall within the top quartile in terms of journal ranking (information on journal ranking is unavailable for approximately 31\% of the papers). The reasons for retractions vary, with approximately 23\% attributed to misconduct, 33\% to plagiarism, 24\% to mistakes, and an additional 20\% to other reasons. For further details on establishing author and paper level characteristics, see the Materials and Methods section ``Creating author and paper level features,'' and for the standardized mean differences and statistical comparisons between the two groups, see Supplementary Table 1.

These descriptive observations suggest that, despite some variability based on author-level characteristics, the careers of authors with retracted publications tend to be cut short, with authors' exit from publishing often occurring around the time of retraction. To confirm this, we create a comparison group for 2,743 retracted authors by matching them with a non-retracted author as leaving scientific publishing can be a result of many reasons other than retraction. We match non-retracted authors exactly on gender, affiliation rank, discipline, number of publications, and number of collaborators at the start of their publishing careers (i.e., in the first year of publishing). We also ensure that both the retracted author and their match had similar careers up to the time of the publication of their retracted paper. This is achieved by confirming that both authors published a paper in the year the retracted paper was released and that their affiliation rank and discipline were the same at that time. Additionally, we ensure that during that period, the matches are similar in terms of the number of collaborators and publications. This process results in a sample of 2,743 retracted authors with suitable matches. Figure 2a visualizes the difference in publishing career length of the retracted authors and their matched counterparts, confirming that retracted authors do leave publishing earlier. For the standardized mean differences and statistical comparisons between our matched sample and filtered sample, see Supplementary Table 2.

Next, we investigate the relationship between attrition and the amount of attention received by the retraction. In particular, heightened levels of attention may bring authors into the spotlight, potentially influencing how the broader scientific community, including individuals who may not have been previously familiar with their scholarly work, perceives them. To probe this relationship, we use a third data set, Altmetric, a database of online mentions of publications, containing a record of more than 191 million mentions for over 35 million research outputs that we merge with RW (see ``Merging RW with Altmetric'' in Materials and Methods). We measure attention using the Altmetric score, tracking it in the six-month period before and after the retraction event as per \cite{peng2022dynamics, abhari2022twitter} (see ``Calculating the Altmetric score'' in Materials and Methods for the calculation of the Altmetric score). In Figure 2b, we present the distribution of the logged average attention received by the retracted papers in the filtered sample during this time window, highlighting that attention peaks during the month of retraction. For a breakdown by social media, news media, blogs, and knowledge repositories see Supplementary Figure 6. We find that while most retracted papers receive no attention, some gain worldwide publicity. More specifically, 64\% of retracted papers receive no attention during their life course, which increases to 75.4\% when considering our time window (papers without attention are not shown in Figure 2b). 
Furthermore, in Figure 2c, we display the gap between the proportion of authors who left publishing among the retracted authors and their matches whom we identified when creating the baseline in Figure 2a. We present this gap at different cutoffs of attention as measured by the Altmetric score. This gap illustrates the possible impact of attention on the likelihood of authors leaving publishing. We find that this gap increases with attention in this sample, suggesting that retracted authors leave even earlier when receiving more attention compared to their matched counterparts. 

We corroborate the result that retractions are a key factor in attrition from publishing using a Cox proportional hazard model estimated on all retracted authors, including those who have not been matched in the previous analysis. In this model, authors leave their publication career as defined above, or are censored in 2020. We include yearly observations for each author from the start of their publishing careers. We control for several time-invariant factors, including gender and cohort, as well as time-varying factors, such as affiliation rank and discipline which are updated annually based on meta-data from authors' publications in MAG. Experience is represented as the cumulative counts of publications, citations, and collaborators, also calculated from MAG. Importantly, we introduce a binary variable ``retracted,'' which remains zero until the year of retraction and switches to one thereafter. Table 1 displays the results from this model, and confirms that experiencing a retraction does indeed precipitate authors' exit from scientific publishing. We perform several robustness analyses using this model. First, to complement literature such as \cite{Lissoni_etal_2013}, which underscores distinctions in author contributions to research articles based on authorship order, we show that the association between retraction and leaving publishing is slightly stronger for first and last authors compared to all authors (descriptively), see Supplementary Table 3. Second, we include authors with multiple retractions whose retractions cluster in a single year, so the model assumption of considering retraction as a single event is most likely to be met and find substantively similar results (see Supplementary Table 4). Most authors with multiple retractions have them clustered within a single year (see Supplementary Figure 7), i.e.,  the majority of authors with multiple retractions are included here.

While retraction may not necessarily result in an author leaving scientific publishing, it can still impact career progression by affecting an author's reputation. Therefore, we extend our analysis to examine how the retraction of a scientific publication influences an author's career, specifically for those who remain in scientific publishing post-retraction. We focus on three main outcomes: (i) the number of collaborators retained post-retraction, (ii) the number of new collaborators gained post-retraction, and (iii) the share of open triads closed (i.e., when an author coauthors with someone who previously worked with their coauthor \cite{Shi_etal_2015}). We perform another matching experiment similar to creating a baseline, pairing retracted authors with ongoing post-retraction publishing careers to comparable non-retracted authors who are not part of their collaboration network. We explicitly exclude all collaborators, not just past ones, to eliminate any negative spillover effects that retractions may cause, as studied in \cite{jin2019reverse} (note that this is a different approach to \cite{Sharma_Mukherjee_2024}). In the matching process, we ensure that each non-retracted author identified as a match shares the following exact characteristics with the retracted author: gender, academic age, affiliation rank at the start of their career, and affiliation rank and scientific discipline at the time of retraction. Additionally, matches were similar (i.e., within 30\% of the values of the retracted author) in terms of the number of publications, citations, and collaborators at the time of retraction, with the closest matches selected based on a theoretically calibrated distance function (see the ``Analytical sample for the matching experiment'' subsection 
in Materials
and Methods). To focus on post-retraction career trajectories, we ensure that both retracted authors and their matches have published at least one paper post-retraction, and we evaluate outcomes in the 5 years after the retraction. This approach aligns the careers of retracted authors with their non-retracted matches, avoids survival bias \cite{Brownetal_2015}, and ensures that all authors have the same amount of time to accumulate and retain collaborators. The matching experiment results in the matching of 2,348 retracted authors, each matched to an average of 1.73 non-retracted authors. For an evaluation of the matched sample, see the ``Analytical sample for the matching experiments'' subsection in Materials and Methods, for more details on calculations and binning decisions, refer to Supplementary Note 2, and for the standardized mean differences and statistical comparisons between authors who stay in academic publishing and those matched, see Supplementary Table 5.

Figure 3 summarizes our findings from the matching experiment. We compare retracted authors to their non-retracted counterparts by averaging across all closest matches when multiple matches are present. We find that authors who have experienced a retraction in their careers tend to gain a significantly higher number of new collaborators and retain significantly more of their previous collaborators, as illustrated in Figures 3a--b. These results are based on two-sided Welch's t-tests to account for unequal variances, and are corroborated using Kolmogorov-Smirnov and Wicoxon signed-rank tests, as detailed in Supplementary Tables 6-8. We also find that these results largely and consistently hold across various factors, including gender, year of retraction, author order on the retracted paper, reason for retraction (mistake, plagiarism, misconduct), type of retraction (author-led vs.\ journal-led), attention (high vs.\ low), discipline, affiliation rank, and the time between the retracted paper's publication and retraction. We find substantively similar differences between the retracted authors and their matches when restricting matches to a 20\% difference in terms of the number of papers, collaborators, and citations prior to retraction, and the same direction using an even stricter restriction (a 10\% difference), see Supplementary Figures 8-9. Furthermore, when looking at only the first and last authors and their matches, and only at retracted authors and matches whose affiliations are in the same country, the results hold (Supplementary Figures 10-11). We do not find credible evidence that retracted and non-retracted authors close a different proportion of open triads with authors who had previously coauthored with their past collaborators, see Figure 3c. 

It is evident that authors who survive a retraction tend to maintain, on average, a greater number of previous collaborators and gain a higher number of new collaborators. However, it is essential to examine the characteristics of these retained and newly formed relationships among retracted authors in comparison to their matched counterparts. Therefore, in our next analysis (Figure 4), we focus on three key characteristics: the academic age (seniority), number of papers (productivity), and number of citations (impact) of the collaborators of both retracted authors and their matched non-retracted counterparts. Additionally, we analyze these characteristics across different groups. Specifically, we consider early-career (0-3 years of experience at the time of retraction), mid-career (4-9 years of experience), and senior (10 or more years of experience) authors separately. We also examine different reasons for retraction---misconduct, plagiarism, and mistake---as well as the level of attention the retraction received (high vs.~low).

Overall, our findings indicate that, although retracted authors have larger collaboration networks, in terms of the seniority and productivity of their retained collaborators, these networks are qualitatively weaker. Importantly, these differences are not a result of differences in the quality of collaboration networks prior to retraction. While these characteristics are not ones we match on, as this would further decrease the sample of retracted authors with suitable matches, we demonstrate that the distribution of these characteristics prior to retraction are the same in terms of seniority and productivity, and are closely similar in terms of impact (Supplementary Figure 12). As for newly gained collaborators, we do not find credible evidence for differences in this respect but retracted authors gain more impactful collaborators overall. 

The direction of these differences generally persists across career stages, but not all differences are significant at the usual levels. We find that early-career retracted authors develop qualitatively different collaboration networks post-retraction compared to their matched counterparts, experiencing a significant loss. Namely, although the collaborators they retain are not less impactful (no credible evidence of a difference), they do retain less senior and less productive collaborators compared to their matched counterparts (Figures 4a-c and Supplementary Table 9). We also find that even senior authors retain less senior collaborators post-retraction compared to their matched pairs (Figure 4a). In terms of collaborators gained, we find that senior retracted authors are not affected when it comes to the age of their collaborators (no credible evidence of a difference), but gain significantly more productive and more impactful collaborators (Figures 4d-f and Supplementary Table 10). All these results are based on two-sided Welch's t-tests to account for unequal variances. 

We observe little heterogeneity in the direction of the differences across retraction reasons; see Figures 4j-o and Supplementary Tables 12-13. Authors retracted due to mistakes do not develop qualitatively different collaboration networks compared to their matched counterparts in terms of the seniority, productivity, and impact of their retained and new collaborators. This lack of credible evidence for these differences suggests that mistakes might be easier to overcome than misconduct or plagiarism (as in those cases the differences in terms of seniority are statistically significant), keeping sample size constraints in mind. In case of misconduct, we find statistically significant differences (based on two-sided Welch's t-test) in the productivity and impact of new collaborators, indicating that those retracted for misconduct build stronger, not weaker, new networks in this regard compared to their matched pairs. Finally, to complement our previous analysis on attention, we observe that authors whose retracted papers receive a high level of attention (Altmetric score $>$ 10) build significantly worse networks in terms of seniority and impact of their retained collaborators. In other words, these retracted authors retain collaborators who are less senior and less impactful compared to their matched counterparts; see Supplementary Figure 13a--f and Supplementary Tables 15 and 16.

We also perform a difference-in-difference analysis to contrast retracted authors and their matches, examining whether the retained collaborators are qualitatively different from those who were lost (Figures 4g--i and p--r, Supplementary Figure 13g--i and Supplementary Tables 11, 14, and 17). This analysis helps determine if, for example, while retracted authors retain more collaborators, those retained are less senior, less productive, and less impactful compared to those they lose relative to their matched counterparts. It is important to note that the initial distribution of all pre-retraction collaborators are identical or very similar (Supplementary Figure 12). Authors who did not retain or lose any collaborators are excluded from this analysis, as meaningful comparisons cannot be made. For each retracted and matched author, we calculate the average of the variable of interest for both retained and lost collaborators and then compute the difference by subtracting the average for lost collaborators from the average for retained collaborators. Positive differences indicate that retained collaborators were more senior, more productive, and more impactful than the ones lost. Overall, this analysis reveals a difference in terms of impact: retracted authors experience a greater relative loss in the impact of their collaborators compared to their matched counterparts. No credible evidence is found for the difference in differences in terms of seniority and productivity (although the differences are in the same direction as impact). The direction of the difference in differences we observe is consistent across career stages as well as for retraction reasons. However, as previously discussed, these differences are generally not statistically significant. Finally, as a robustness check, we repeated the analysis focusing only on retracted authors who were first or last authors. Although the reduced sample size resulted in a loss of statistical significance, we found that the direction of the differences, while not significant in most cases, remained largely consistent (see Supplementary Figure 14). 

In addition to examining these qualitative differences, we explore whether retracted authors develop collaboration networks with collaborators who are more physically distant post-retraction compared to pre-retraction, and how this compares to their matched counterparts. We also analyze whether this shift is evident in the fields they publish in post-retraction compared to pre-retraction (see Supplementary Note 3 for details on the distance calculation). Our findings, as illustrated in Supplementary Figure 16, indicate that both groups expand their collaboration networks with more distant collaborators over time. This trend is likely influenced by authors and their collaborators moving between institutions. Specifically, we observe that on average, post-retraction collaborators of retracted authors are approximately 30 kilometers more distant compared to those of their matches. However, we do not consider this difference to be particularly meaningful given the global scale of science production. In terms of disciplinary focus, our analysis reveals that both retracted authors and their matched counterparts undergo a divergence in the fields they publish in over time. Interestingly, this shift is strikingly similar for both retracted and non-retracted authors, suggesting that retracted authors do not move from their original fields in a manner different from their non-retracted counterparts post-retraction.

In summary, the career impact of retractions may not be fully understood without considering attrition, i.e., leaving scientific careers where publishing is integral to success. We find that experiencing a retraction is associated with an earlier exit from publishing. However, the results of the matching experiment suggest that retracted authors who continue to publish do not suffer a reduction in the size of their collaboration networks. On average, these authors actually build larger collaboration networks compared to their counterparts by retaining more collaborators and establishing more new collaborations. This increase in size, however, is accompanied by qualitative differences: retracted authors tend to retain collaborators who are less senior, less productive, and less impactful, which is balenced by gaining more impactful collaborators. We do not find significant and consistent heterogeneity by career stage and reason for retraction, though the relatively smaller and statistically insignificant difference in cases of mistakes may suggest that mistakes are easier to overcome compared to misconduct or plagiarism.

\section*{Discussion}

Retractions can have significant consequences for authors' careers, leading to their departure from scientific publishing. In this study, we conducted an analysis utilizing data from Retraction Watch, Microsoft Academic Graph and Altmetric, identifying and examining an analytical sample of around 4,500 retracted papers involving over 14,500 authors. Our findings reveal that: 1) around 45.9\% of authors left their publishing careers around the time of retraction, 2) authors who left exhibited shorter pre-retraction careers, had fewer citations, collaborators, and publications compared to those who stayed, 3) higher attention precipitated retracted authors' exit compared to their similar counterparts, 4) retracted authors who stayed post-retraction formed larger collaboration networks, retaining more collaborators and gaining more new ones, and 5) overall they built qualitatively weaker collaboration networks in terms of their coauthors seniority, productivity whom they retained, but gained mode impactful new collaborators compared to their similar counterparts.

It is important to acknowledge that our study is not without limitations. First, online attention, as measured by the Altmetric score, captures the volume rather than the quality of attention and lacks a nuanced description of its specific sources beyond categories of platforms. Additionally, the score fails to reveal how relevant the coverage may be for retracted authors, which, if explored, could yield further insights into our findings. Moreover, retained and new collaborators may be qualitatively different beyond the aspects we explore. Second, our matching analysis reveals that retracted authors matched to similar non-retracted counterparts were on average, more junior compared to the average retracted author both when constructing matches for the baseline and for retracted authors who published after retraction. Therefore, it is possible that our estimates represent a lower bound of a difference, assuming that more senior authors possess greater resources to further develop their networks. Conversely, they may also represent an upper bound, assuming that more established authors receive less benefit of the doubt when assessing their culpability compared to their early-career counterparts. Third, throughout we document average effects which might mask tremendous author-level heterogeneity that we are unable to explore, including their workplace (academic and/or research intuition or industry) and other factors, e.g., their teaching load and service obligations. Fourth, in addition to these characteristics, retracted authors also vary in their awareness of the issues about their paper that led to the retraction, and the actions that colleagues and mentors might have taken to shield (or chastise) retracted authors. Future work may incorporate additional author-level detail that allows authors on retracted papers to be treated differently depending on their involvement behind the reason of retraction. Some prior work has leveraged retraction notices in smaller samples in this way, which reveal such heterogeneity most of the time in terms of misconduct, but not in terms of mistakes  \cite{Mongeon_Lariviere_2016}. Fifth, while we incorporated some authors with multiple retractions in robustness analyses, our modeling assumptions for this study treated retraction as a single event. Future work should integrate better multiple retractions to see if they have compound linear or non-linear effects.

Some considerations fall outside of the scope of the paper. For instance, self-retraction might signal integrity, which could be a factor contributing to why some retracted authors develop larger collaboration networks. It is possible that these authors become more cautious in their future endeavors to avoid a second retraction, making them desirable collaborators. It is also plausible that they change how agentically they search for new collaborators and cultivate already existing relationships to compensate for the assumed (and empirically documented) negative impact of retractions. The role of the scientific community is similarly under-explored. Specifically, some retracted authors might receive support from their colleagues, presumably in cases when their papers are retracted for mistakes, or instances when retractions are viewed as ``honorable'' or the ``right thing to do.'' Bringing these authors on as collaborators may be one way to show them support. These assertions, we hope, would form the starting point of future work.

Our study serves as an initial step in documenting how important institutions of science, such as retractions that serve a key role in policing the content of the canon, impact the careers of scientists. Future research should complement our work by exploring how authors navigate retractions and the micro-mechanisms underlying the strategies employed by retracted and non-retracted authors when seeking collaborators. It would also be valuable to investigate whether retracted and non-retracted scientists are sought after for similar opportunities, particularly when retracted authors' work was not retracted due to misconduct. Furthermore, the role of online attention in these matters also deserves further exploration, as it becomes intertwined with the names of authors whose work is discussed, extending beyond the scientific content of papers and encompassing a broader set of issues.

\section*{Materials and Methods}

\subsection*{Data sources} 

The analyses presented in this paper rely on three  data sets: 
\begin{enumerate}
 \item \emph{Retraction Watch (RW)} \cite{retractionwatch} is the largest publicly available data set of retracted articles, obtained on the 18$^{\mathrm{th}}$ of May 2021. At the time it contained 26,504 retracted papers published in 5,844 journals and conferences. The earliest publication record goes back to the year 1753, whereas the latest record is in 2021. The data set consists of articles classified by a combination of 104 reasons for retraction.
 
 \item \emph{Microsoft Academic Graph (MAG)} \cite{wang2020microsoft} is one of the largest data sets for scientific publication records. We collected this data set on the 30$^{\mathrm{th}}$ of July 2021. It contained, at the time, approximately 263 million publications, authored by approximately 271.5 million authors, with the earliest publication record in 1800.

 \item \emph{Altmetric} \cite{adie2013altmetric} is a database of online mentions of publications. It contains a record of more than 191 million mentions for over 35 million research outputs. It uses unique identifiers (e.g., Digital Object Identifier or DOI, and PubMedID) to match attention to research across several social media platforms, blogs, news sites, and knowledge repositories. 
\end{enumerate}

\subsection*{Merging RW and MAG} 

To merge MAG and RW we are using a two-step approach. Merging these two data sets takes place after the filtering of RW by eliminating bulk retractions and retractions prior to 1990 and post-2015. As both MAG and RW provide the DOI of the publication record, we start with these identifiers, as it is a persistent identifier unique to each document on the web. However, as not all records in MAG and RW have a DOI, we also identify papers in RW and MAG with the exact same title. Out of 6,704 in RW after filtering, we merge 2,646 papers to MAG. In order to increase the size of our data set we can analyze, we merge the rest of the publication titles in RW using fuzzymatcher~\cite{fuzzymatcher} which employs probabilistic record linkage \cite{sayers2016probabilistic}, to find similar titles based on Levenshtein distance. We validate the robustness of our fuzzy matching, by randomly sampling 100 retracted papers and manually checking the accuracy of the merge. Out of 100 sampled papers, 99 in RW were linked to the correct entry in MAG. As a result of this second step, we additionally merge 3,542 retractions resulting in a total of 6,188 (92\%) retracted papers in RW linked to their corresponding entry in MAG. Lastly, we filter papers with authors with multiple retractions and data with missing fields (see Supplementary Note 1). This resulted in the final ``filtered'' sample. 

\subsection*{Creating author and paper level features}

We use RW and MAG to create features for authors and papers. Here, we discuss the features that require additional data collection and calculations, such as gender, scientific discipline, type of retraction, and venue ranking. 

\noindent \textbf{Gender.} To identify the perceived gender of authors, we use \href{https://genderize.io/}{Genderize.io} to map author first names to gender. Genderize.io returns the \emph{probability} indicating the certainty of the assigned gender. We exclude all authors whose gender could not be identified with $>0.5$ probability. We validate the name-based gender identified by Genderize.io by comparing the agreement (or concordance) of its labels against another classifier, Ethnea. Ethnea is a name-based gender and ethnicity classifier specifically designed for bibliographic records. We compare the labels of 31,907 authors in RW for whom ``male'' or ``female'' labels were available using both Genderize.io and Ethnea. We find that the assignment of these labels agreed for 31,028 (i.e., 97\%) retracted authors with a Cohen's $\kappa$ score of 0.93 showing an almost perfect level of agreement \cite{mchugh2012interrater}. Our approach is in line with prior research which uses similar name-based gender classifiers \cite{alshebli2018preeminence, lee2019homophily, squazzoni2021peer, morgan2021unequal, liu2022gender, peng2020author}, however, automated classifiers, such as the ones here have significant shortcomings. They do not rely on self-identification, and therefore could misgender authors. Annotations are performed on the basis of historical name–gender associations to assign male or female to an author, recognizing that there are expansive identities beyond this limiting binary that our approach can not explore.

\noindent \textbf{Scientific discipline.} To assign a scientific discipline to every author, we utilize the fields in MAG that span more than 520,000 hierarchically structured fields. For every paper $p$ and field $f$, MAG specifies a confidence score, denoted by $score(p,f) \in [0,1]$, which indicates the level of confidence that $p$ is associated with $f$. The aforementioned hierarchy contains 19 top-level fields, which we refer to as ``disciplines.'' These fields are Art, Biology, Business, Chemistry, Computer Science, Economics, Engineering, Environmental Science, Geography, Geology, History , Materials Science, Mathematics, Medicine, Philosophy, Physics, Political Science, Psychology, and Sociology. Almost every field, $f$, has at least one ancestor that is a discipline. Let $D(f)$ denote the set of all disciplines that are ancestors of $f$. For any given paper, $p$, the set of disciplines associated with $p$ is denoted by $D(p)$, and is computed as follows:
\begin{equation*}
D(p) = \Big\{D(f): f\in \underset{f}{\operatorname{argmax}} \text{ } \mathit{score}(p, f)\Big\}.
\end{equation*}

For any given author, $a$, let $P(a)$ denote the set of papers authored by $a$. We compute the discipline(s) of $a$ as follows:
\begin{equation*}
D(a) = \Big\{f\in \underset{f}{\operatorname{argmax}} \big|\big\{p\in P(a): D(p) = f\big\}\big| \Big\},
\end{equation*} 

where $\left|\cdot\right|$ denotes the set cardinality operator. In other words, $a$ is associated with the most frequent discipline(s) amongst all papers authored by $a$ up to and including the retraction year. 

\noindent \textbf{Retraction reasons.} In order to identify the reason for retraction, we manually extracted the retraction notes of 1,250 retracted papers. The reasons for retraction can be classified into four broad categories: (i) misconduct; (ii) plagiarism (note that some prior research considers plagiarism as misconduct, for example \cite{gaudino2021trends}); (iii) mistake; and (iv) other. Every retraction note was evaluated by multiple annotators. We started with two annotators and assigned additional annotators up to 5 until a majority reason was reached. If a majority reason was not reached, the reason was classified as ``ambiguous.'' Finally, if no reason was provided for the retraction, then it was classified as ``unknown.'' The final distribution of the reasons for retraction of the annotated papers is: 251 (20\%) misconduct, 311 (25\%) plagiarism, 347 (28\%) mistake, 170 (14\%) other, 121 (10\%) unknown, and 50 (4\%) ambiguous. Note that the reason ``plagiarism'' includes plagiarising others' work, as well as prior work by the authors. Based on a random sample of 100 retraction notes, 50 referred to taking someone else's work without proper reference, 30 referred to lacking citations or quotes from the authors' own work, and 20 did not include information whose work has been plagiarized. Therefore, 30\%-50\% of this category is self-plagiarism.

We use the manually annotated papers to automatically code reasons of retraction for the rest of the papers in RW. We do so using a label propagation algorithm. There are 104 unique reasons for retractions provided in RW. Each retracted paper is associated with one or more of these reasons. We map the 104 reasons to a majority coarser class of plagiarism, misconduct, mistake, and other in our analysis. Then we use this mapping to annotate the rest of the papers without labels using the majority class (see Supplementary Figure 17 for a more detailed visualization of the label propagation algorithm). The final distribution of the reasons for retraction after label propagation for the filtered sample is as follows: 1106 (24\%) misconduct, 1513 (33\%) plagiarism, 1078 (24\%) mistake, 498 (11\%) unknown, 334 (7\%) other, and 49 (1\%) ambiguous. These numbers are comparable to those reported about a decade ago in a sample of papers indexed in PubMed \cite{Fang_etal_2012}. In our analysis, we merge all three of the other, unknown, and ambiguous categories in the ``other'' category.

\noindent \textbf{Type of retraction.} Using the manually extracted retraction notes, we also identify whether the retraction was author-led or journal-led. The breakdown of the different types of retractions is as follows: 604 (48\%) author-led, 499 (40\%) journal-led, 119 (10\%) unknown, and 28 (2\%) ambiguous. These data are only available for the manually annotated papers.

\noindent \textbf{Journal and conference ranking.} 
To identify the ranking for the venue (journal or conference) of retracted papers, we utilize the database of SCImago journal rankings (SJR)~\cite{scimagoJournalRank}. For a given journal or conference year, the SJR score is computed as the average number of weighted citations received by the articles published in the venue during the past three years \cite{guerrero2012further}. Based on this score, for each subject area, a quartile is also assigned to each journal. SJR provides rankings from 1999 to 2020. We use the year of publication of the retracted article to identify the SJR score and quartile ranking of the venue. Out of the 4,578 papers and 14,579 authors in the filtered sample, we were able to identify the rankings for 3,129 (68.3\%) papers and 10,379 (71.2\%) authors. Note that papers prior to 1999 do not have this information, nor do papers whose venues were not featured in SJR.

\subsection*{Merging RW and Altmetric}

For each paper in RW, we use the associated DOI or the PubMedID to query the Altmetric API. Out of the 26,504 retracted papers in RW, we are able to identify 11,265 (42.5\%) papers with online presence based on their unique identifiers. There are 15,239 papers for which an Altmetric entry could not be located, however, these papers and their respective authors are also part of our analysis. 

\subsection*{Calculating the Altmetric score} 

The Altmetric score is a weighted count of the attention a research output receives from different online sources (e.g., Twitter (now X), news, etc.). The Altmetric API, however, only provides the current cumulative Altmetric score for a given record, and does not give the breakdown or a customized score for a given time window. Since we focus on the 6 months before and after retraction to isolate the attention that the retraction likely garnered, we compute this score using the methodology detailed by Altmetric on their webpage~\cite{altmetricScoreCalculation}. While the algorithm Altmetric uses to compute its score is proprietary \cite{trueger2015altmetric}, the description allows us to closely estimate it. We compute the Spearman correlation of the available cumulative Altmetric score against our computed score for the complete time window using our methodology. For the 11,265 retracted papers for which an Altmetric record (and score) is available, the Spearman correlation is high ($\rho=0.96$). We compute the Altmetric score for both the retracted paper and its respective retraction note (most retraction notes and papers have separate DOIs) and aggregate these by taking their sum. We assign papers that are not indexed in Altmetric an attention score of zero. We validate this choice by randomly sampling 100 of these papers and manually searching for mentions of them on Google and on Twitter using the title and the DOI of the paper. Out of this 100, 96 have not received any attention. The remaining papers only garnered one mention on average. For this reason, we treat the papers that do not have mentions on Altmetric as having an attention score of zero.

\subsection*{Analytical sample for the matching experiments}

In this paper we report the results of two matching experiments. First, we create a baseline of attrition of authors comparable to retracted authors. Second, we create a comparable set of authors  to retracted authors with post-retraction careers. The procedures we employ in both cases are similar, except that in the second case we create matched sets using more confounders. For this reason, we describe our process in detail for the creation of this second sample only to reduce redundancy. 

We use our filtered sample of 4,578 papers and 14,579 authors to generate the analytical sample for the matching experiment. Of the 14,579 authors, we find suitable matches for 2,348 authors (16.1\%). These matches are established on the basis of a three-step process. First we perform exact matching on gender, academic age, affiliation rank at the start of the career, as well as affiliation rank and scientific discipline at the time of retraction. Second, we pick a threshold (30\%) within which we accept matches on the remaining characteristics: number of papers, number of collaborators, and number of citations pre-retraction; i.e., these characteristics of the match ought to be within 30\% of the same for the retracted author. Third, we use a calibrated distance function to achieve balance on the three latter characteristics, giving more weight to the number of pre-retraction collaborators, our most important confounder. Specifically, we identify the closest match for each author using the lowest weighted Euclidean distance that minimizes the standardized mean difference for the number of papers, the number of citations, and the number of collaborators between the author and the match, calculated over the set of potential matches identified in the second step. We repeat these steps using a 20\% and a 10\% threshold, and present robustness analyses with this threshold in Supplementary Figures 8-9.

For each author $a$, let $p_a$, $c_a$ and $o_a$ denote the standardized number of papers, the number of citations and the number of collaborators of $a$ by the year of retraction. We choose the closest match $m$ for each author by minimizing the following distance function:

\begin{equation*}
    \mathcal{D} = \sqrt{w_{papers} (p_a-p_m)^2 + w_{citations} (c_a-c_m)^2 + w_{collaborators} (o_a-o_m)^2},
\end{equation*}

where $w_{papers} = 0.1$, $w_{citations}=0.1$, and $w_{collaborators}=0.8$, denote the weights determined empirically by minimizing the standardized mean differences.

If the collaboration year of a collaborator is missing, we cannot place them on the authors' career timeline. In other words, we cannot identify whether that collaboration occurred pre- or post-retraction. As such, in the case of a missing collaboration year, we remove the author and their corresponding matches from our analysis altogether. We carry out our analysis on retracted authors who authored at least one paper post their retraction year in the 5-year window following retraction. All matched authors meet the same criteria. 

The ``matched sample'' of these authors who stayed in scientific publishing is different from those in the filtered sample in the following important ways: the matched sample is younger, has fewer papers, fewer citations and fewer collaborators on average, and slightly more likely to be a middle author, also more likely to be in institutions ranked 101-1000 (see Supplementary Figure 5). In sum, the matched authors are lower status, on average, compared to the non-matched filtered sample. These differences are essential to consider when evaluating our inferences. We calculate standardized mean differences (SMDs) \cite{flury1986standard} for our matched sample and find that these values are 0.036, 0.017, 0.053, respectively for the number of papers, citations, and collaborators. For the rest of the characteristics the SMDs are 0, as the matches are exact. These statistics give us confidence that retracted authors are matched to non-retracted authors with similar career trajectories up to the time of retraction. 

\section*{Data Availability}

The Microsoft Academic Graph (MAG) dataset can be downloaded from the following \textcolor{blue}{\href{https://www.microsoft.com/en-us/research/project/microsoft-academic-graph/}{website}}. The Retraction Watch (RW) database can be accessed from their \textcolor{blue}{\href{https://retractionwatch.com/}{website}}. Access to Altmetric API can be requested from their \textcolor{blue}{\href{https://www.altmetric.com/solutions/altmetric-api/}{website}}. The processed data necessary to reproduce main plots and statistical analyses are freely available at \href{https://github.com/samemon/retraction_effects_on_academic_careers}{here}.

\section*{Code Availability}
The code necessary to reproduce the main plots and tables is available for download \textcolor{blue}{\href{https://github.com/samemon/retraction_effects_on_academic_careers}{here}}.

\section*{Acknowledgements}

We gratefully acknowledge support and resources from the High Performance Computing Center at New York University Abu Dhabi. We thank Ivan Oransky and Adam Marcus, co-founders of Retraction Watch \cite{oransky2012retraction}, as well as The Center For Scientific Integrity, the parent organization of RW, for diligently maintaining a curated list of scientific retractions, and making it freely available to researchers. We also thank Altmetric.com and Microsoft Academic Graph for providing the data used in this study. We thank Sarhana Adhikari, Alex Chae, Aasharya Dutt, Rhiane Kall, Danish Khan, Rhythm Kukreja, Ritin Malhotra and Zaeem Shahzad for finding and annotating the retraction notices. We thank Peter Bearman, Sanjeev Goyal, Byungkyu Lee, Fengyuan (Michael) Liu, Minsu Park, Nicholas Weber, and the participants of the Workshop on the Frontiers of Network Science 2023 for thoughtful comments and suggestions.

\section*{Author Contributions Statement}
S.A.M., K.M., and B.A. conceived and designed the study. S.A.M., K.M., and B.A. were responsible for the methodology. S.A.M., and B.A.\ collected and processed the data. S.A.M., K.M., and B.A.\ conducted the analysis. S.A.M., K.M., and B.A.\ did the visualization. K.M., and B.A.\ supervised the study. S.A.M., K.M., and B.A.\ wrote the methods section. S.A.M., K.M., and B.A.\ created the Supplementary Information. K.M., and B.A.\ wrote the main text of the paper.

\section*{Conflict of Interest}
The authors, B.A. and K.M. acknowledge that their study was inspired by a personal experience, experiencing a retraction of one of their papers, which galvanized the research questions asked in this paper, namely, how retractions influence the careers of authors of scientific papers? S.A.M.\ declares no conflicts of interest.
\newpage
\section*{Tables}
\begin{table}[H]
{\fontsize{10.0}{10.0}\selectfont{ 
\caption{\textbf{Cox proportional hazard model~\cite{Cox_1972} of attrition.} In the table below, we present four models. The models differ in how authors' experience is measured using (1) number of papers, (2) logged number of citations, (3) logged number of collaborators, and (4) all experiences put together respectively. All models incorporate clustered standard errors at the author-level. Controls for author's scientific discipline are included as categorical variables, but are not shown.}
\label{maintab:cox_model}
\begin{center}
\begin{tabular}{@{\extracolsep{5pt}}lD{.}{.}{-3} D{.}{.}{-3} D{.}{.}{-3} D{.}{.}{-3} }
\\[-1.8ex]\hline
\hline \\[-1.8ex]
& \multicolumn{4}{c}{\textit{Outcome: Left publishing}} \
\cr \cline{2-5}\\[-1.8ex]
& \multicolumn{1}{c}{\hspace{10pt}(1)} & \multicolumn{1}{c}{\hspace{10pt}(2)} & \multicolumn{1}{c}{\hspace{10pt}(3)}& \multicolumn{1}{c}{\hspace{10pt}(4)} \\
\hline \\[-1.8ex]
\textbf{Retracted} & 1.75\textbf{***} & 1.77\textbf{***} & 1.74\textbf{***} & 1.74\textbf{***}\\
& (0.04) & (0.04) & (0.04) & (0.04)\\
\\
\textbf{Gender} (reference: Male)\\
Female & -0.03 & -0.00 & -0.02 & -0.03\\
& (0.03) & (0.03) & (0.03) & (0.03)\\
\\
\textbf{Affiliation Rank} & 0.00\textbf{**} & 0.00\textbf{**} & 0.00\textbf{*} & 0.00\textbf{*}\\
& (0.00) & (0.00) & (0.00) & (0.00)\\
\\
\textbf{Cohort Year} & 0.02\textbf{***} & 0.02\textbf{***} & 0.03\textbf{***} & 0.03\textbf{***}\\
& (0.00) & (0.00) & (0.00) & (0.00) \\
\\
\textbf{Author's Experience}\\
\# Publications & -0.01\textbf{***} & & & -0.01\textbf{**}\\
& (0.00) & & & (0.00)\\
log(\# Citations) & & -0.09\textbf{***} & & 0.02\\
& & (0.01) & & (0.02)\\
log(\# Collaborators) & & & -0.28\textbf{***} & -0.24\textbf{***}\\
& & &(0.02)& (0.02)\\
\hline \\[-1.8ex]
 Number of events & \multicolumn{1}{c}{\hspace{10pt}5,225 } & \multicolumn{1}{c}{\hspace{10pt}5,225 } & \multicolumn{1}{c}{\hspace{10pt}5,225 } & \multicolumn{1}{c}{\hspace{10pt}5,225 }\\
 Number of subject-periods (n) & \multicolumn{1}{c}{\hspace{10pt}238,925}& \multicolumn{1}{c}{\hspace{10pt}238,925}&\multicolumn{1}{c}{\hspace{10pt}238,925}&\multicolumn{1}{c}{\hspace{10pt}238,925}\\
 %Number of events &  \multicolumn{1}{c}{\hspace{10pt}5,225 } & &\\
\hline
\hline \\[-1.8ex]
\textbf{*}p$<$0.05; \textbf{**}p$<$0.01; \textbf{***}p$<$0.001
\end{tabular}
\end{center}
}}
\end{table}
\newpage
\section*{Figures}

% \vspace{-30cm}
\thispagestyle{empty} % Removes page number only for this page

\begin{figure}[htbp]
  \centering
\includegraphics[scale=0.9]{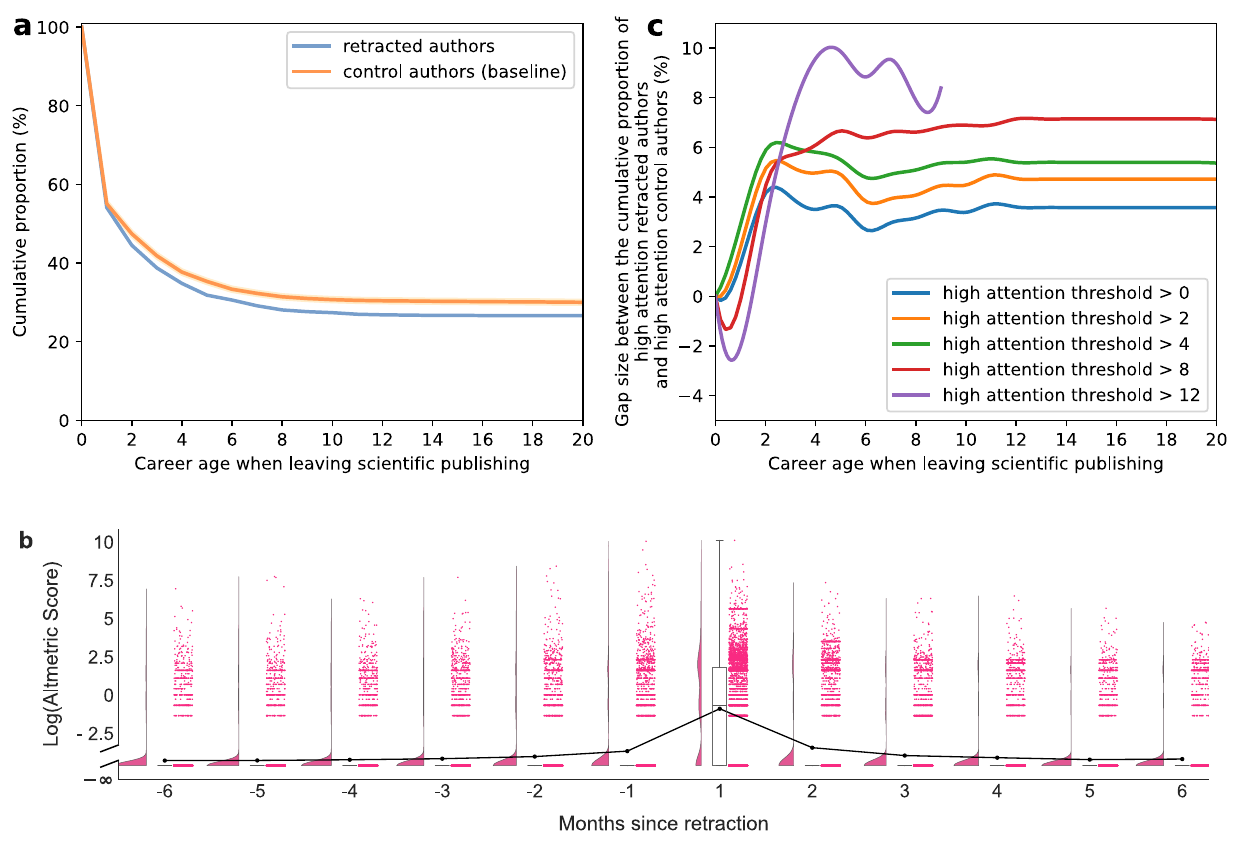}
  \caption{\textbf{Retraction and attrition, and the role of attention}
  \textbf{(a)} Cumulative visualization of the difference in publishing career length between retracted authors with known attrition years and their matched non-retracted counterparts.
  \textbf{(b)} Raincloud plot \cite{allen2019raincloud} showing the distribution of logged Altmetric score 6 months pre- and post-retraction. The x-axis represents monthly windows between the retraction and attention, with 0 being the day of retraction (not displayed), -1 the month right before, and 1 is the month after. The y-axis shows the logged Altmetric score for a paper in the given month. Note, that Altmetric scores [0, 1] are frequent, e.g., 1 tweet results in a score of 0.25. The boxes extend from the lower to upper quartile values of the data, with a line at the median; whiskers extend to the minimum and maximum values within 1.5 times the interquartile range (IQR) of the lower and upper quartiles, respectively. The black trend line represents the average logged Altmetric score.  Total number of papers being plotted is 6,507. Papers with no attention within the 12-month window are excluded (based on \cite{abhari2022twitter}). Comparison across months shows retracted papers receive the most attention within one month of retraction.
  \textbf{(c)} Gap between the cumulative proportion of high attention retracted authors who left publishing and their matched counterparts, shown at different cutoffs of attention as measured by the Altmetric score.
}
\label{mainfig:figure2}
\end{figure}

\begin{figure}
  \centering
  \includegraphics[width=0.99\textwidth,keepaspectratio]{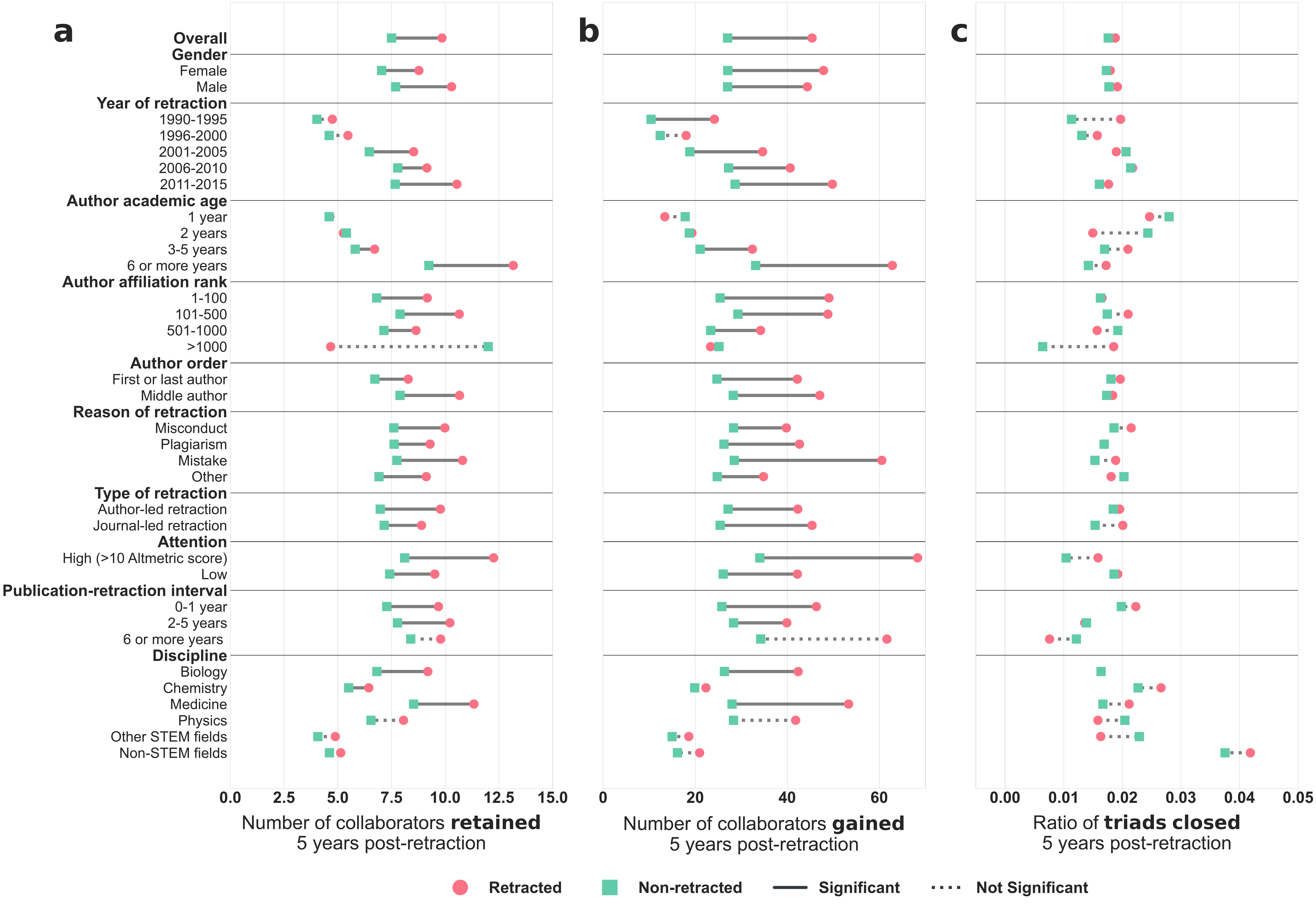}
  \caption{\textbf{Analyzing collaborator retention, gain, and triadic closure among retracted authors who stayed in scientific publishing post-retraction.} The figure shows the difference between \textbf{(a)} the numbers of collaborators retained, \textbf{(b)} the numbers of collaborators gained, and \textbf{(c)} the proportions of triads closed 5 years post-retraction for the ($N=2,348$) authors who were retracted (red circle), and their matched non-retracted pairs (green square). These are further stratified by gender, year of retraction, academic age, author order, reason of retraction, type of retraction, and discipline. Data are presented as mean values. Solid line represents statistically significant difference using two-sided Welch's t-test (assuming unequal variances). Supplementary Tables 6–8 present the 95\% confidence intervals ($CI_{95\%}$), as well as results from additional non-parametric tests. 
}
\label{mainfig:matching}
\end{figure}

\clearpage
\thispagestyle{empty}
\begin{figure}
\centering
\includegraphics[scale=0.65]{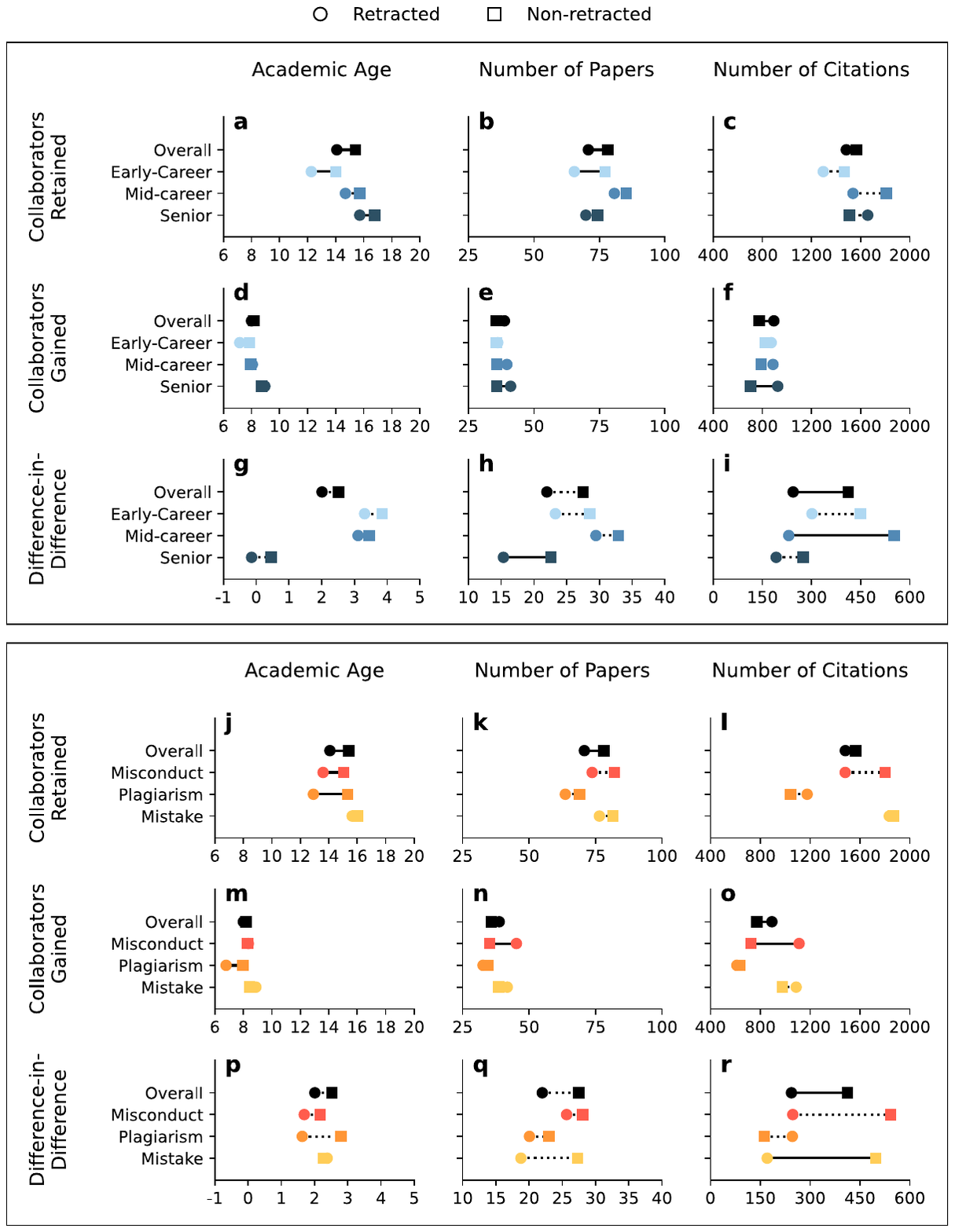}
\caption{\textbf{Comparison of characteristics of collaborators retained and gained by retracted and non-retracted authors, stratified by seniority ($N=1,187$) and by reasons of retraction ($N=1,143$).} \textbf{(a-c)} display the comparison based on academic age, number of papers, and number of citations for retained collaborators of retracted and non-retracted authors of different age groups. \textbf{(d-f)} show the comparison for collaborators gained. \textbf{(g-i)} illustrate the results of the difference-in-difference analysis comparing the difference of collaborators retained and collaborators lost. \textbf{(j-r)} show similar comparison for collaborators of retracted and non-retracted authors stratified by reasons of retraction. }
\label{mainfig:collaborator_exploration}
\end{figure}

\clearpage

\newpage

\clearpage

% \vspace*{-1cm}
% \bibliography{naturebib}
% \bibliographystyle{naturemag}

\clearpage

\part*{Supplementary Material} % <--- Start of the Supplementary Material Part (unnumbered)
\label{part:supplementary_material} % <--- Generate Table of Contents for this Part ONLY

% Corrected code to reset and prefix figures, tables, notes
\renewcommand{\figurename}{}       % Set figurename to empty string - IMPORTANT!
\setcounter{figure}{0}
\renewcommand{\thefigure}{Supplementary Figure \arabic{figure}}

\renewcommand{\tablename}{}       % Set tablename to empty string - IMPORTANT!
\setcounter{table}{0}
\renewcommand{\thetable}{Supplementary Table \arabic{table}}

\section*{Supplementary Figures}
\addcontentsline{toc}{section}{Supplementary Figures}

\begin{figure}[htbp]
\centering
\includegraphics[width=0.95\textwidth,height=\textheight,keepaspectratio]{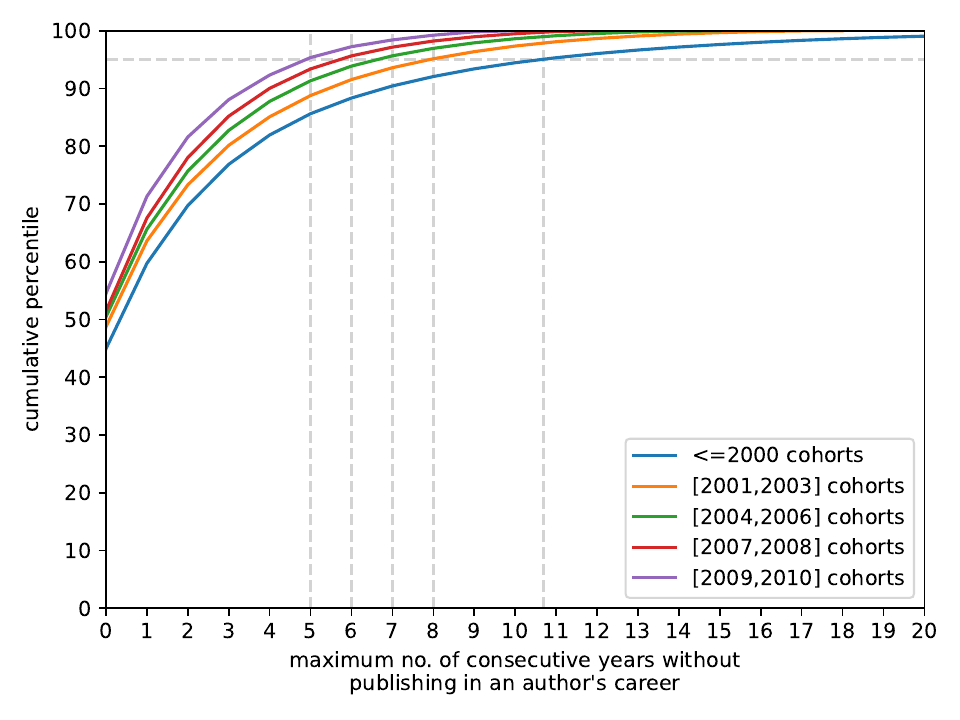}
\caption{
\textbf{Distribution of the longest gap found in the publication career of all STEM scientists in Microsoft Academic Graph up to the 2010 cohort.}
}
\label{supplementaryfig:gap_allmag}
\end{figure}

\begin{figure}[htbp]
\centering
\includegraphics[width=0.95\textwidth,height=\textheight,keepaspectratio]{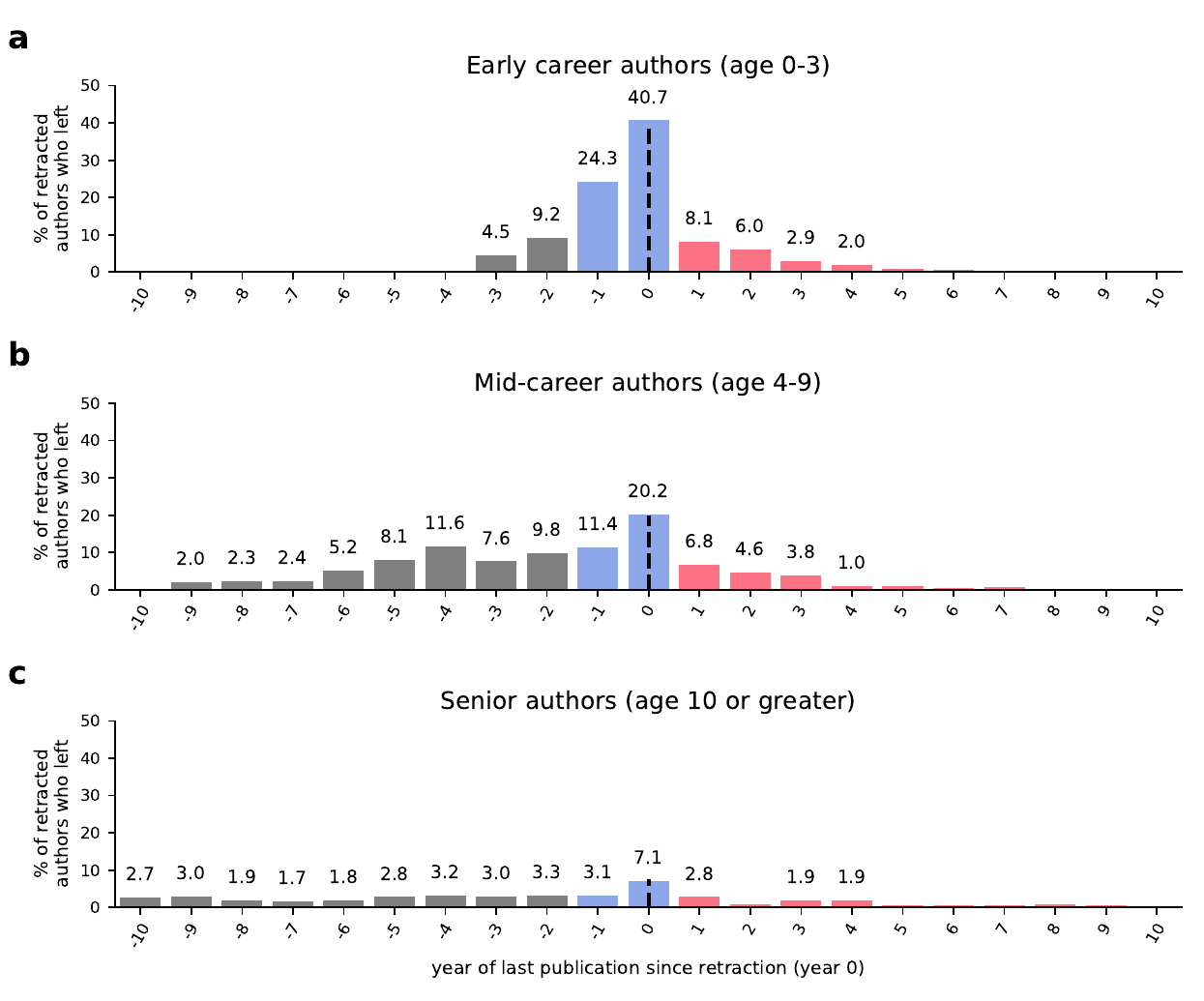}
\caption{
\textbf{Percentage of retracted authors who left (blue)
versus those who did not (red) stratified by age.} \textbf{(a)} shows early-career authors (age 0-3), \textbf{(b)} shows mid-career authors (age 4-9), and \textbf{(c)} shows senior authors (age $\geq$ 10).
}
\label{supplementaryfig:attrition_barplot_byAge}
\end{figure}

\begin{figure}[htbp]
\centering
\includegraphics[width=0.95\textwidth,height=\textheight,keepaspectratio]{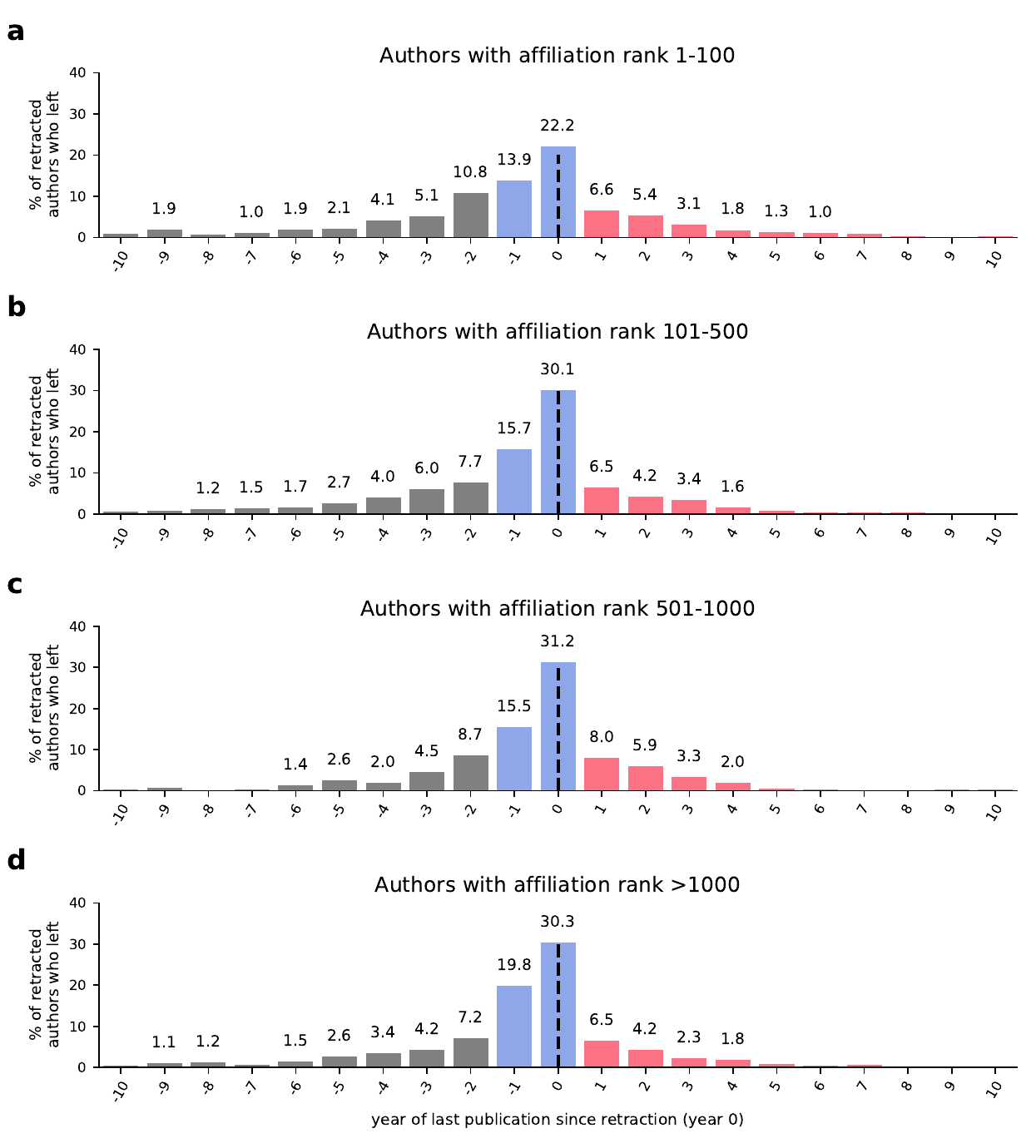}
\caption{
\textbf{Percentage of retracted authors who left (blue)
versus those who did not (red) stratified by affiliation rank at the time of retraction.} \textbf{(a)} shows authors with affiliation rank between 1 and 100, \textbf{(b)} with affiliation rank between 101 and 500, \textbf{(c)} with affiliation rank between 501 and 1000, and  \textbf{(d)} with affiliation rank greater than 1000.
}
\label{supplementaryfig:attrition_barplot_byAffRank}
\end{figure}

\begin{figure}[htbp]
\centering
\includegraphics[width=0.95\textwidth,height=\textheight,keepaspectratio]{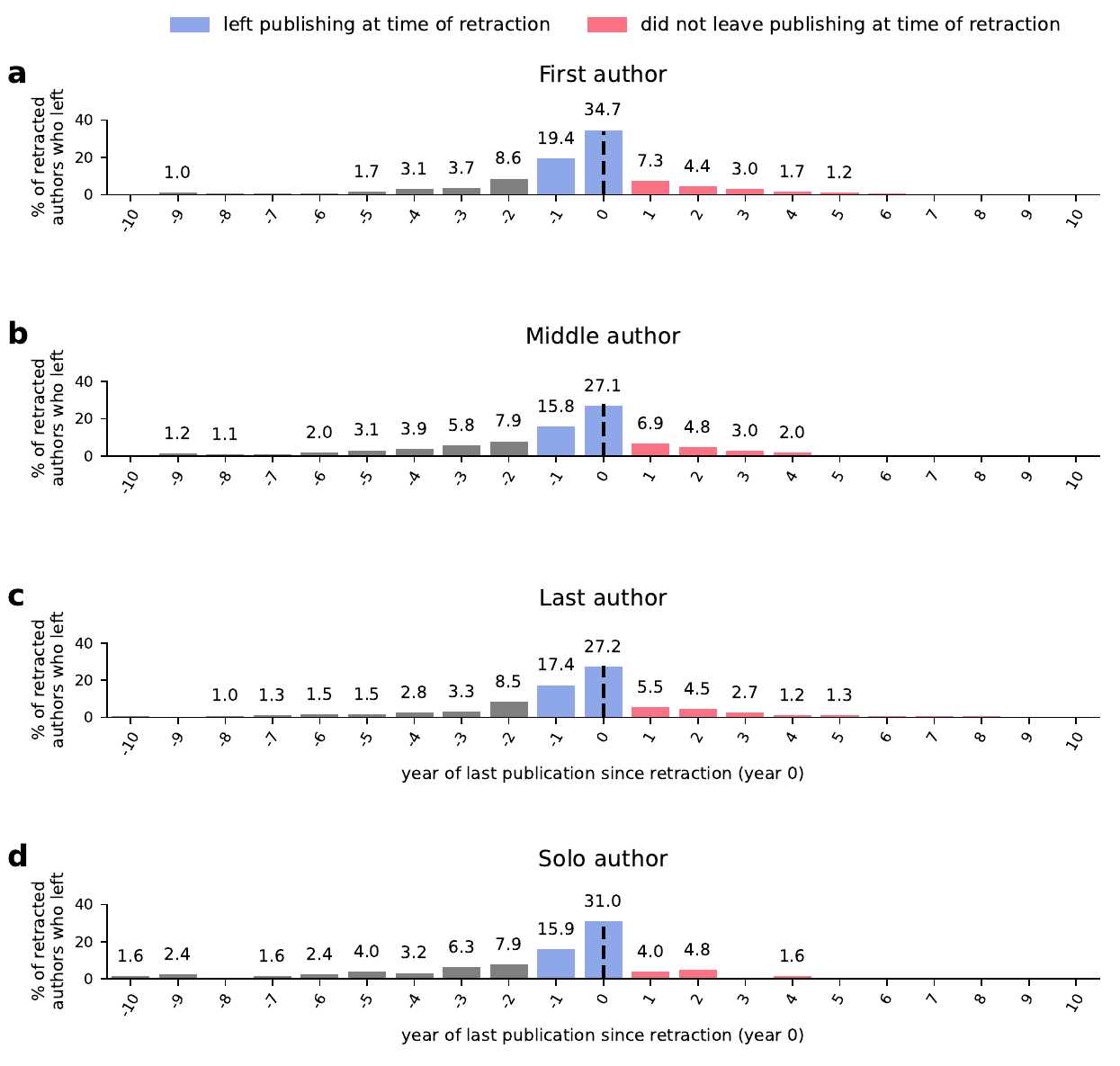}
\caption{
\textbf{Percentage of retracted authors who left (blue)
versus those who did not (red) stratified by author order in the paper.} \textbf{(a)} shows first authors, \textbf{(b)} shows middle authors, \textbf{(c)} shows last authors, and \text{(d)} shows solo authors.
}
\label{supplementaryfig:attrition_barplot_byAuthorOrder}
\end{figure}

\begin{figure}[htbp]
\centering
\includegraphics[width=0.95\textwidth,height=\textheight,keepaspectratio]{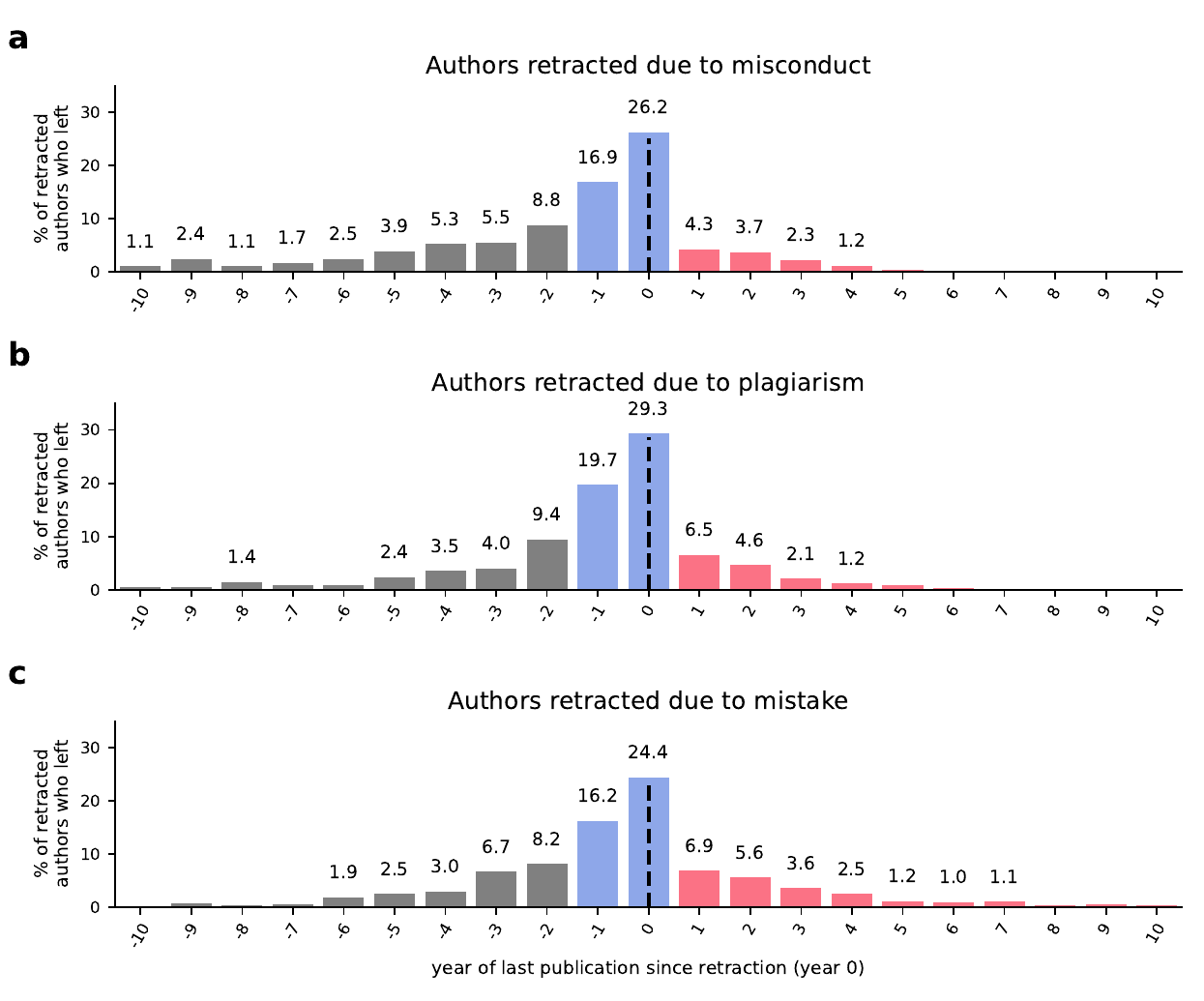}
\caption{
\textbf{Percentage of retracted authors who left (blue)
versus those who did not (red) stratified by reason of retraction.} \textbf{(a)} shows authors retracted due to misconduct, \textbf{(b)} shows authors retracted due to plagiarism, and \textbf{(c)} shows authors retracted due to mistake.
}
\label{supplementaryfig:attrition_barplot_byReason}
\end{figure}

\begin{figure}[htbp]
\centering
\includegraphics[width=0.7\textwidth,height=\textheight,keepaspectratio]{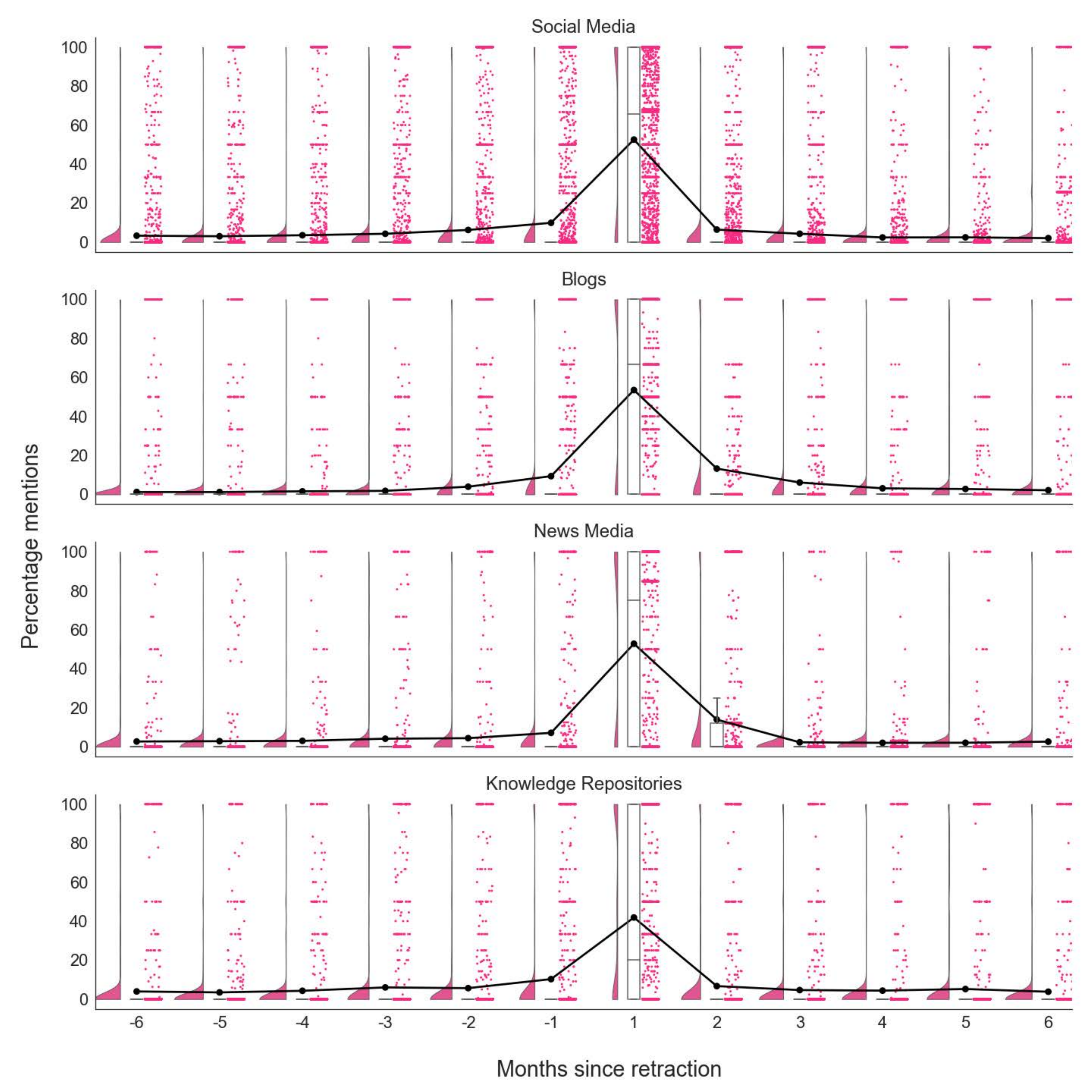}
\caption{
\textbf{Raincloud plot \cite{allen2019raincloud} showing the normalized distribution of mentions 6 months pre- and post-retraction across different types of platforms.} The x-axis represents time from the month of retraction (0). The y-axis shows the percentage of mentions for a given paper in the given month where the denominator is all mentions in the 12-months window displayed \cite{abhari2022twitter}. The black line shows the average. All papers that do not receive any attention within the 12-months window are excluded. Papers receive the most attention within 1 month of the retraction across a variety of platforms: (i) social media (this includes Twitter, Facebook, Google+, LinkedIn, Pinterest, Reddit, and videos); (ii) news media; (iii) blogs; and (iv) knowledge repositories (this includes Wikipedia, patents, F1000, Q\&A, and peer reviews), a classification based on \cite{peng2022dynamics} studying the dynamics of cross-platform attention to retracted papers.
}
\label{supplementaryfig:altmetric6months}
\end{figure}

\clearpage
\newpage

\begin{figure}[htbp]
\centering
\includegraphics[width=0.9\textwidth,keepaspectratio]{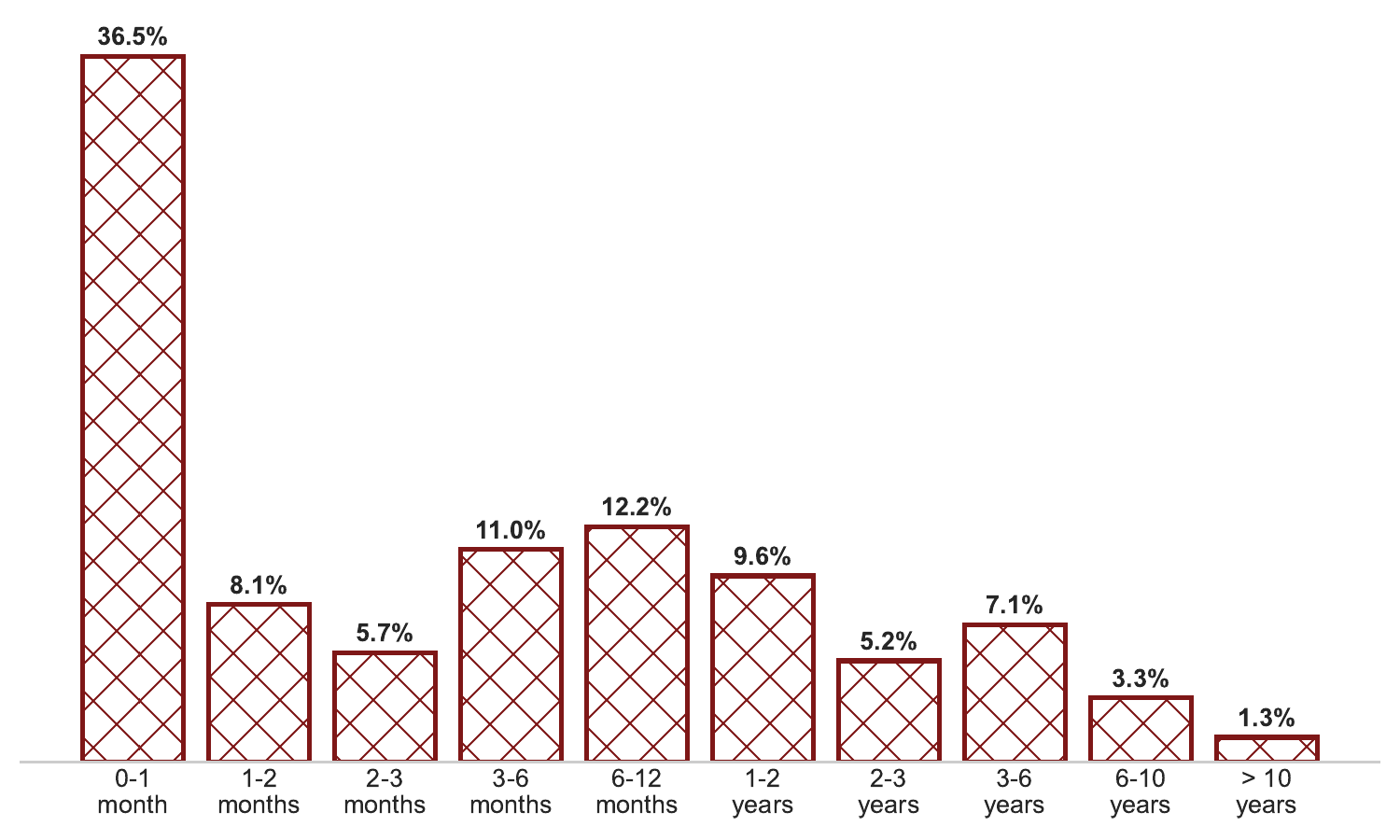}
\caption{
\textbf{Percentage distribution of time interval between consecutive retraction for authors with multiple retractions.}
}
\label{supplementaryfig:multiple_retractions}
\end{figure}

\begin{figure}
  \centering
  \includegraphics[width=0.95\textwidth,keepaspectratio]{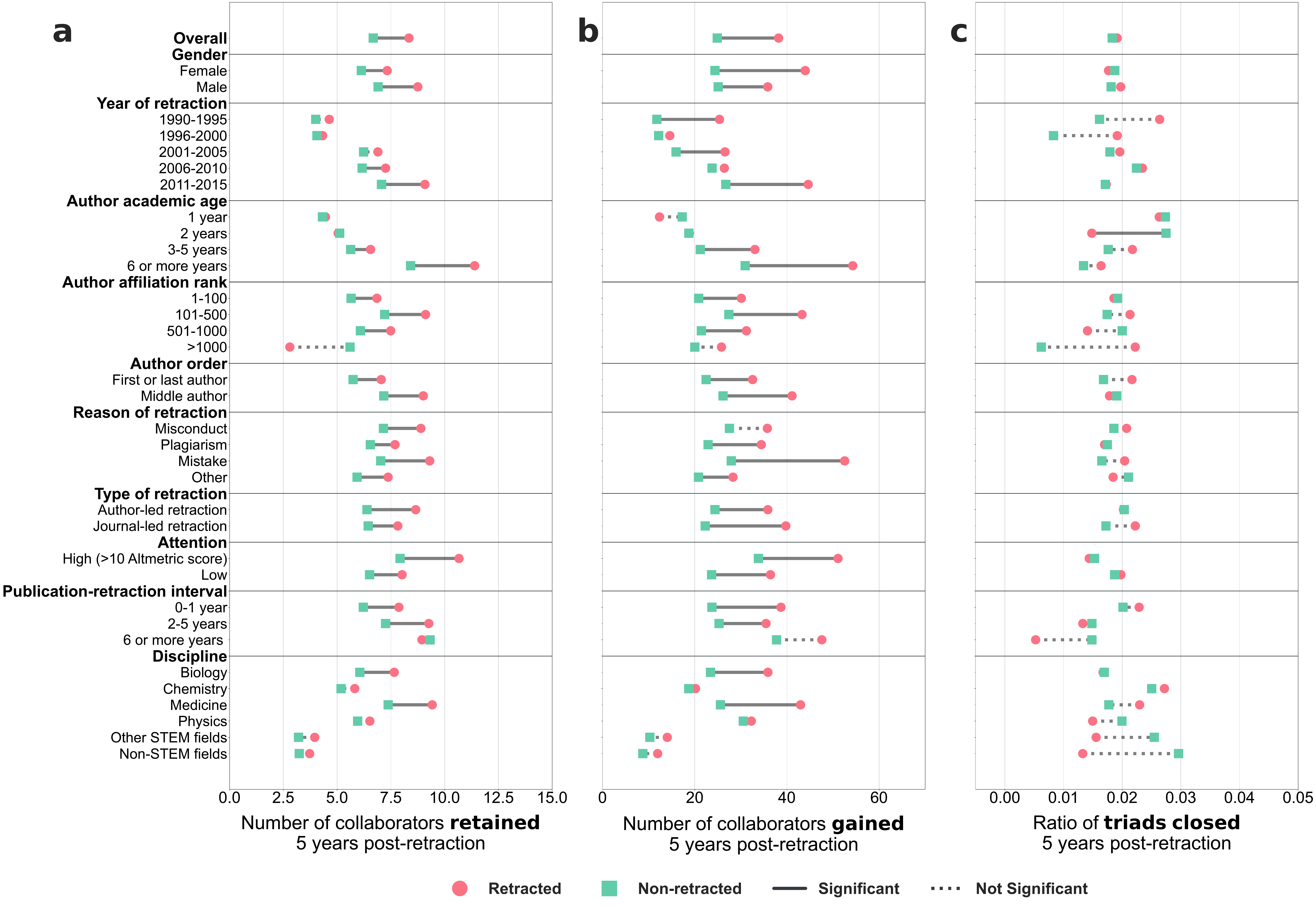}
  \caption{\textbf{Robustness analysis for analyzing collaborator retention, gain, and triadic closure among retracted authors who stayed in scientific publishing post-retraction using a 20\% threshold for percentage difference in papers, citations, and collaborators between retracted and non-retracted scientists.} The figure shows the difference between \textbf{(a)} the numbers of collaborators retained, \textbf{(b)} the numbers of collaborators gained, and \textbf{(c)} the proportions of triads closed 5 years post-retraction for the ($N=1,700$) authors who were retracted (red circle), and their matched non-retracted pairs (green square). These are further stratified by gender, year of retraction, academic age, author order, reason of retraction, type of retraction, and discipline. Data are presented as mean values. Solid line represents statistically significant difference using two-sided Welch's t-test (assuming unequal variances).
}
\label{supplementaryfig:matching_analysis_20perc}
\end{figure}

\begin{figure}
  \centering
  \includegraphics[width=0.95\textwidth,keepaspectratio]{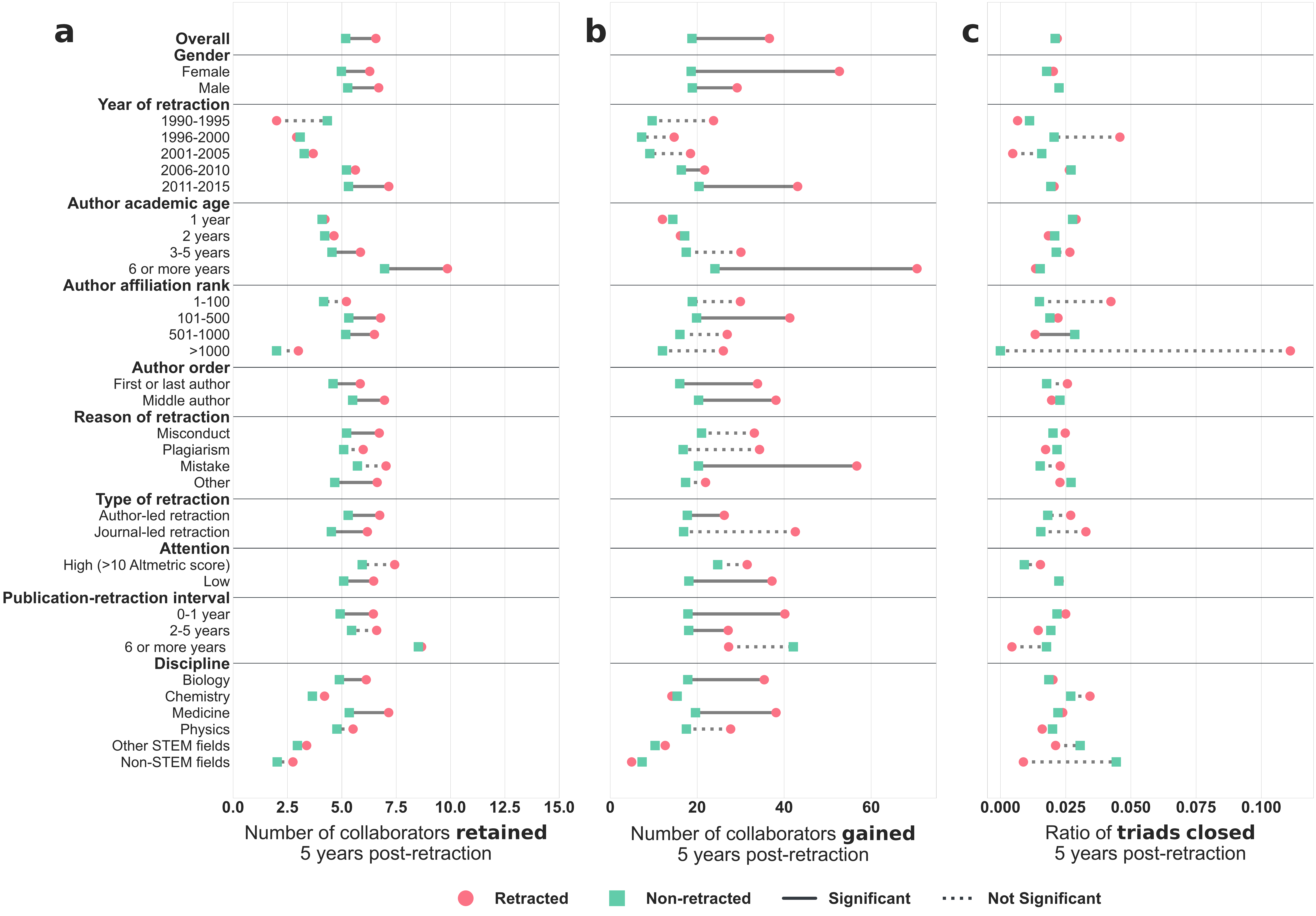}
  \caption{\textbf{Robustness analysis for analyzing collaborator retention, gain, and triadic closure among retracted authors who stayed in scientific publishing post-retraction using a 10\% threshold for percentage difference in papers, citations, and collaborators between retracted and non-retracted scientists.} The figure shows the difference between \textbf{(a)} the numbers of collaborators retained, \textbf{(b)} the numbers of collaborators gained, and \textbf{(c)} the proportions of triads closed 5 years post-retraction for the ($N=751$) authors who were retracted (red circle), and their matched non-retracted pairs (green square). These are further stratified by gender, year of retraction, academic age, author order, reason of retraction, type of retraction, and discipline. Data are presented as mean values. Solid line represents statistically significant difference using two-sided Welch's t-test (assuming unequal variances).
}
\label{supplementaryfig:matching_analysis_10perc}
\end{figure}

\begin{figure}
  \centering
  \includegraphics[width=0.95\textwidth,keepaspectratio]{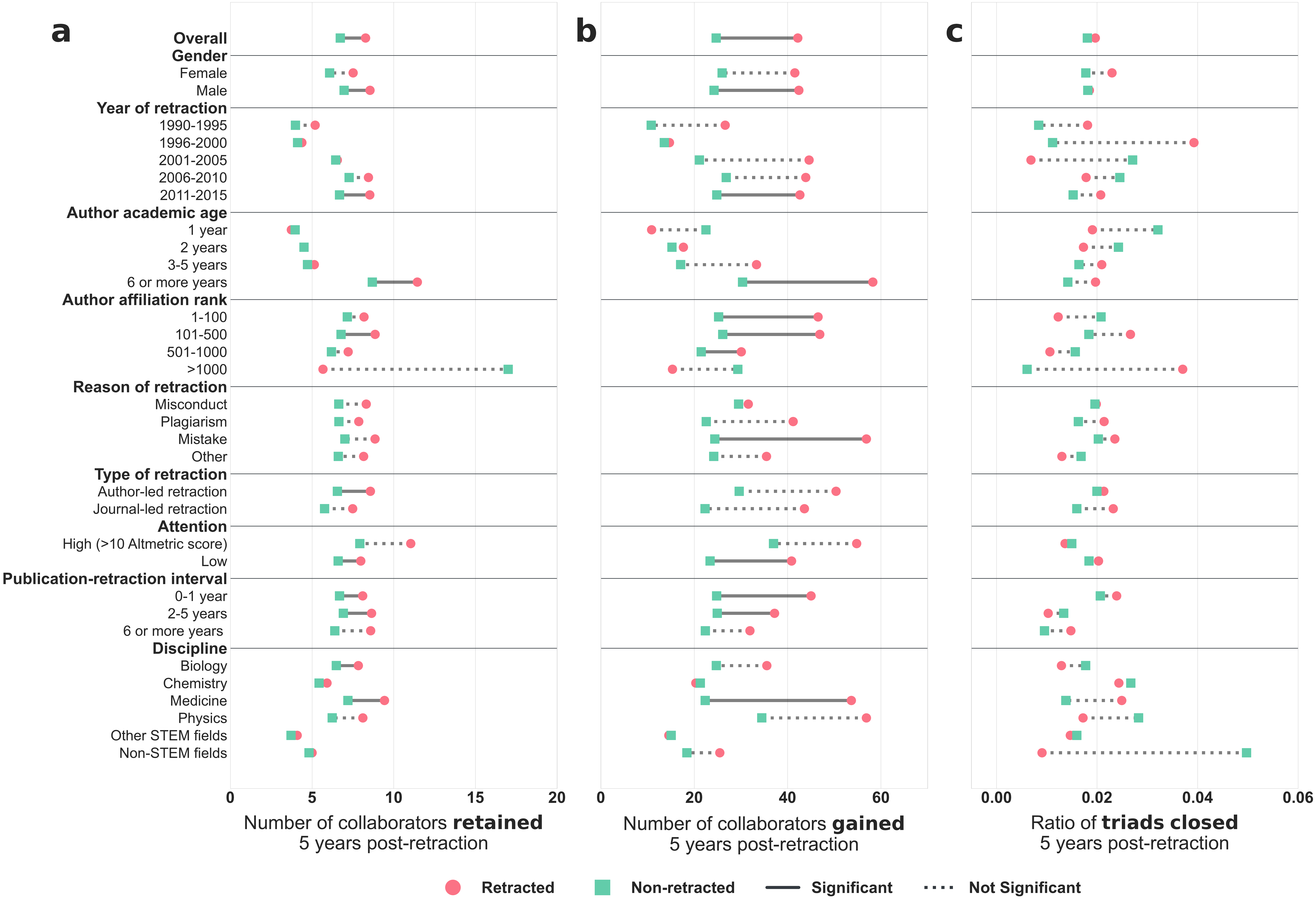}
  \caption{\textbf{Robustness analysis for analyzing collaborator retention, gain, and triadic closure among retracted authors who stayed in scientific publishing post-retraction and were the first or last authors on their retracted papers.} The figure shows the difference between \textbf{(a)} the numbers of collaborators retained, \textbf{(b)} the numbers of collaborators gained, and \textbf{(c)} the proportions of triads closed 5 years post-retraction for the ($N=803$) authors who were retracted (red circle), and their matched non-retracted pairs (green square). These are further stratified by gender, year of retraction, academic age, author order, reason of retraction, type of retraction, and discipline. Data are presented as mean values. Solid line represents statistically significant difference using two-sided Welch's t-test (assuming unequal variances).
}
\label{supplementaryfig:matching_analysis_firstlast}
\end{figure}

\begin{figure}
  \centering
  \includegraphics[width=0.95\textwidth,keepaspectratio]{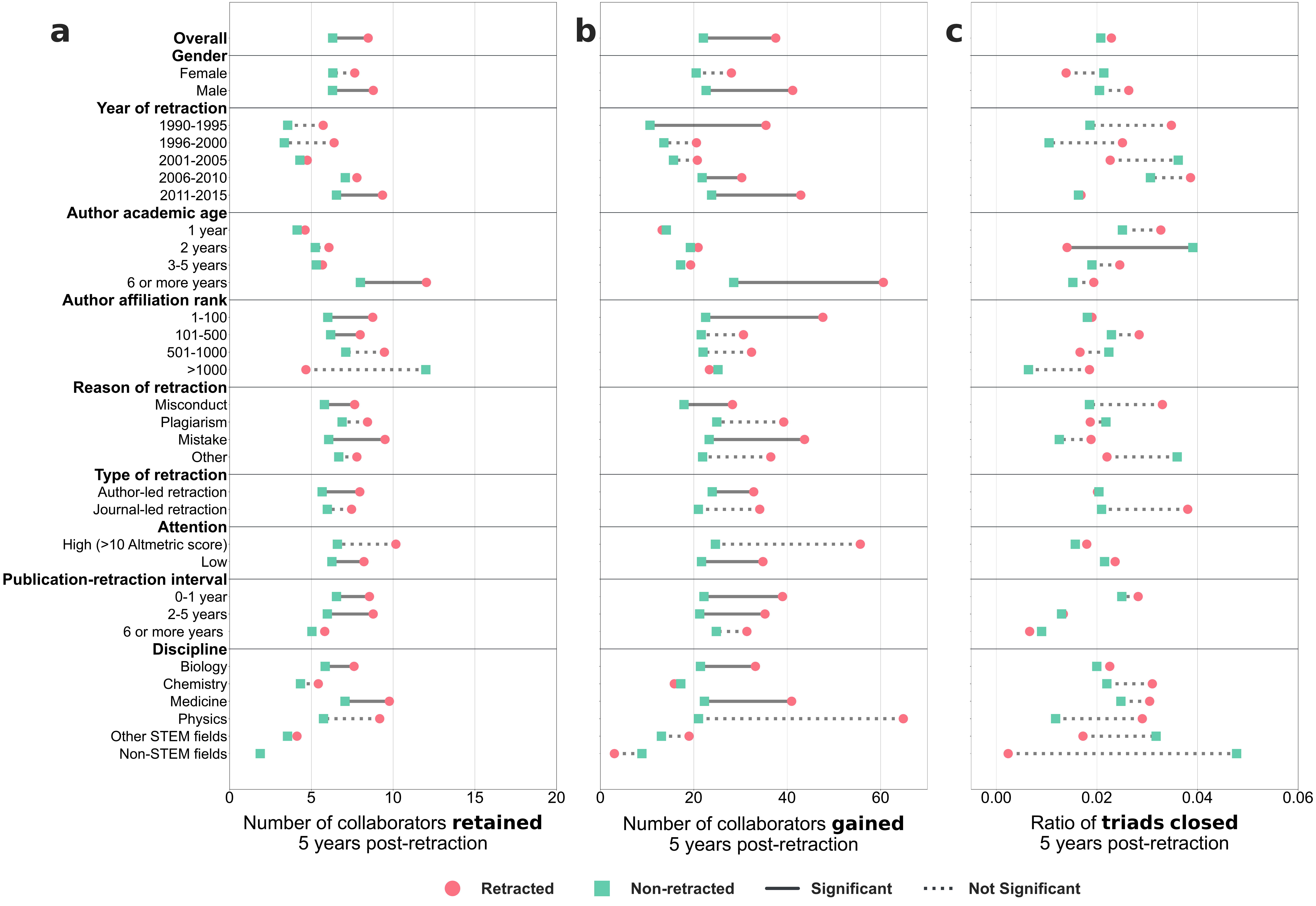}
  \caption{\textbf{Robustness analysis for analyzing collaborator retention, gain, and triadic closure among retracted authors who stayed in scientific publishing post-retraction and whose matches were in the same country.} The figure shows the difference between \textbf{(a)} the numbers of collaborators retained, \textbf{(b)} the numbers of collaborators gained, and \textbf{(c)} the proportions of triads closed 5 years post-retraction for the ($N=632$) authors who were retracted (red circle), and their matched non-retracted pairs (green square). These are further stratified by gender, year of retraction, academic age, author order, reason of retraction, type of retraction, and discipline. Data are presented as mean values. Solid line represents statistically significant difference using two-sided Welch's t-test (assuming unequal variances).
}
\label{supplementaryfig:matching_analysis_samecountry}
\end{figure}

\begin{figure}[htbp]
    \centering
    \includegraphics[width=0.72\textwidth,keepaspectratio]{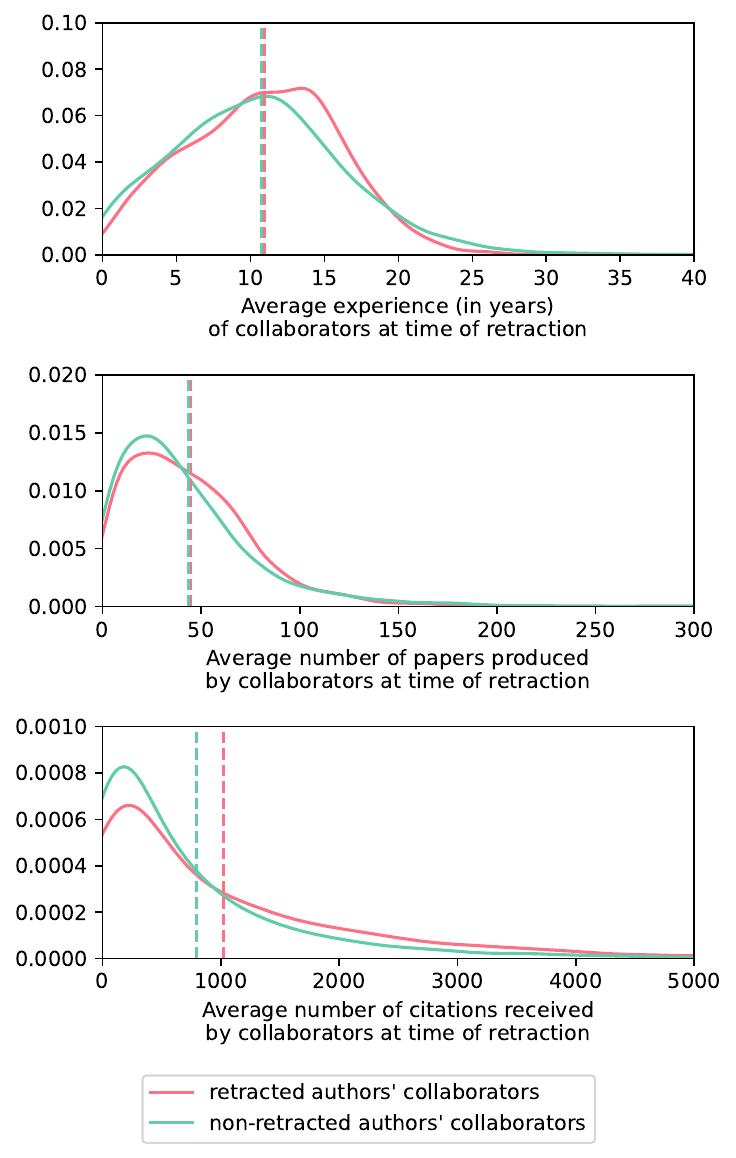}
    \caption{\textbf{Comparison of collaborator characteristics between retracted authors and their matched equivalents.} The plots illustrate the average experience, cumulative productivity, and cumulative citations of collaborators for retracted authors versus their matched counterparts at the time of retraction. No significant differences were found between the two groups in terms of collaborator experience and productivity distributions using a t-test and a Kolmogorov-Smirnov test. Additionally, the citation distributions are very similar.}
    \label{supplementaryfig:pre-retraction_distributions}
\end{figure}

\begin{figure}[htbp]
\centering
\begin{subfigure}[t]{\textwidth}
  \centering
  \includegraphics[width=0.90\textwidth,keepaspectratio]{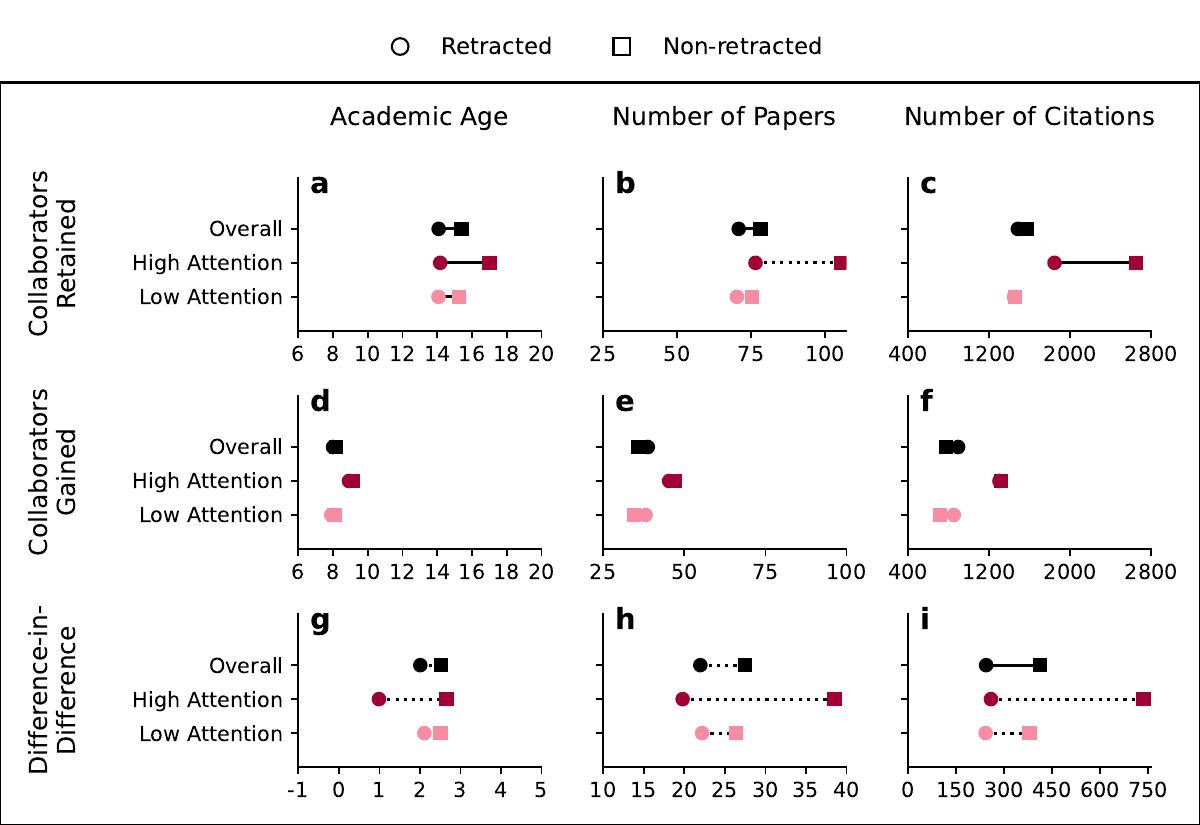}
\end{subfigure}

\caption{\textbf{Comparison of characteristics of collaborators retained and gained by retracted authors ($N=1,143$) and non-retracted authors, stratified by attention.} \textbf{(a-c)} display the comparison based on academic age, number of papers, and number of citations for retained collaborators of retracted and non-retracted authors of different age groups. \textbf{(d-f)} show the comparison for collaborators gained. \textbf{(g-i)} illustrate the results of the difference-in-difference analysis comparing the difference of collaborators retained and collaborators lost.}
\label{supplementaryfig:collaborator_exploration_attention}
\end{figure}

\begin{figure}[htbp]
\centering
\begin{subfigure}[t]{0.95\textwidth}
  \centering
  \includegraphics[width=0.90\textwidth]{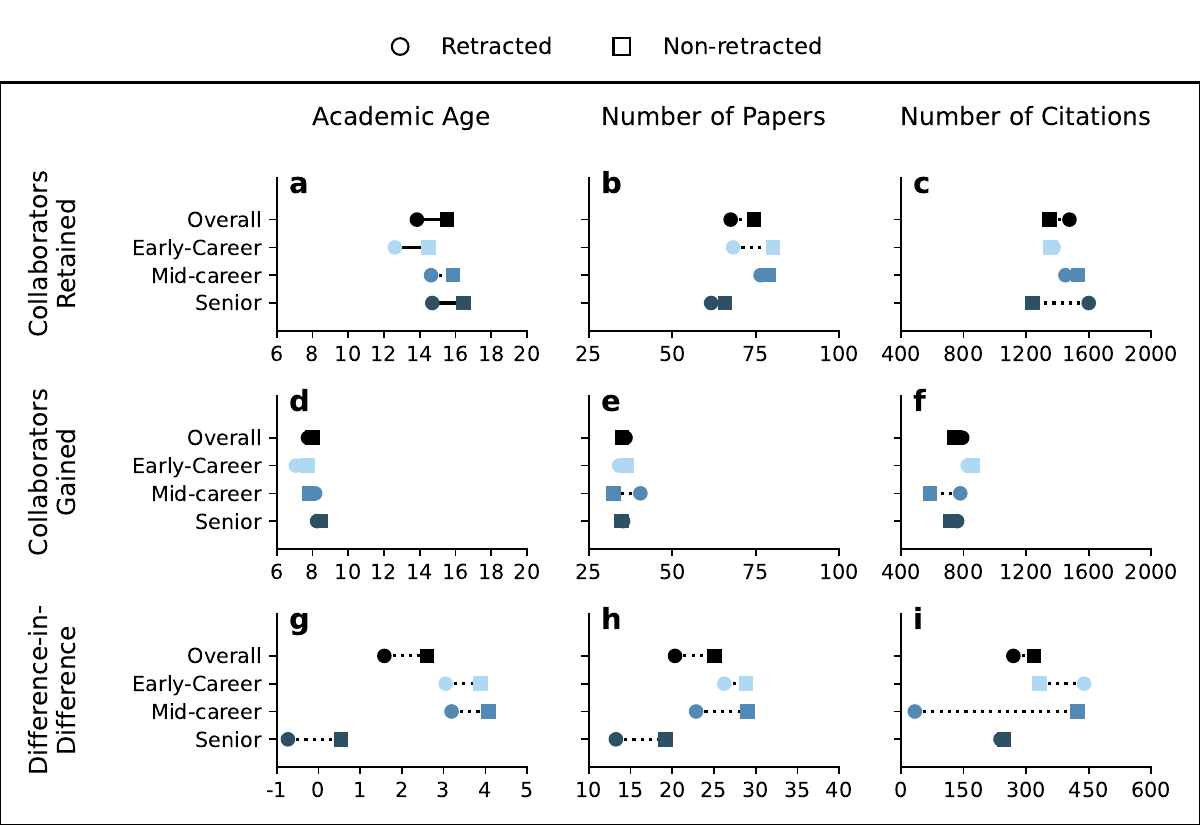}
\end{subfigure}

\vspace{-1.5\baselineskip}

\begin{subfigure}[t]{0.95\textwidth}
  \centering
  \includegraphics[width=0.90\textwidth,keepaspectratio]{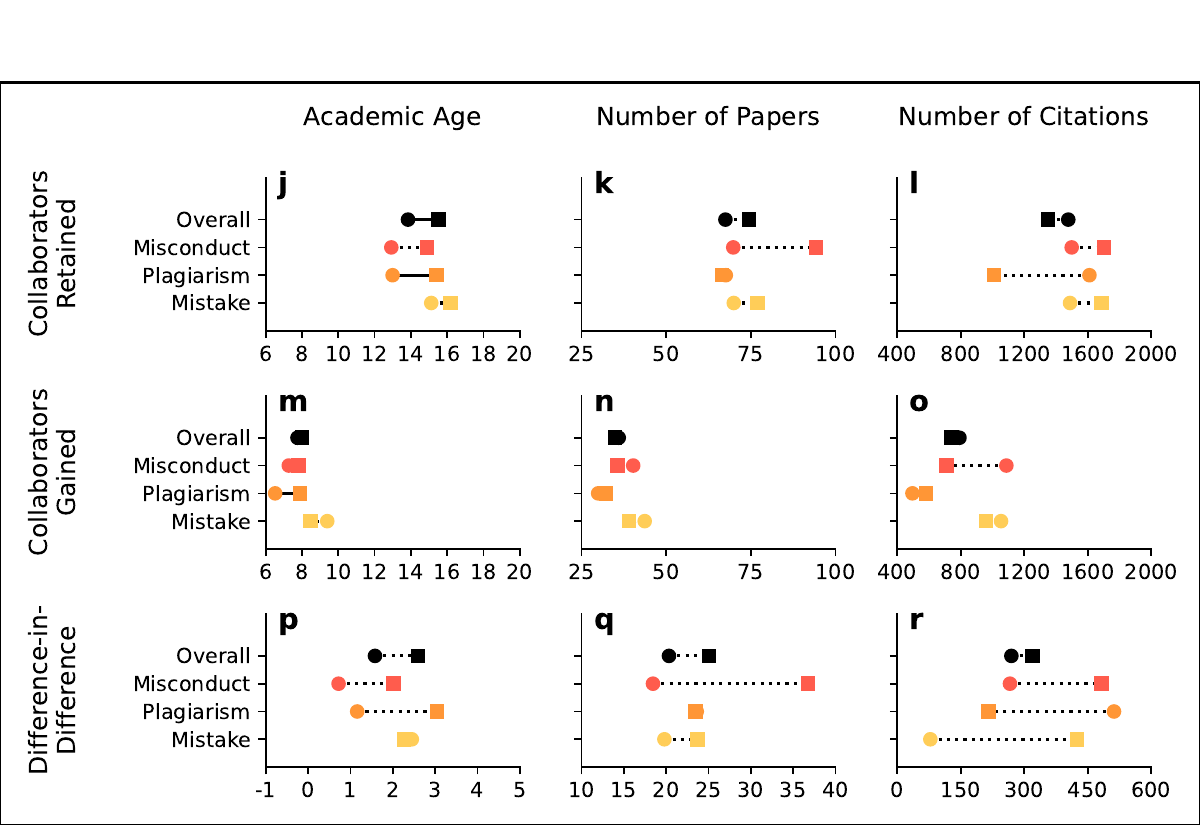}
\end{subfigure}

\caption{\textbf{Comparison of characteristics of collaborators retained and gained by retracted and non-retracted authors, stratified by seniority ($N=400$) and by reasons of retraction ($N=302$) for first and last authors only.} \textbf{(a-c)} display the comparison based on academic age, number of papers, and number of citations for retained collaborators of retracted and non-retracted authors of different age groups. \textbf{(d-f)} show the comparison for collaborators gained. \textbf{(g-i)} illustrate the results of the difference-in-difference analysis comparing the difference of collaborators retained and collaborators lost. \textbf{(j-r)} show similar comparison for collaborators of retracted and non-retracted authors stratified by reasons of retraction. }
\label{supplementaryfig:collaborator_exploration}
\end{figure}

\begin{figure}
  \centering
  \includegraphics[width=0.95\textwidth,keepaspectratio]{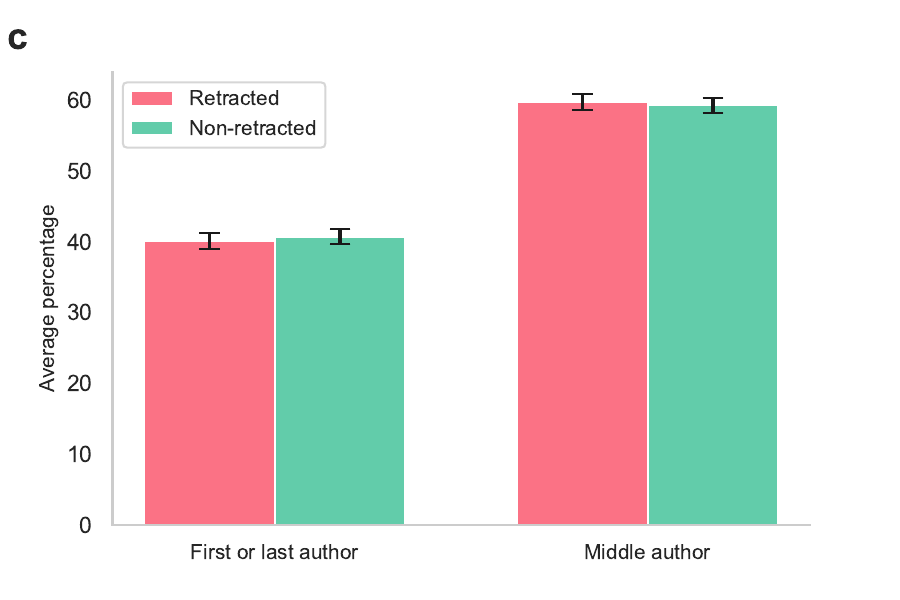}
  \caption{\textbf{Comparison of authorship order share between ($N=2,348$) retracted authors and their matched equivalents.} Bars represent mean authorship order share, with error bars indicating the 95\% confidence intervals. No significant differences were found between the two groups using a two-sided Welch's t-tests $t(5787.64)= 1.086, p = 0.278$ for the first or last author category.  
}
\label{supplementaryfig:authororderanalysis}
\end{figure}

\begin{figure}[H]
\centering
\begin{subfigure}[t]{0.95\textwidth}
  \centering
  \includegraphics[width=0.95\textwidth,keepaspectratio]{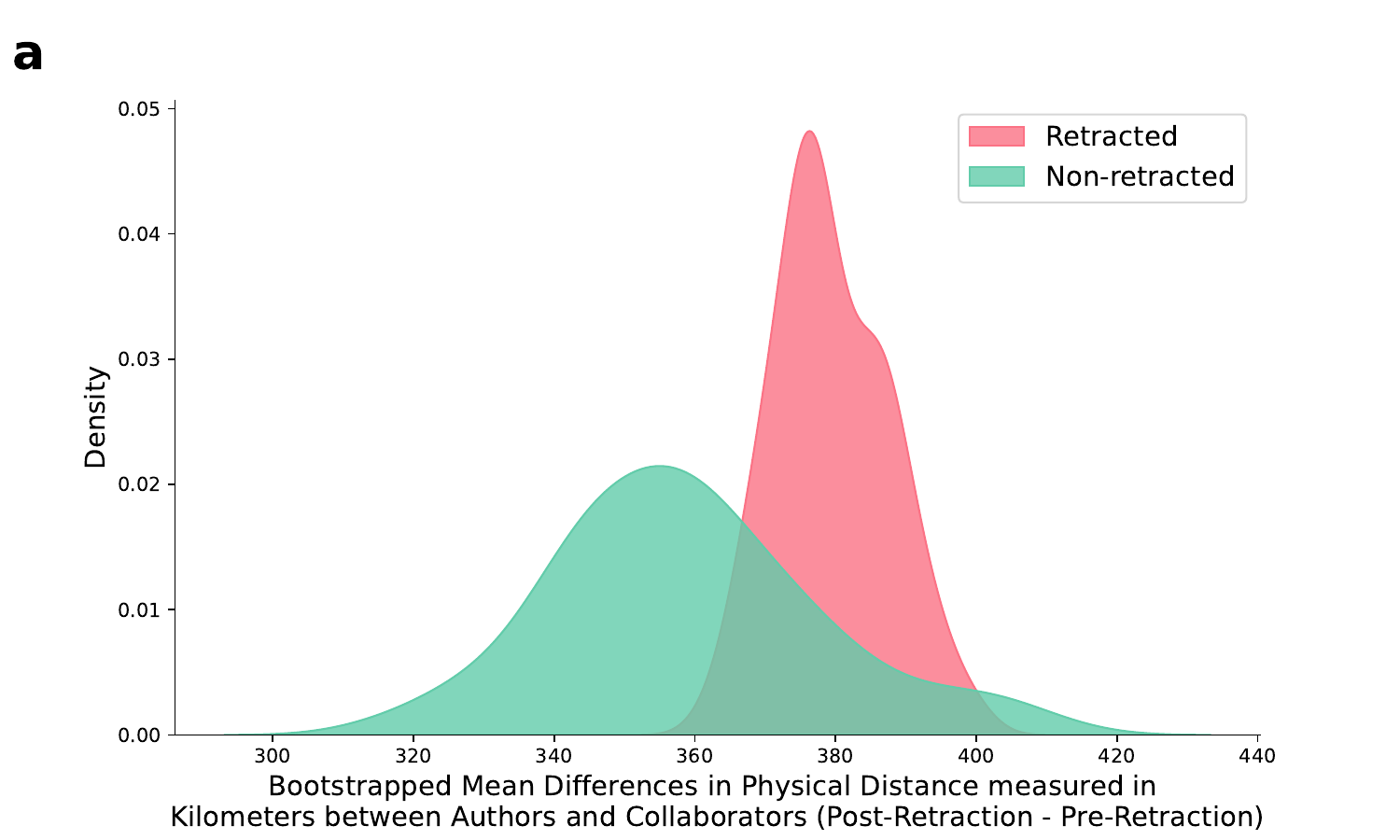}
\end{subfigure}

\vspace{-0.7\baselineskip}

\begin{subfigure}[t]{0.95\textwidth}
  \centering
  \includegraphics[width=0.95\textwidth,keepaspectratio]{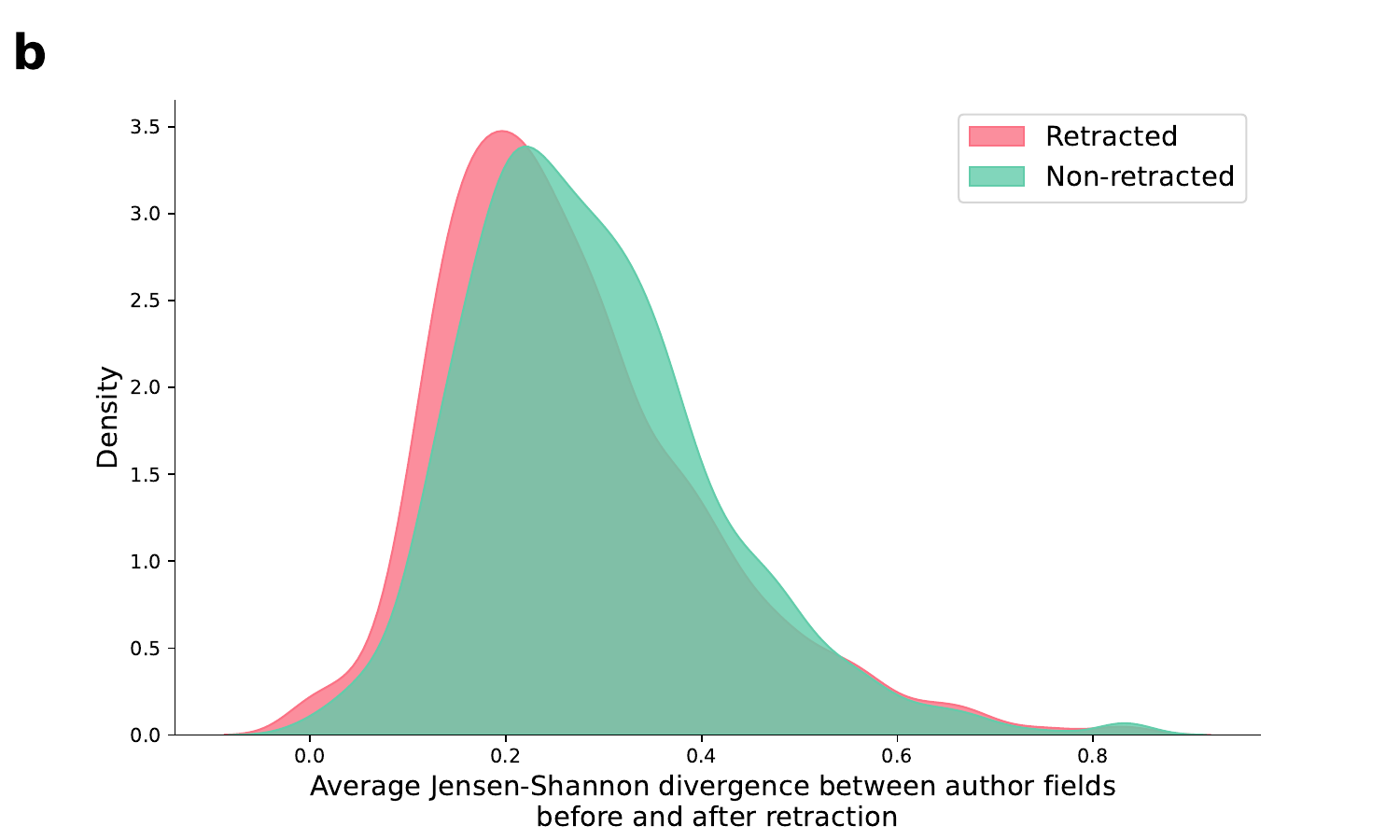}
\end{subfigure}

\caption{\textbf{Physical distance and disciplinary shift in collaboration networks post-retraction.} The plots illustrate the changes in the physical distance of collaboration networks and the shift in publication fields for retracted authors compared to their matched counterparts, pre- and post-retraction. (a) Illustrates the average differences in physical distance between authors ($N=1,224$) and collaborators. (b) displays the divergence in publication fields over time ($N=2,126$).}
\label{suppfig:prepostanalysis}
\end{figure}

\begin{figure}[htbp]
  \centering\includegraphics[width=\textwidth,height=0.9\textheight,keepaspectratio]{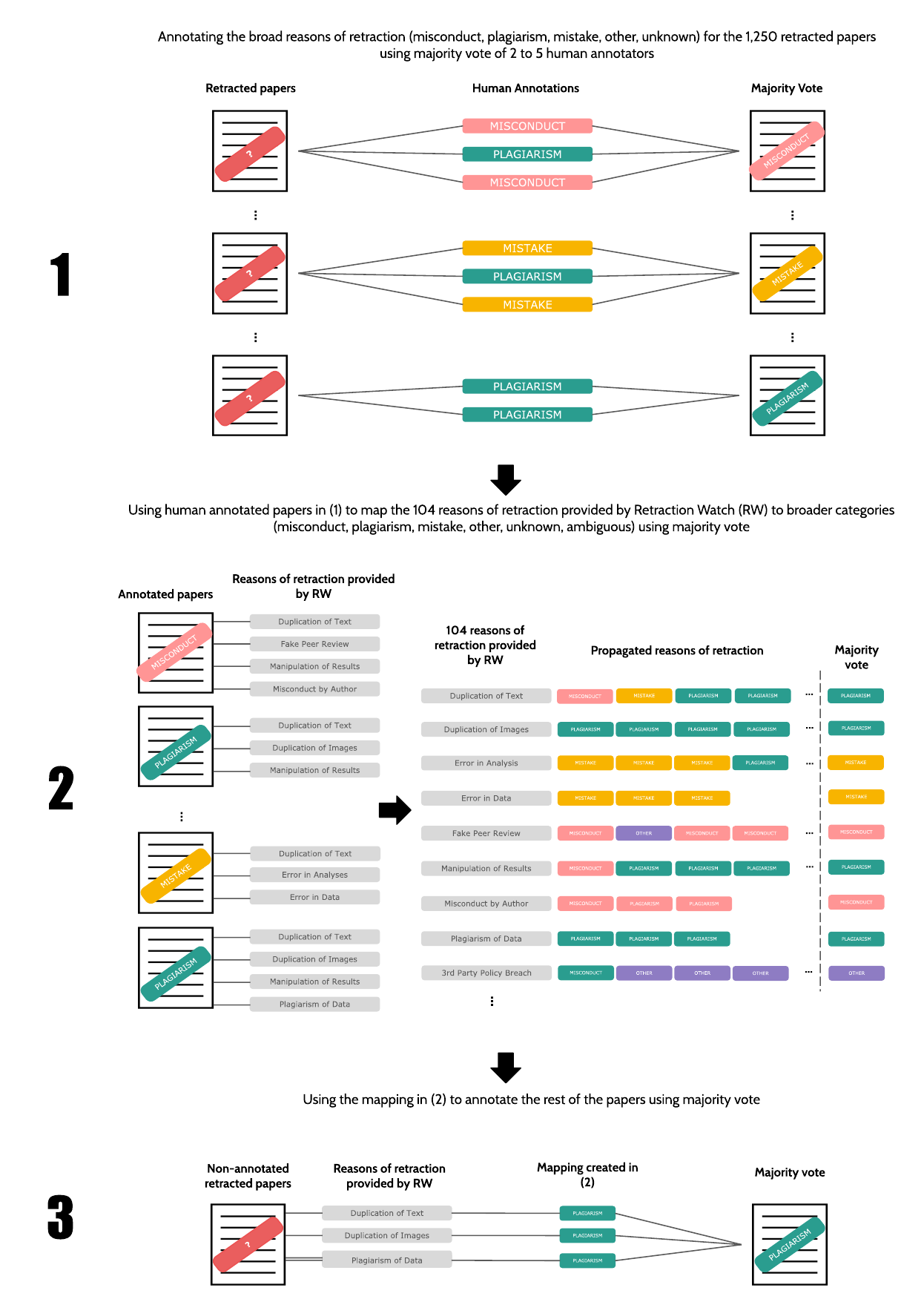}\caption{\textbf{Annotating reasons of retraction using label propagation.} This diagram shows the process we employ to annotate reasons of retraction for the non-annotated papers. Step (1) shows the process used to annotate papers manually; Steps (2) and (3) illustrate the label propagation algorithm.}
\label{supplementaryfig:reasonpropagation}
\end{figure}

\clearpage

\section*{Supplementary Tables}
\addcontentsline{toc}{section}{Supplementary Tables}

\begin{table}[H]
\centering
\caption{\textbf{Standardized mean differences between authors who left academic publishing and those who stayed.} $a$ denotes the proportion of authors who stayed, and $a'$ denotes the proportion of authors who left academic publishing. For continuous variables, $a$ and $a'$ represent the mean. $p$ represents the p-value of a two-sided chi-square test for proportions and a two-sided Welch's t-test for experience variables.}
\resizebox{\textwidth}{!}{%
\begin{tabular}{lccccc}
\toprule
\multicolumn{2}{c}{\textbf{Category}} & \textbf{$a$ $(n=10,342)$} & \textbf{$a'$ $(n=2,400)$} & \textbf{Standardized Mean Difference} & \textbf{$p$} \\
\cmidrule(lr){1-2} \cmidrule(lr){3-3} \cmidrule(lr){4-4} \cmidrule(lr){5-5} \cmidrule(lr){6-6}
\multirow{2}{*}{Attention} & High attention & 0.04 & 0.02 & -0.14 & $<.001$ \\
& Low attention & 0.96 & 0.98 & 0.14 &  \\
\midrule
Academic age & & 13.44 & 2.46 & 1.36 & $<.001$ \\
\# papers & & 69.29 & 4.76 & 0.81 & $<.001$ \\
\# citations & & 1180.05 & 43.7 & 0.52 & $<.001$ \\
\# collaborators & & 197.56 & 13.63 & 0.43 & $<.001$ \\
\midrule
\multirow{2}{*}{Gender} & Female & 0.25 & 0.3 & 0.1 & $<.001$ \\
& Male & 0.75 & 0.7 & -0.1 &  \\
\midrule
\multirow{4}{*}{Affiliation Rank} & 1-100 & 0.24 & 0.13 & -0.27 & $<.001$ \\
& 101-500 & 0.31 & 0.3 & -0.02 &  \\
& 501-1000 & 0.13 & 0.15 & 0.04 &  \\
& $>1001$ & 0.32 & 0.42 & 0.21 &  \\
\midrule
\multirow{4}{*}{Reason} & Misconduct & 0.22 & 0.22 & -0.01 & $<.001$ \\
& Plagiarism & 0.28 & 0.33 & 0.13 &  \\
& Mistake & 0.3 & 0.23 & -0.15 &  \\
& Other & 0.2 & 0.22 & 0.03 &  \\
\midrule
\multirow{2}{*}{Author order} & First or last author & 0.39 & 0.45 & 0.14 & $<.001$ \\
& Middle author & 0.61 & 0.55 & -0.14 &  \\
\midrule
\multirow{5}{*}{Year of retraction} & 1990-1995 & 0.02 & 0.03 & 0.08 & $<.001$ \\
& 1996-2000 & 0.03 & 0.02 & -0.01 &  \\
& 2001-2005 & 0.06 & 0.05 & -0.04 &  \\
& 2006-2010 & 0.26 & 0.28 & 0.04 &  \\
& 2011-2015 & 0.64 & 0.62 & -0.03 &  \\
\midrule
\multirow{2}{*}{Venue} & Journal & 0.97 & 0.94 & -0.15 & $<.001$ \\
& Conference & 0.01 & 0.04 & 0.19 &  \\
\midrule
\multirow{4}{*}{Journal Ranking} & Q1 & 0.74 & 0.62 & -0.26 & $<.001$ \\
& Q2 & 0.2 & 0.27 & 0.17 &  \\
& Q3 & 0.06 & 0.1 & 0.16 &  \\
& Q4 & 0.01 & 0.02 & 0.04 &  \\
\midrule
\multirow{4}{*}{Discipline} & Biology & 0.38 & 0.29 & -0.18 & $<.001$ \\
& Chemistry & 0.15 & 0.2 & 0.14 &  \\
& Medicine & 0.3 & 0.23 & -0.17 & $<.001$ \\
& Physics & 0.06 & 0.08 & 0.07 &  \\
& Other-STEM fields & 0.11 & 0.18 & 0.22 &  \\
& Non-STEM fields & 0.01 & 0.02 & 0.12 &  \\
\bottomrule
\end{tabular}
}
\label{supplementarytab:SMDs_attrited_nonattrited}
\end{table}

\begin{table}[H]
\centering
\caption{\textbf{Standardized mean differences between authors in the filtered sample and those that were matched early-career}. $a$ denotes the proportion of authors who stayed within the filtered sample, and $m$ denotes the proportion of authors who were matched. For continuous variables, $a$ and $m$ represent the mean. $p$ represents the p-value of a two-sided chi-square test for proportions and a two-sided Welch's t-test for experience variables.}
\resizebox{\textwidth}{!}{%
\begin{tabular}{lccccc}
\toprule
\multicolumn{2}{c}{\textbf{Category}} & \textbf{$a$ $(n=14,579)$} & \textbf{$m$ $(n=2,743)$} & \textbf{Standardized Mean Difference} & \textbf{$p$} \\
\cmidrule(lr){1-2} \cmidrule(lr){3-3} \cmidrule(lr){4-4} \cmidrule(lr){5-5} \cmidrule(lr){6-6}
\multirow{2}{*}{Attention} & High attention & 0.04 & 0.03 & -0.05 & $0.039$ \\
& Low attention & 0.96 & 0.97 & 0.05 &  \\
\midrule
Academic age && 11.68 & 3.27 & 0.98 & $<.001$ \\
% \midrule
\# papers && 53.11 & 5.63 & 0.66 & $<.001$ \\
\# citations && 895.63 & 67.89 & 0.43 & $<.001$ \\
\# collaborators && 151.31 & 16.71 & 0.36 & $<.001$ \\
\midrule
\multirow{2}{*}{Gender} & Female & 0.26 & 0.29 & 0.05 & $0.018$ \\
& Male & 0.74 & 0.71 & -0.05 &  \\
\midrule
\multirow{4}{*}{Affiliation Rank} & 1-100 & 0.22 & 0.19 & -0.07 & $<.001$ \\
& 101-500 & 0.31 & 0.36 & 0.1 &  \\
& 501-1000 & 0.14 & 0.14 & 0.01 &  \\
& $>1000$ & 0.34 & 0.31 & -0.05 &  \\
\midrule
\multirow{4}{*}{Reason} & Misconduct & 0.23 & 0.26 & 0.08 & $<.001$ \\
& Plagiarism & 0.29 & 0.28 & -0.02 &  \\
& Mistake & 0.28 & 0.25 & -0.08 &  \\
& Other & 0.2 & 0.2 & 0.02 &  \\
\midrule
\multirow{2}{*}{Author order} & First or last author & 0.39 & 0.41 & 0.02 & $0.28$ \\
& Middle author & 0.61 & 0.59 & -0.02 &  \\
\midrule
\multirow{5}{*}{Year of retraction} & 1990-1995 & 0.02 & 0.03 & 0.07 & $<.001$ \\
& 1996-2000 & 0.03 & 0.03 & 0.02 &  \\
& 2001-2005 & 0.06 & 0.04 & -0.06 &  \\
& 2006-2010 & 0.26 & 0.27 & 0.03 &  \\
& 2011-2015 & 0.64 & 0.62 & -0.03 &  \\
\midrule
\multirow{4}{*}{Discipline} & Biology & 0.35 & 0.31 & -0.09 & $<.001$ \\
& Chemistry & 0.17 & 0.22 & 0.12 &  \\
& Medicine & 0.28 & 0.23 & -0.11 &  \\
& Physics & 0.07 & 0.08 & 0.04 &  \\
& Non-STEM fields & 0.01 & 0.02 & 0.04 &  \\
& Other STEM fields & 0.13 & 0.15 & 0.08 &  \\
\bottomrule
\end{tabular}
}
\label{supplementarytab:SMDs_filtered_earlymatched}
\end{table}

\begin{table}[H]
{\fontsize{10.0}{10.0}\selectfont{
\caption{\textbf{Cox proportional hazard model of attrition focusing on first and last authors.} In the table below, we present three models. The models differ in how authors' experience is measured using (1) number of papers, (2) logged number of citations, (3) logged number of collaborators, and (4) all experiences put together respectively. All models incorporate clustered standard errors at the author-level. Controls for author's scientific discipline are included as categorical variables, but are not shown.}
\label{supplementarytab:cox_model_firstlast}
\begin{center}
\begin{tabular}{@{\extracolsep{5pt}}lD{.}{.}{-3} D{.}{.}{-3} D{.}{.}{-3} D{.}{.}{-3} }
\\[-1.8ex]\hline
\hline \\[-1.8ex]
& \multicolumn{4}{c}{\textit{Outcome: Left publishing}} \
\cr \cline{2-5}\\[-1.8ex]
& \multicolumn{1}{c}{\hspace{10pt}(1)} & \multicolumn{1}{c}{\hspace{10pt}(2)} & \multicolumn{1}{c}{\hspace{10pt}(3)}& \multicolumn{1}{c}{\hspace{10pt}(4)} \\
\hline \\[-1.8ex]
\textbf{Retracted} & 1.81\textbf{***} & 1.83\textbf{***} & 1.79\textbf{***} & 1.80\textbf{***}\\
& (0.06) & (0.06) & (0.06) & (0.06)\\
\\
\textbf{Gender} (reference: Male)\\
Female & 0.00 & 0.02 & -0.01 & -0.01\\
& (0.05) & (0.05) & (0.05) & (0.05)\\
\\
\textbf{Affiliation Rank} & 0.00\textbf{*} & 0.00 & 0.00 & 0.00\\
& (0.00) & (0.00) & (0.00) & (0.00)\\
\\
\textbf{Cohort Year} & 0.02\textbf{***} & 0.02\textbf{***} & 0.03\textbf{***} & 0.02\textbf{***}\\
& (0.00) & (0.00) & (0.00) & (0.00) \\
\\
\textbf{Author's Experience}\\
\# Publications & -0.01\textbf{**} & & & -0.00\\
& (0.00) & & & (0.00)\\
log(\# Citations) & & -0.14\textbf{***} & & -0.02\\
& & (0.02) & & (0.02)\\
log(\# Collaborators) & & & -0.34\textbf{***} & -0.31\textbf{***}\\
& & &(0.03)& (0.03)\\
\hline \\[-1.8ex]
 Number of events & \multicolumn{1}{c}{\hspace{10pt}2,161 } & \multicolumn{1}{c}{\hspace{10pt}2,161 } & \multicolumn{1}{c}{\hspace{10pt}2,161 } & \multicolumn{1}{c}{\hspace{10pt}2,161 }\\
 Number of subject-periods (n) & \multicolumn{1}{c}{\hspace{10pt}93,954}& \multicolumn{1}{c}{\hspace{10pt}93,954}&\multicolumn{1}{c}{\hspace{10pt}93,954}&\multicolumn{1}{c}{\hspace{10pt}93,954}\\
\hline
\hline \\[-1.8ex]
\textbf{*}p$<$0.05; \textbf{**}p$<$0.01; \textbf{***}p$<$0.001
\end{tabular}
\end{center}
}}
\end{table}

\begin{table}[H]
{\fontsize{10.0}{10.0}\selectfont{
\caption{\textbf{Cox proportional hazard model of attrition including authors with multiple retractions.} In the table below, we present three models. The models differ in how authors' experience is measured using (1) number of papers, (2) logged number of citations, (3) logged number of collaborators, and (4) all experiences put together respectively. All models incorporate clustered standard errors at the author-level. Controls for author's scientific discipline are included as categorical variables, but are not shown.}
\label{supplementarytab:cox_model_multret}
\begin{center}
\begin{tabular}{@{\extracolsep{5pt}}lD{.}{.}{-3} D{.}{.}{-3} D{.}{.}{-3} D{.}{.}{-3} }
\\[-1.8ex]\hline
\hline \\[-1.8ex]
& \multicolumn{4}{c}{\textit{Outcome: Left publishing}} \
\cr \cline{2-5}\\[-1.8ex]
& \multicolumn{1}{c}{\hspace{10pt}(1)} & \multicolumn{1}{c}{\hspace{10pt}(2)} & \multicolumn{1}{c}{\hspace{10pt}(3)}& \multicolumn{1}{c}{\hspace{10pt}(4)} \\
\hline \\[-1.8ex]
\textbf{Retracted} & 1.77\textbf{***} & 1.78\textbf{***} & 1.76\textbf{***} & 1.76\textbf{***}\\
& (0.04) & (0.04) & (0.04) & (0.04)\\
\\
\textbf{Gender} (reference: Male)\\
Female & -0.01 & 0.01 & -0.00 & -0.01\\
& (0.03) & (0.03) & (0.03) & (0.03)\\
\\
\textbf{Affiliation Rank} & 0.00\textbf{*} & 0.00\textbf{**} & 0.00 & 0.00\\
& (0.00) & (0.00) & (0.00) & (0.00)\\
\\
\textbf{Cohort Year} & 0.02\textbf{***} & 0.02\textbf{***} & 0.03\textbf{***} & 0.03\textbf{***}\\
& (0.00) & (0.00) & (0.00) & (0.00) \\
\\
\textbf{Author's Experience}\\
\# Publications & -0.01\textbf{***} & & & -0.01\textbf{***}\\
& (0.00) & & & (0.00)\\
log(\# Citations) & & -0.07\textbf{***} & & 0.05\textbf{**}\\
& & (0.01) & & (0.01)\\
log(\# Collaborators) & & & -0.29\textbf{***} & -0.25\textbf{***}\\
& & &(0.02)& (0.02)\\
\hline \\[-1.8ex]
 Number of events & \multicolumn{1}{c}{\hspace{10pt}5,749 } & \multicolumn{1}{c}{\hspace{10pt}5,749 } & \multicolumn{1}{c}{\hspace{10pt}5,749 } & \multicolumn{1}{c}{\hspace{10pt}5,749 }\\
 Number of subject-periods (n) & \multicolumn{1}{c}{\hspace{10pt}264,138}& \multicolumn{1}{c}{\hspace{10pt}264,138}&\multicolumn{1}{c}{\hspace{10pt}264,138}&\multicolumn{1}{c}{\hspace{10pt}264,138}\\
\hline
\hline \\[-1.8ex]
\textbf{*}p$<$0.05; \textbf{**}p$<$0.01; \textbf{***}p$<$0.001
\end{tabular}
\end{center}
}}
\end{table}

\begin{table}[H]
\centering
\caption{\textbf{Standardized mean differences between authors who stayed within academic publishing and those that were matched}. $a$ denotes the proportion of authors who stayed within the filtered sample, and $m$ denotes the proportion of authors who were matched. For continuous variables, $a$ and $m$ represent the mean. $p$ represents the p-value of a two-sided chi-square test for proportions and a two-sided Welch's t-test for experience variables.}
\resizebox{\textwidth}{!}{%
\begin{tabular}{lccccc}
\toprule
\multicolumn{2}{c}{\textbf{Category}} & \textbf{$a$ $(n=10,342)$} & \textbf{$m$ $(n=2,348)$} & \textbf{Standardized Mean Difference} & \textbf{$p$} \\
\cmidrule(lr){1-2} \cmidrule(lr){3-3} \cmidrule(lr){4-4} \cmidrule(lr){5-5} \cmidrule(lr){6-6}
\multirow{2}{*}{Attention} & High attention & 0.04 & 0.03 & -0.04 & $0.1$ \\
& Low attention & 0.96 & 0.97 & 0.04 &  \\
\midrule
Academic age & & 13.44 & 8.55 & 0.53 & $<.001$ \\
% \midrule
\# papers & & 69.29 & 24.66 & 0.54 & $<.001$ \\
\# citations & & 1180.05 & 270.85 & 0.41 & $<.001$ \\
\# collaborators & & 197.56 & 57.26 & 0.33 & $<.001$ \\
\midrule
\multirow{2}{*}{Gender} & Female & 0.25 & 0.29 & 0.08 & $<.001$ \\
& Male & 0.75 & 0.71 & -0.08 &  \\
\midrule
\multirow{4}{*}{Affiliation Rank} & 1-100 & 0.24 & 0.21 & -0.06 & $<.001$ \\
& 101-500 & 0.31 & 0.55 & 0.49 & \\
& 501-1000 & 0.13 & 0.23 & 0.26 &  \\
& $>$1000 & 0.24 & 0.21 & -0.06 &  \\
\midrule
\multirow{4}{*}{Reason} & Misconduct & 0.22 & 0.23 & 0.02 & $0.11$ \\
& Plagiarism & 0.28 & 0.26 & -0.03 &  \\
& Mistake & 0.3 & 0.29 & -0.03 &  \\
& Other & 0.2 & 0.22 & 0.04 &  \\
\midrule
\multirow{2}{*}{Author order} & First or last author & 0.39 & 0.34 & -0.09 & $<.001$ \\
& Middle author & 0.61 & 0.66 & 0.09 &  \\
\midrule
\multirow{5}{*}{Year of retraction} & 1990-1995 & 0.02 & 0.02 & 0.01 & $0.569$ \\
& 1996-2000 & 0.03 & 0.02 & -0.04 &  \\
& 2001-2005 & 0.06 & 0.06 & -0.01 &  \\
& 2006-2010 & 0.26 & 0.26 & 0.0 &  \\
& 2011-2015 & 0.64 & 0.64 & 0.01 &  \\
\midrule
\multirow{4}{*}{Discipline} & Biology & 0.38 & 0.36 & -0.03 & $<.001$ \\
& Chemistry & 0.15 & 0.18 & 0.07 &  \\
& Medicine & 0.3 & 0.31 & 0.02 &  \\
& Physics & 0.06 & 0.06 & -0.02 &  \\
& Non-STEM fields & 0.01 & 0.0 & -0.04 &  \\
& Other STEM fields & 0.11 & 0.09 & -0.06 &  \\
\bottomrule
\end{tabular}
}
\label{supplementarytab:SMDs_nonattrited_matched}
\end{table}

\begin{table}[H]
{\fontsize{9}{9}\selectfont{
\caption{\textbf{Results of matching analysis on \textit{collaborator retention} for the authors who stayed}. $r$ and $r'$ represent retracted and non-retracted authors, respectively. $n_r$ represents the number of retracted authors. Each retracted author is compared to a closest average matched non-retracted author in terms of the collaborators retained 5 years post retraction. $\mu_r$ and $\mu_{r'}$ are the average number of collaborators retained for retracted and non-retracted authors respectively. $\mu_{\delta}$ is the average relative gain of $r$ over $r'$. $CI_{95\%}$ is the 95\% confidence interval for $\mu_\delta$. $p_t$, $p_{ks}$, and $p_w$ are the p-values for the two-sided Welch t-test, Kolmogorov–Smirnov test, and Wilcoxon signed ranked test, respectively.}
\label{supplementarytab:collaborator_retention_nonattrited}
\begin{center}
\begin{tabular}{lcccccccc}
\toprule
Group & $n_r$ & $\mu_r$ & $\mu_{r'}$ & $\mu_{\delta}$ & $CI_{95\%}$ & $p_t$ & $p_{ks}$ & $p_w$ \\
\midrule
\textbf{Overall} & \hspace*{0.1cm}$2,348$ & \hspace*{0.1cm}$9.847$ & \hspace*{0.1cm}$7.497$ & \hspace*{0.1cm}$2.349$ & [1.931, 2.768] & $<.001$ & $<.001$ & $<.001$ \\
\textbf{Gender} &  &  &  &  &  &  &  &  \\
Male & \hspace*{0.1cm}$1,661$ & \hspace*{0.1cm}$10.291$ & \hspace*{0.1cm}$7.686$ & \hspace*{0.1cm}$2.605$ & [2.085, 3.125] & $<.001$ & $<.001$ & $<.001$ \\
Female & \hspace*{0.1cm}$687$ & \hspace*{0.1cm}$8.771$ & \hspace*{0.1cm}$7.040$ & \hspace*{0.1cm}$1.732$ & [1.050, 2.413] & $<.001$ & \hspace*{0.1cm}$0.029$ & $<.001$ \\
\textbf{Year of retraction} &  &  &  &  &  &  &  &  \\
1990-1995 & \hspace*{0.1cm}$43$ & \hspace*{0.1cm}$4.744$ & \hspace*{0.1cm}$4.027$ & \hspace*{0.1cm}$0.717$ & [-1.201, 2.635] & \hspace*{0.1cm}$0.543$ & \hspace*{0.1cm}$0.451$ & \hspace*{0.1cm}$0.830$ \\
1996-2000 & \hspace*{0.1cm}$49$ & \hspace*{0.1cm}$5.469$ & \hspace*{0.1cm}$4.596$ & \hspace*{0.1cm}$0.873$ & [-1.238, 2.985] & \hspace*{0.1cm}$0.550$ & \hspace*{0.1cm}$0.964$ & \hspace*{0.1cm}$0.775$ \\
2001-2005 & \hspace*{0.1cm}$136$ & \hspace*{0.1cm}$8.529$ & \hspace*{0.1cm}$6.462$ & \hspace*{0.1cm}$2.067$ & [0.301, 3.834] & \hspace*{0.1cm}$0.043$ & \hspace*{0.1cm}$0.381$ & \hspace*{0.1cm}$0.219$ \\
2006-2010 & \hspace*{0.1cm}$614$ & \hspace*{0.1cm}$9.145$ & \hspace*{0.1cm}$7.787$ & \hspace*{0.1cm}$1.358$ & [0.663, 2.052] & \hspace*{0.1cm}$0.013$ & \hspace*{0.1cm}$0.168$ & \hspace*{0.1cm}$0.040$ \\
2011-2015 & \hspace*{0.1cm}$1,506$ & \hspace*{0.1cm}$10.540$ & \hspace*{0.1cm}$7.666$ & \hspace*{0.1cm}$2.874$ & [2.315, 3.433] & $<.001$ & $<.001$ & $<.001$ \\
\textbf{Author academic age} &  &  &  &  &  &  &  &  \\
1 year & \hspace*{0.1cm}$302$ & \hspace*{0.1cm}$4.629$ & \hspace*{0.1cm}$4.596$ & \hspace*{0.1cm}$0.033$ & [-0.382, 0.448] & \hspace*{0.1cm}$0.917$ & \hspace*{0.1cm}$0.006$ & \hspace*{0.1cm}$0.820$ \\
2 years & \hspace*{0.1cm}$230$ & \hspace*{0.1cm}$5.257$ & \hspace*{0.1cm}$5.401$ & \hspace*{0.1cm}$-0.144$ & [-0.799, 0.510] & \hspace*{0.1cm}$0.744$ & \hspace*{0.1cm}$0.294$ & \hspace*{0.1cm}$0.240$ \\
3-5 years & \hspace*{0.1cm}$526$ & \hspace*{0.1cm}$6.711$ & \hspace*{0.1cm}$5.811$ & \hspace*{0.1cm}$0.900$ & [0.391, 1.410] & \hspace*{0.1cm}$0.014$ & \hspace*{0.1cm}$0.002$ & $<.001$ \\
6 or more years & \hspace*{0.1cm}$1,290$ & \hspace*{0.1cm}$13.165$ & \hspace*{0.1cm}$9.238$ & \hspace*{0.1cm}$3.927$ & [3.221, 4.633] & $<.001$ & $<.001$ & $<.001$ \\
\textbf{Author affiliation rank} &  &  &  &  &  &  &  &  \\
1-100 & \hspace*{0.1cm}$501$ & \hspace*{0.1cm}$9.162$ & \hspace*{0.1cm}$6.807$ & \hspace*{0.1cm}$2.354$ & [1.491, 3.218] & $<.001$ & \hspace*{0.1cm}$0.014$ & $<.001$ \\
101-500 & \hspace*{0.1cm}$1,293$ & \hspace*{0.1cm}$10.649$ & \hspace*{0.1cm}$7.893$ & \hspace*{0.1cm}$2.756$ & [2.144, 3.368] & $<.001$ & $<.001$ & $<.001$ \\
501-1000 & \hspace*{0.1cm}$548$ & \hspace*{0.1cm}$8.637$ & \hspace*{0.1cm}$7.145$ & \hspace*{0.1cm}$1.492$ & [0.791, 2.193] & \hspace*{0.1cm}$0.004$ & \hspace*{0.1cm}$0.042$ & $<.001$ \\
$>1000$ & \hspace*{0.1cm}$6$ & \hspace*{0.1cm}$4.667$ & \hspace*{0.1cm}$12.000$ & \hspace*{0.1cm}$-7.333$ & [-19.494, 4.827] & \hspace*{0.1cm}$0.334$ & \hspace*{0.1cm}$0.931$ & \hspace*{0.1cm}$0.080$ \\
\textbf{Author order} &  &  &  &  &  &  &  &  \\
First or last author & \hspace*{0.1cm}$803$ & \hspace*{0.1cm}$8.274$ & \hspace*{0.1cm}$6.726$ & \hspace*{0.1cm}$1.548$ & [0.932, 2.164] & $<.001$ & \hspace*{0.1cm}$0.128$ & $<.001$ \\
Middle author & \hspace*{0.1cm}$1,545$ & \hspace*{0.1cm}$10.664$ & \hspace*{0.1cm}$7.898$ & \hspace*{0.1cm}$2.766$ & [2.217, 3.315] & $<.001$ & $<.001$ & $<.001$ \\
\textbf{Reason of retraction} &  &  &  &  &  &  &  &  \\
Misconduct & \hspace*{0.1cm}$543$ & \hspace*{0.1cm}$9.980$ & \hspace*{0.1cm}$7.608$ & \hspace*{0.1cm}$2.372$ & [1.516, 3.227] & $<.001$ & \hspace*{0.1cm}$0.011$ & $<.001$ \\
Plagiarism & \hspace*{0.1cm}$611$ & \hspace*{0.1cm}$9.293$ & \hspace*{0.1cm}$7.616$ & \hspace*{0.1cm}$1.677$ & [0.913, 2.441] & \hspace*{0.1cm}$0.003$ & \hspace*{0.1cm}$0.073$ & \hspace*{0.1cm}$0.003$ \\
Mistake & \hspace*{0.1cm}$674$ & \hspace*{0.1cm}$10.806$ & \hspace*{0.1cm}$7.748$ & \hspace*{0.1cm}$3.058$ & [2.231, 3.884] & $<.001$ & $<.001$ & $<.001$ \\
Other & \hspace*{0.1cm}$520$ & \hspace*{0.1cm}$9.115$ & \hspace*{0.1cm}$6.917$ & \hspace*{0.1cm}$2.198$ & [1.289, 3.108] & $<.001$ & \hspace*{0.1cm}$0.048$ & $<.001$ \\
\textbf{Type of retraction} &  &  &  &  &  &  &  &  \\
Author-led retraction & \hspace*{0.1cm}$586$ & \hspace*{0.1cm}$9.773$ & \hspace*{0.1cm}$6.972$ & \hspace*{0.1cm}$2.801$ & [1.928, 3.674] & $<.001$ & $<.001$ & $<.001$ \\
Journal-led retraction & \hspace*{0.1cm}$404$ & \hspace*{0.1cm}$8.896$ & \hspace*{0.1cm}$7.154$ & \hspace*{0.1cm}$1.742$ & [0.952, 2.532] & \hspace*{0.1cm}$0.004$ & \hspace*{0.1cm}$0.159$ & $<.001$ \\
\textbf{Attention} &  &  &  &  &  &  &  &  \\
High ($>10$ Altmetric score) & \hspace*{0.1cm}$288$ & \hspace*{0.1cm}$12.250$ & \hspace*{0.1cm}$8.106$ & \hspace*{0.1cm}$4.144$ & [2.671, 5.617] & $<.001$ & \hspace*{0.1cm}$0.006$ & $<.001$ \\
Low & \hspace*{0.1cm}$2,060$ & \hspace*{0.1cm}$9.511$ & \hspace*{0.1cm}$7.412$ & \hspace*{0.1cm}$2.099$ & [1.669, 2.528] & $<.001$ & $<.001$ & $<.001$ \\
\textbf{Publication-retraction interval} &  &  &  &  &  &  &  &  \\
0-1 year & \hspace*{0.1cm}$1,506$ & \hspace*{0.1cm}$9.683$ & \hspace*{0.1cm}$7.282$ & \hspace*{0.1cm}$2.401$ & [1.887, 2.915] & $<.001$ & $<.001$ & $<.001$ \\
2-5 years & \hspace*{0.1cm}$694$ & \hspace*{0.1cm}$10.216$ & \hspace*{0.1cm}$7.774$ & \hspace*{0.1cm}$2.442$ & [1.655, 3.228] & $<.001$ & \hspace*{0.1cm}$0.003$ & $<.001$ \\
6 or more years  & \hspace*{0.1cm}$148$ & \hspace*{0.1cm}$9.784$ & \hspace*{0.1cm}$8.392$ & \hspace*{0.1cm}$1.392$ & [-0.410, 3.194] & \hspace*{0.1cm}$0.221$ & \hspace*{0.1cm}$0.983$ & \hspace*{0.1cm}$0.162$ \\
\textbf{Discipline} &  &  &  &  &  &  &  &  \\
Biology & \hspace*{0.1cm}$1,057$ & \hspace*{0.1cm}$9.187$ & \hspace*{0.1cm}$6.819$ & \hspace*{0.1cm}$2.368$ & [1.809, 2.928] & $<.001$ & $<.001$ & $<.001$ \\
Chemistry & \hspace*{0.1cm}$409$ & \hspace*{0.1cm}$6.440$ & \hspace*{0.1cm}$5.507$ & \hspace*{0.1cm}$0.933$ & [0.354, 1.513] & \hspace*{0.1cm}$0.035$ & \hspace*{0.1cm}$0.027$ & \hspace*{0.1cm}$0.013$ \\
Medicine & \hspace*{0.1cm}$890$ & \hspace*{0.1cm}$11.333$ & \hspace*{0.1cm}$8.525$ & \hspace*{0.1cm}$2.807$ & [2.026, 3.589] & $<.001$ & \hspace*{0.1cm}$0.001$ & $<.001$ \\
Physics & \hspace*{0.1cm}$157$ & \hspace*{0.1cm}$8.057$ & \hspace*{0.1cm}$6.538$ & \hspace*{0.1cm}$1.520$ & [-0.324, 3.363] & \hspace*{0.1cm}$0.191$ & \hspace*{0.1cm}$0.390$ & \hspace*{0.1cm}$0.646$ \\
Other STEM fields & \hspace*{0.1cm}$177$ & \hspace*{0.1cm}$4.881$ & \hspace*{0.1cm}$4.068$ & \hspace*{0.1cm}$0.813$ & [0.075, 1.551] & \hspace*{0.1cm}$0.123$ & \hspace*{0.1cm}$0.003$ & \hspace*{0.1cm}$0.035$ \\
Non-STEM fields & \hspace*{0.1cm}$30$ & \hspace*{0.1cm}$5.133$ & \hspace*{0.1cm}$4.612$ & \hspace*{0.1cm}$0.521$ & [-1.019, 2.061] & \hspace*{0.1cm}$0.816$ & \hspace*{0.1cm}$0.999$ & \hspace*{0.1cm}$0.455$ \\
\bottomrule
\end{tabular}

\end{center}
}}
\end{table}

\begin{table}[H]
{\fontsize{9}{9}\selectfont{
\caption{\textbf{Results of matching analysis on \textit{collaborator gained} for the authors who stayed}. $r$ and $r'$ represent retracted and non-retracted authors, respectively. $n_r$ represents the number of retracted authors. Each retracted author is compared to a closest average matched non-retracted author in terms of the collaborators gained 5 years post retraction. $\mu_r$ and $\mu_{r'}$ are the average number of collaborators gained for retracted and non-retracted authors respectively. $\mu_{\delta}$ is the average relative gain of $r$ over $r'$. $CI_{95\%}$ is the 95\% confidence interval for $\mu_\delta$. $p_t$, $p_{ks}$, and $p_w$ are the p-values for the two-sided Welch t-test, Kolmogorov–Smirnov test, and Wilcoxon signed ranked test, respectively.}
\label{supplementarytab:collaborator_gain_nonattrited}
\begin{center}
\begin{tabular}{lcccccccc}
\toprule
Group & $n_r$ & $\mu_r$ & $\mu_{r'}$ & $\mu_{\delta}$ & $CI_{95\%}$ & $p_t$ & $p_{ks}$ & $p_w$ \\
\midrule
\textbf{Overall} & \hspace*{0.1cm}$2,348$ & \hspace*{0.1cm}$45.385$ & \hspace*{0.1cm}$27.051$ & \hspace*{0.1cm}$18.335$ & [12.831, 23.838] & $<.001$ & $<.001$ & $<.001$ \\
\textbf{Gender} &  &  &  &  &  &  &  &  \\
Male & \hspace*{0.1cm}$1,661$ & \hspace*{0.1cm}$44.358$ & \hspace*{0.1cm}$27.044$ & \hspace*{0.1cm}$17.314$ & [12.032, 22.596] & $<.001$ & $<.001$ & $<.001$ \\
Female & \hspace*{0.1cm}$687$ & \hspace*{0.1cm}$47.869$ & \hspace*{0.1cm}$27.067$ & \hspace*{0.1cm}$20.802$ & [6.965, 34.639] & \hspace*{0.1cm}$0.004$ & \hspace*{0.1cm}$0.029$ & \hspace*{0.1cm}$0.568$ \\
\textbf{Year of retraction} &  &  &  &  &  &  &  &  \\
1990-1995 & \hspace*{0.1cm}$43$ & \hspace*{0.1cm}$24.163$ & \hspace*{0.1cm}$10.421$ & \hspace*{0.1cm}$13.742$ & [3.403, 24.081] & \hspace*{0.1cm}$0.013$ & \hspace*{0.1cm}$0.120$ & \hspace*{0.1cm}$0.009$ \\
1996-2000 & \hspace*{0.1cm}$49$ & \hspace*{0.1cm}$18.020$ & \hspace*{0.1cm}$12.388$ & \hspace*{0.1cm}$5.632$ & [-1.273, 12.537] & \hspace*{0.1cm}$0.209$ & \hspace*{0.1cm}$0.997$ & \hspace*{0.1cm}$0.409$ \\
2001-2005 & \hspace*{0.1cm}$136$ & \hspace*{0.1cm}$34.640$ & \hspace*{0.1cm}$18.808$ & \hspace*{0.1cm}$15.831$ & [5.628, 26.034] & \hspace*{0.1cm}$0.008$ & \hspace*{0.1cm}$0.020$ & $<.001$ \\
2006-2010 & \hspace*{0.1cm}$614$ & \hspace*{0.1cm}$40.632$ & \hspace*{0.1cm}$27.240$ & \hspace*{0.1cm}$13.392$ & [2.300, 24.485] & \hspace*{0.1cm}$0.021$ & \hspace*{0.1cm}$0.085$ & \hspace*{0.1cm}$0.052$ \\
2011-2015 & \hspace*{0.1cm}$1,506$ & \hspace*{0.1cm}$49.790$ & \hspace*{0.1cm}$28.670$ & \hspace*{0.1cm}$21.120$ & [13.890, 28.351] & $<.001$ & $<.001$ & $<.001$ \\
\textbf{Author academic age} &  &  &  &  &  &  &  &  \\
1 year & \hspace*{0.1cm}$302$ & \hspace*{0.1cm}$13.397$ & \hspace*{0.1cm}$17.843$ & \hspace*{0.1cm}$-4.446$ & [-9.225, 0.333] & \hspace*{0.1cm}$0.069$ & $<.001$ & \hspace*{0.1cm}$0.012$ \\
2 years & \hspace*{0.1cm}$230$ & \hspace*{0.1cm}$19.291$ & \hspace*{0.1cm}$18.710$ & \hspace*{0.1cm}$0.582$ & [-4.576, 5.740] & \hspace*{0.1cm}$0.827$ & \hspace*{0.1cm}$0.040$ & \hspace*{0.1cm}$0.358$ \\
3-5 years & \hspace*{0.1cm}$526$ & \hspace*{0.1cm}$32.418$ & \hspace*{0.1cm}$21.063$ & \hspace*{0.1cm}$11.356$ & [2.245, 20.467] & \hspace*{0.1cm}$0.018$ & \hspace*{0.1cm}$0.252$ & \hspace*{0.1cm}$0.156$ \\
6 or more years & \hspace*{0.1cm}$1,290$ & \hspace*{0.1cm}$62.814$ & \hspace*{0.1cm}$33.135$ & \hspace*{0.1cm}$29.679$ & [20.532, 38.826] & $<.001$ & $<.001$ & $<.001$ \\
\textbf{Author affiliation rank} &  &  &  &  &  &  &  &  \\
1-100 & \hspace*{0.1cm}$501$ & \hspace*{0.1cm}$49.046$ & \hspace*{0.1cm}$25.416$ & \hspace*{0.1cm}$23.630$ & [10.972, 36.287] & $<.001$ & \hspace*{0.1cm}$0.006$ & $<.001$ \\
101-500 & \hspace*{0.1cm}$1,293$ & \hspace*{0.1cm}$48.810$ & \hspace*{0.1cm}$29.254$ & \hspace*{0.1cm}$19.555$ & [11.322, 27.789] & $<.001$ & \hspace*{0.1cm}$0.004$ & $<.001$ \\
501-1000 & \hspace*{0.1cm}$548$ & \hspace*{0.1cm}$34.201$ & \hspace*{0.1cm}$23.366$ & \hspace*{0.1cm}$10.835$ & [4.089, 17.581] & \hspace*{0.1cm}$0.002$ & \hspace*{0.1cm}$0.021$ & \hspace*{0.1cm}$0.041$ \\
$>1000$ & \hspace*{0.1cm}$6$ & \hspace*{0.1cm}$23.333$ & \hspace*{0.1cm}$25.167$ & \hspace*{0.1cm}$-1.833$ & [-26.997, 23.331] & \hspace*{0.1cm}$0.908$ & \hspace*{0.1cm}$0.931$ & \hspace*{0.1cm}$1.000$ \\
\textbf{Author order} &  &  &  &  &  &  &  &  \\
First or last author & \hspace*{0.1cm}$803$ & \hspace*{0.1cm}$42.182$ & \hspace*{0.1cm}$24.703$ & \hspace*{0.1cm}$17.478$ & [8.676, 26.281] & $<.001$ & \hspace*{0.1cm}$0.004$ & $<.001$ \\
Middle author & \hspace*{0.1cm}$1,545$ & \hspace*{0.1cm}$47.050$ & \hspace*{0.1cm}$28.271$ & \hspace*{0.1cm}$18.780$ & [11.771, 25.788] & $<.001$ & $<.001$ & $<.001$ \\
\textbf{Reason of retraction} &  &  &  &  &  &  &  &  \\
Misconduct & \hspace*{0.1cm}$543$ & \hspace*{0.1cm}$39.792$ & \hspace*{0.1cm}$28.321$ & \hspace*{0.1cm}$11.471$ & [2.523, 20.418] & \hspace*{0.1cm}$0.018$ & \hspace*{0.1cm}$0.078$ & \hspace*{0.1cm}$0.241$ \\
Plagiarism & \hspace*{0.1cm}$611$ & \hspace*{0.1cm}$42.646$ & \hspace*{0.1cm}$26.257$ & \hspace*{0.1cm}$16.390$ & [6.325, 26.454] & \hspace*{0.1cm}$0.002$ & \hspace*{0.1cm}$0.028$ & \hspace*{0.1cm}$0.005$ \\
Mistake & \hspace*{0.1cm}$674$ & \hspace*{0.1cm}$60.518$ & \hspace*{0.1cm}$28.493$ & \hspace*{0.1cm}$32.025$ & [17.633, 46.416] & $<.001$ & $<.001$ & $<.001$ \\
Other & \hspace*{0.1cm}$520$ & \hspace*{0.1cm}$34.831$ & \hspace*{0.1cm}$24.787$ & \hspace*{0.1cm}$10.044$ & [3.544, 16.544] & \hspace*{0.1cm}$0.005$ & \hspace*{0.1cm}$0.165$ & \hspace*{0.1cm}$0.257$ \\
\textbf{Type of retraction} &  &  &  &  &  &  &  &  \\
Author-led retraction & \hspace*{0.1cm}$586$ & \hspace*{0.1cm}$42.282$ & \hspace*{0.1cm}$27.133$ & \hspace*{0.1cm}$15.149$ & [6.979, 23.318] & $<.001$ & \hspace*{0.1cm}$0.017$ & $<.001$ \\
Journal-led retraction & \hspace*{0.1cm}$404$ & \hspace*{0.1cm}$45.374$ & \hspace*{0.1cm}$25.408$ & \hspace*{0.1cm}$19.966$ & [4.239, 35.694] & \hspace*{0.1cm}$0.016$ & \hspace*{0.1cm}$0.588$ & \hspace*{0.1cm}$0.360$ \\
\textbf{Attention} &  &  &  &  &  &  &  &  \\
High ($>10$ Altmetric score) & \hspace*{0.1cm}$288$ & \hspace*{0.1cm}$68.292$ & \hspace*{0.1cm}$34.011$ & \hspace*{0.1cm}$34.280$ & [12.821, 55.740] & \hspace*{0.1cm}$0.003$ & \hspace*{0.1cm}$0.028$ & $<.001$ \\
Low & \hspace*{0.1cm}$2,060$ & \hspace*{0.1cm}$42.183$ & \hspace*{0.1cm}$26.078$ & \hspace*{0.1cm}$16.105$ & [10.594, 21.617] & $<.001$ & $<.001$ & $<.001$ \\
\textbf{Publication-retraction interval} &  &  &  &  &  &  &  &  \\
0-1 year & \hspace*{0.1cm}$1,506$ & \hspace*{0.1cm}$46.317$ & \hspace*{0.1cm}$25.764$ & \hspace*{0.1cm}$20.554$ & [13.043, 28.065] & $<.001$ & $<.001$ & $<.001$ \\
2-5 years & \hspace*{0.1cm}$694$ & \hspace*{0.1cm}$39.898$ & \hspace*{0.1cm}$28.317$ & \hspace*{0.1cm}$11.581$ & [6.742, 16.419] & $<.001$ & \hspace*{0.1cm}$0.005$ & $<.001$ \\
6 or more years  & \hspace*{0.1cm}$148$ & \hspace*{0.1cm}$61.635$ & \hspace*{0.1cm}$34.209$ & \hspace*{0.1cm}$27.426$ & [-8.528, 63.379] & \hspace*{0.1cm}$0.144$ & \hspace*{0.1cm}$0.224$ & \hspace*{0.1cm}$0.944$ \\
\textbf{Discipline} &  &  &  &  &  &  &  &  \\
Biology & \hspace*{0.1cm}$1,057$ & \hspace*{0.1cm}$42.362$ & \hspace*{0.1cm}$26.336$ & \hspace*{0.1cm}$16.026$ & [8.028, 24.023] & $<.001$ & \hspace*{0.1cm}$0.015$ & \hspace*{0.1cm}$0.002$ \\
Chemistry & \hspace*{0.1cm}$409$ & \hspace*{0.1cm}$22.323$ & \hspace*{0.1cm}$19.876$ & \hspace*{0.1cm}$2.447$ & [-0.318, 5.212] & \hspace*{0.1cm}$0.185$ & \hspace*{0.1cm}$0.430$ & \hspace*{0.1cm}$0.272$ \\
Medicine & \hspace*{0.1cm}$890$ & \hspace*{0.1cm}$53.309$ & \hspace*{0.1cm}$27.996$ & \hspace*{0.1cm}$25.313$ & [14.999, 35.627] & $<.001$ & $<.001$ & $<.001$ \\
Physics & \hspace*{0.1cm}$157$ & \hspace*{0.1cm}$41.809$ & \hspace*{0.1cm}$28.343$ & \hspace*{0.1cm}$13.466$ & [-7.046, 33.979] & \hspace*{0.1cm}$0.216$ & \hspace*{0.1cm}$0.472$ & \hspace*{0.1cm}$0.451$ \\
Other STEM fields & \hspace*{0.1cm}$177$ & \hspace*{0.1cm}$18.610$ & \hspace*{0.1cm}$14.991$ & \hspace*{0.1cm}$3.619$ & [-0.204, 7.442] & \hspace*{0.1cm}$0.124$ & \hspace*{0.1cm}$0.639$ & \hspace*{0.1cm}$0.246$ \\
Non-STEM fields & \hspace*{0.1cm}$30$ & \hspace*{0.1cm}$20.967$ & \hspace*{0.1cm}$16.131$ & \hspace*{0.1cm}$4.836$ & [-4.175, 13.847] & \hspace*{0.1cm}$0.614$ & \hspace*{0.1cm}$1.000$ & \hspace*{0.1cm}$0.381$ \\
\bottomrule
\end{tabular}

\end{center}
}}
\end{table}

\begin{table}[H]
{\fontsize{9}{9}\selectfont{
\caption{\textbf{Results of matching analysis on \textit{triadic closure} for the authors who stayed}. $r$ and $r'$ represent retracted and non-retracted authors respectively. $n_r$ represents the number of retracted authors. Each retracted author is compared to a closest average matched non-retracted author in terms of the ratio of the triads closed 5 years post-retraction. $\mu_r$ and $\mu_{r'}$ are the average ratio of the triads closed for retracted and non-retracted authors respectively. $\mu_{\delta}$ is the average relative gain of $r$ over $r'$. $CI_{95\%}$ is the 95\% confidence interval for $\mu_\delta$. $p_t$, $p_{ks}$, and $p_w$ are the p-values for the two-sided Welch t-test, Kolmogorov–Smirnov test, and Wilcoxon signed ranked test, respectively.}
\label{supplementarytab:collaborator_triadicclosure_nonattrited}
\begin{center}
\begin{tabular}{lcccccccc}
\toprule
Group & $n_r$ & $\mu_r$ & $\mu_{r'}$ & $\mu_{\delta}$ & $CI_{95\%}$ & $p_t$ & $p_{ks}$ & $p_w$ \\
\midrule
\textbf{Overall} & \hspace*{0.1cm}$2,348$ & \hspace*{0.1cm}$0.019$ & \hspace*{0.1cm}$0.018$ & \hspace*{0.1cm}$0.001$ & [-0.002, 0.005] & \hspace*{0.1cm}$0.503$ & \hspace*{0.1cm}$0.588$ & \hspace*{0.1cm}$0.161$ \\
\textbf{Gender} &  &  &  &  &  &  &  &  \\
Male & \hspace*{0.1cm}$1,661$ & \hspace*{0.1cm}$0.019$ & \hspace*{0.1cm}$0.018$ & \hspace*{0.1cm}$0.001$ & [-0.003, 0.005] & \hspace*{0.1cm}$0.499$ & \hspace*{0.1cm}$0.918$ & \hspace*{0.1cm}$0.472$ \\
Female & \hspace*{0.1cm}$687$ & \hspace*{0.1cm}$0.018$ & \hspace*{0.1cm}$0.017$ & \hspace*{0.1cm}$0.001$ & [-0.006, 0.008] & \hspace*{0.1cm}$0.845$ & \hspace*{0.1cm}$0.487$ & \hspace*{0.1cm}$0.137$ \\
\textbf{Year of retraction} &  &  &  &  &  &  &  &  \\
1990-1995 & \hspace*{0.1cm}$43$ & \hspace*{0.1cm}$0.020$ & \hspace*{0.1cm}$0.011$ & \hspace*{0.1cm}$0.008$ & [-0.022, 0.039] & \hspace*{0.1cm}$0.571$ & \hspace*{0.1cm}$0.994$ & \hspace*{0.1cm}$0.594$ \\
1996-2000 & \hspace*{0.1cm}$49$ & \hspace*{0.1cm}$0.016$ & \hspace*{0.1cm}$0.013$ & \hspace*{0.1cm}$0.003$ & [-0.021, 0.026] & \hspace*{0.1cm}$0.817$ & \hspace*{0.1cm}$0.705$ & \hspace*{0.1cm}$0.372$ \\
2001-2005 & \hspace*{0.1cm}$136$ & \hspace*{0.1cm}$0.019$ & \hspace*{0.1cm}$0.021$ & \hspace*{0.1cm}$-0.002$ & [-0.017, 0.014] & \hspace*{0.1cm}$0.835$ & \hspace*{0.1cm}$0.929$ & \hspace*{0.1cm}$0.581$ \\
2006-2010 & \hspace*{0.1cm}$614$ & \hspace*{0.1cm}$0.022$ & \hspace*{0.1cm}$0.021$ & \hspace*{0.1cm}$0.000$ & [-0.007, 0.008] & \hspace*{0.1cm}$0.935$ & \hspace*{0.1cm}$0.866$ & \hspace*{0.1cm}$0.195$ \\
2011-2015 & \hspace*{0.1cm}$1,506$ & \hspace*{0.1cm}$0.018$ & \hspace*{0.1cm}$0.016$ & \hspace*{0.1cm}$0.002$ & [-0.002, 0.006] & \hspace*{0.1cm}$0.447$ & \hspace*{0.1cm}$0.754$ & \hspace*{0.1cm}$0.520$ \\
\textbf{Author academic age} &  &  &  &  &  &  &  &  \\
1 year & \hspace*{0.1cm}$302$ & \hspace*{0.1cm}$0.025$ & \hspace*{0.1cm}$0.028$ & \hspace*{0.1cm}$-0.003$ & [-0.015, 0.008] & \hspace*{0.1cm}$0.593$ & $<.001$ & $<.001$ \\
2 years & \hspace*{0.1cm}$230$ & \hspace*{0.1cm}$0.015$ & \hspace*{0.1cm}$0.024$ & \hspace*{0.1cm}$-0.009$ & [-0.019, 0.000] & \hspace*{0.1cm}$0.054$ & \hspace*{0.1cm}$0.031$ & \hspace*{0.1cm}$0.015$ \\
3-5 years & \hspace*{0.1cm}$526$ & \hspace*{0.1cm}$0.021$ & \hspace*{0.1cm}$0.017$ & \hspace*{0.1cm}$0.004$ & [-0.003, 0.011] & \hspace*{0.1cm}$0.276$ & \hspace*{0.1cm}$0.359$ & \hspace*{0.1cm}$0.104$ \\
6 or more years & \hspace*{0.1cm}$1,290$ & \hspace*{0.1cm}$0.017$ & \hspace*{0.1cm}$0.014$ & \hspace*{0.1cm}$0.003$ & [-0.002, 0.008] & \hspace*{0.1cm}$0.206$ & \hspace*{0.1cm}$0.161$ & \hspace*{0.1cm}$0.220$ \\
\textbf{Author affiliation rank} &  &  &  &  &  &  &  &  \\
1-100 & \hspace*{0.1cm}$501$ & \hspace*{0.1cm}$0.017$ & \hspace*{0.1cm}$0.016$ & \hspace*{0.1cm}$0.000$ & [-0.007, 0.008] & \hspace*{0.1cm}$0.942$ & \hspace*{0.1cm}$0.461$ & \hspace*{0.1cm}$0.797$ \\
101-500 & \hspace*{0.1cm}$1,293$ & \hspace*{0.1cm}$0.021$ & \hspace*{0.1cm}$0.017$ & \hspace*{0.1cm}$0.004$ & [-0.002, 0.009] & \hspace*{0.1cm}$0.177$ & \hspace*{0.1cm}$0.247$ & \hspace*{0.1cm}$0.216$ \\
501-1000 & \hspace*{0.1cm}$548$ & \hspace*{0.1cm}$0.016$ & \hspace*{0.1cm}$0.019$ & \hspace*{0.1cm}$-0.004$ & [-0.009, 0.002] & \hspace*{0.1cm}$0.237$ & \hspace*{0.1cm}$0.345$ & \hspace*{0.1cm}$0.318$ \\
$>1000$ & \hspace*{0.1cm}$6$ & \hspace*{0.1cm}$0.019$ & \hspace*{0.1cm}$0.006$ & \hspace*{0.1cm}$0.012$ & [-0.039, 0.064] & \hspace*{0.1cm}$0.546$ & \hspace*{0.1cm}$0.931$ & \hspace*{0.1cm}$0.715$ \\
\textbf{Author order} &  &  &  &  &  &  &  &  \\
First or last author & \hspace*{0.1cm}$803$ & \hspace*{0.1cm}$0.020$ & \hspace*{0.1cm}$0.018$ & \hspace*{0.1cm}$0.002$ & [-0.005, 0.008] & \hspace*{0.1cm}$0.641$ & \hspace*{0.1cm}$0.330$ & \hspace*{0.1cm}$0.195$ \\
Middle author & \hspace*{0.1cm}$1,545$ & \hspace*{0.1cm}$0.018$ & \hspace*{0.1cm}$0.017$ & \hspace*{0.1cm}$0.001$ & [-0.003, 0.005] & \hspace*{0.1cm}$0.629$ & \hspace*{0.1cm}$0.962$ & \hspace*{0.1cm}$0.416$ \\
\textbf{Reason of retraction} &  &  &  &  &  &  &  &  \\
Misconduct & \hspace*{0.1cm}$543$ & \hspace*{0.1cm}$0.022$ & \hspace*{0.1cm}$0.019$ & \hspace*{0.1cm}$0.003$ & [-0.006, 0.012] & \hspace*{0.1cm}$0.524$ & \hspace*{0.1cm}$0.855$ & \hspace*{0.1cm}$0.386$ \\
Plagiarism & \hspace*{0.1cm}$611$ & \hspace*{0.1cm}$0.017$ & \hspace*{0.1cm}$0.017$ & \hspace*{0.1cm}$0.000$ & [-0.006, 0.006] & \hspace*{0.1cm}$0.996$ & \hspace*{0.1cm}$0.372$ & \hspace*{0.1cm}$0.243$ \\
Mistake & \hspace*{0.1cm}$674$ & \hspace*{0.1cm}$0.019$ & \hspace*{0.1cm}$0.015$ & \hspace*{0.1cm}$0.004$ & [-0.003, 0.010] & \hspace*{0.1cm}$0.262$ & \hspace*{0.1cm}$0.698$ & \hspace*{0.1cm}$0.520$ \\
Other & \hspace*{0.1cm}$520$ & \hspace*{0.1cm}$0.018$ & \hspace*{0.1cm}$0.020$ & \hspace*{0.1cm}$-0.002$ & [-0.009, 0.005] & \hspace*{0.1cm}$0.551$ & \hspace*{0.1cm}$0.485$ & \hspace*{0.1cm}$0.153$ \\
\textbf{Type of retraction} &  &  &  &  &  &  &  &  \\
Author-led retraction & \hspace*{0.1cm}$586$ & \hspace*{0.1cm}$0.020$ & \hspace*{0.1cm}$0.018$ & \hspace*{0.1cm}$0.001$ & [-0.005, 0.008] & \hspace*{0.1cm}$0.760$ & \hspace*{0.1cm}$0.991$ & \hspace*{0.1cm}$0.587$ \\
Journal-led retraction & \hspace*{0.1cm}$404$ & \hspace*{0.1cm}$0.020$ & \hspace*{0.1cm}$0.015$ & \hspace*{0.1cm}$0.005$ & [-0.005, 0.014] & \hspace*{0.1cm}$0.324$ & \hspace*{0.1cm}$0.038$ & \hspace*{0.1cm}$0.189$ \\
\textbf{Attention} &  &  &  &  &  &  &  &  \\
High ($>10$ Altmetric score) & \hspace*{0.1cm}$288$ & \hspace*{0.1cm}$0.016$ & \hspace*{0.1cm}$0.010$ & \hspace*{0.1cm}$0.005$ & [-0.001, 0.012] & \hspace*{0.1cm}$0.109$ & \hspace*{0.1cm}$0.191$ & \hspace*{0.1cm}$0.020$ \\
Low & \hspace*{0.1cm}$2,060$ & \hspace*{0.1cm}$0.019$ & \hspace*{0.1cm}$0.019$ & \hspace*{0.1cm}$0.001$ & [-0.003, 0.004] & \hspace*{0.1cm}$0.756$ & \hspace*{0.1cm}$0.150$ & \hspace*{0.1cm}$0.023$ \\
\textbf{Publication-retraction interval} &  &  &  &  &  &  &  &  \\
0-1 year & \hspace*{0.1cm}$1,506$ & \hspace*{0.1cm}$0.022$ & \hspace*{0.1cm}$0.020$ & \hspace*{0.1cm}$0.002$ & [-0.002, 0.007] & \hspace*{0.1cm}$0.330$ & \hspace*{0.1cm}$0.016$ & \hspace*{0.1cm}$0.019$ \\
2-5 years & \hspace*{0.1cm}$694$ & \hspace*{0.1cm}$0.014$ & \hspace*{0.1cm}$0.014$ & \hspace*{0.1cm}$-0.000$ & [-0.005, 0.005] & \hspace*{0.1cm}$0.911$ & \hspace*{0.1cm}$0.278$ & \hspace*{0.1cm}$0.317$ \\
6 or more years  & \hspace*{0.1cm}$148$ & \hspace*{0.1cm}$0.008$ & \hspace*{0.1cm}$0.012$ & \hspace*{0.1cm}$-0.005$ & [-0.013, 0.004] & \hspace*{0.1cm}$0.296$ & \hspace*{0.1cm}$0.890$ & \hspace*{0.1cm}$0.403$ \\
\textbf{Discipline} &  &  &  &  &  &  &  &  \\
Biology & \hspace*{0.1cm}$1,057$ & \hspace*{0.1cm}$0.016$ & \hspace*{0.1cm}$0.016$ & \hspace*{0.1cm}$0.000$ & [-0.005, 0.005] & \hspace*{0.1cm}$0.994$ & \hspace*{0.1cm}$0.436$ & \hspace*{0.1cm}$0.131$ \\
Chemistry & \hspace*{0.1cm}$409$ & \hspace*{0.1cm}$0.027$ & \hspace*{0.1cm}$0.023$ & \hspace*{0.1cm}$0.004$ & [-0.006, 0.014] & \hspace*{0.1cm}$0.464$ & \hspace*{0.1cm}$0.070$ & \hspace*{0.1cm}$0.202$ \\
Medicine & \hspace*{0.1cm}$890$ & \hspace*{0.1cm}$0.021$ & \hspace*{0.1cm}$0.017$ & \hspace*{0.1cm}$0.004$ & [-0.002, 0.011] & \hspace*{0.1cm}$0.159$ & \hspace*{0.1cm}$0.999$ & \hspace*{0.1cm}$0.894$ \\
Physics & \hspace*{0.1cm}$157$ & \hspace*{0.1cm}$0.016$ & \hspace*{0.1cm}$0.020$ & \hspace*{0.1cm}$-0.005$ & [-0.016, 0.007] & \hspace*{0.1cm}$0.449$ & \hspace*{0.1cm}$0.121$ & \hspace*{0.1cm}$0.084$ \\
Other STEM fields & \hspace*{0.1cm}$177$ & \hspace*{0.1cm}$0.016$ & \hspace*{0.1cm}$0.023$ & \hspace*{0.1cm}$-0.007$ & [-0.018, 0.005] & \hspace*{0.1cm}$0.282$ & \hspace*{0.1cm}$0.885$ & \hspace*{0.1cm}$0.218$ \\
Non-STEM fields & \hspace*{0.1cm}$30$ & \hspace*{0.1cm}$0.042$ & \hspace*{0.1cm}$0.038$ & \hspace*{0.1cm}$0.004$ & [-0.075, 0.084] & \hspace*{0.1cm}$0.910$ & \hspace*{0.1cm}$0.393$ & \hspace*{0.1cm}$0.360$ \\
\bottomrule
\end{tabular}

\end{center}
}}
\end{table}

% Tables for collaborator characteristics

% Table showing retention 

\begin{table}[htbp]
\centering
\caption{Characterizing differences in \textbf{collaborators retained} between retracted ($a$) and non-retracted ($a'$) authors stratified by academic age. $p$-Value is calculated using two-sided Welch's t-test; *$p<0.05$; **$p<0.01$; ***$p<0.001$.}
\label{tab:collabAnalysisByAge_retained}
\begin{adjustbox}{max width=.95\textwidth}
\begin{tabular}{|c|c|c|c|c|c|c|c|c|c|}
\hline
\multirow{2}{*}{\textbf{Metric}} & &
\multicolumn{2}{c|}{\textbf{Overall}} & 
\multicolumn{2}{c|}{\textbf{Early-career}} & \multicolumn{2}{c|}{\textbf{Mid-career}} & \multicolumn{2}{c|}{\textbf{Senior}}\\
\cline{3-10}
 & & \textbf{$a$} & \textbf{$a'$} & \textbf{$a$} & \textbf{$a'$} & \textbf{$a$} & \textbf{$a'$} & \textbf{$a$} & \textbf{$a'$} \\
\hline

\multirow{3}{*}{\shortstack{Academic\\ Age}} & $\mu$ & 14.08$^{***}$ & 15.41$^{***}$ & 12.26$^{**}$ & 14.01$^{**}$ & 14.69 & 15.73 & 15.71$^{*}$ & 16.79$^{*}$\\ 

 & $M$ & 14.1 & 14.6 & 11.33 & 13.12 & 14.62 & 15.24 & 15.91 & 15.85\\ 

 & $\sigma$ & 7.44 & 7.09 & 8.55 & 6.91 & 6.81 & 6.89 & 5.93 & 7.14\\ 

\hline
\multirow{3}{*}{\shortstack{Number of\\ Papers}} & $\mu$ & 70.78$^{*}$ & 78.15$^{*}$ & 65.43$^{*}$ & 77.13$^{*}$ & 80.72 & 85.26 & 69.83 & 74.28\\ 

 & $M$ & 52.2 & 60.5 & 43.75 & 60.0 & 59.06 & 64.11 & 61.0 & 57.81\\ 

 & $\sigma$ & 76.66 & 81.68 & 71.35 & 92.06 & 103.82 & 83.75 & 56.88 & 66.04\\ 

\hline
\multirow{3}{*}{\shortstack{Number of\\ Citations}} & $\mu$ & 1482.94 & 1566.66 & 1295.48 & 1467.54 & 1537.61 & 1809.91 & 1657.42 & 1507.48\\ 

 & $M$ & 684.75 & 776.79 & 458.75 & 776.79 & 796.29 & 857.97 & 866.66 & 760.78\\ 

 & $\sigma$ & 3112.48 & 2614.25 & 2334.67 & 2439.67 & 2111.18 & 3249.68 & 4262.91 & 2277.16\\ 

\hline
 
\end{tabular}
\end{adjustbox}
\end{table}

\begin{table}[htbp]
\centering
\caption{Characterizing differences in \textbf{ collaborators gained} between retracted ($a$) and non-retracted ($a'$) authors stratified by academic age. $p$-Value is calculated using two-sided Welch's t-test; *$p<0.05$; **$p<0.01$; ***$p<0.001$.}
\label{tab:collabAnalysisByAge_gained}
\begin{adjustbox}{max width=.95\textwidth}
\begin{tabular}{|c|c|c|c|c|c|c|c|c|c|}
\hline
\multirow{2}{*}{\textbf{Metric}} & &
\multicolumn{2}{c|}{\textbf{Overall}} & 
\multicolumn{2}{c|}{\textbf{Early-career}} & \multicolumn{2}{c|}{\textbf{Mid-career}} & \multicolumn{2}{c|}{\textbf{Senior}}\\
\cline{3-10}
 & & \textbf{$a$} & \textbf{$a'$} & \textbf{$a$} & \textbf{$a'$} & \textbf{$a$} & \textbf{$a'$} & \textbf{$a$} & \textbf{$a'$} \\
\hline

\multirow{3}{*}{\shortstack{Academic\\ Age}} & $\mu$ & 8.0 & 8.18 & 7.16 & 7.86 & 8.06 & 7.95 & 8.95 & 8.72\\ 

 & $M$ & 7.58 & 7.8 & 5.93 & 7.59 & 7.78 & 7.57 & 8.74 & 8.49\\ 

 & $\sigma$ & 5.79 & 5.0 & 6.86 & 5.35 & 5.13 & 4.9 & 4.56 & 4.58\\ 

\hline
\multirow{3}{*}{\shortstack{Number of\\ Papers}} & $\mu$ & 38.76 & 35.73 & 36.21 & 35.6 & 39.7 & 35.77 & 41.12$^{*}$ & 35.84$^{*}$\\ 

 & $M$ & 28.0 & 28.56 & 20.49 & 28.83 & 29.24 & 27.57 & 34.32 & 29.51\\ 

 & $\sigma$ & 45.09 & 33.12 & 52.73 & 31.86 & 47.32 & 39.86 & 31.48 & 29.12\\ 

\hline
\multirow{3}{*}{\shortstack{Number of\\ Citations}} & $\mu$ & 893.43$^{*}$ & 773.14$^{*}$ & 871.58 & 822.8 & 886.13 & 788.61 & 924.74$^{**}$ & 702.8$^{**}$\\ 

 & $M$ & 386.88 & 386.71 & 246.0 & 399.92 & 396.1 & 302.32 & 569.13 & 398.92\\ 

 & $\sigma$ & 1509.15 & 1231.03 & 1775.04 & 1199.4 & 1575.87 & 1536.14 & 1047.52 & 1001.46\\ 

\hline

\end{tabular}
\end{adjustbox}
\end{table}

\begin{table}[htbp]
\centering
\caption{Characterizing differences in \textbf{collaborators retained} vs.\ \textbf{collaborators lost} between retracted ($a$) and non-retracted ($a'$) authors stratified by academic age. $\mu$ represents the average of the difference between retained collaborators and lost collaborators, $M$ represents the median of the difference between retained collaborators and lost collaborators. $p$-Value is calculated using two-sided Welch's t-test; *$p<0.05$; **$p<0.01$; ***$p<0.001$.}
\label{tab:collabAnalysisByAge_diffdiff}
\begin{adjustbox}{max width=.95\textwidth}
\begin{tabular}{|c|c|c|c|c|c|c|c|c|c|}
\hline
\multirow{2}{*}{\textbf{Metric}} & &
\multicolumn{2}{c|}{\textbf{Overall}} & 
\multicolumn{2}{c|}{\textbf{Early-career}} & \multicolumn{2}{c|}{\textbf{Mid-career}} & \multicolumn{2}{c|}{\textbf{Senior}}\\
\cline{3-10}
 & & \textbf{$a$} & \textbf{$a'$} & \textbf{$a$} & \textbf{$a'$} & \textbf{$a$} & \textbf{$a'$} & \textbf{$a$} & \textbf{$a'$} \\
\hline

\multirow{3}{*}{\shortstack{Academic\\ Age}} & $\mu$ & 2.01 & 2.52 & 3.31 & 3.85 & 3.11 & 3.45 & -0.14 & 0.45\\ 

 & $M$ & 1.44 & 2.05 & 2.45 & 3.18 & 2.31 & 3.22 & 0.14 & 0.18\\ 

 & $\sigma$ & 7.0 & 6.89 & 8.31 & 6.81 & 6.19 & 6.57 & 5.3 & 6.7\\ 

\hline
\multirow{3}{*}{\shortstack{Number of\\ Papers}} & $\mu$ & 21.98 & 27.5 & 23.28 & 28.55 & 29.47 & 32.91 & 15.35$^{*}$ & 22.6$^{*}$\\ 

 & $M$ & 11.13 & 15.58 & 10.95 & 17.21 & 13.36 & 22.07 & 9.73 & 11.3\\ 

 & $\sigma$ & 65.32 & 68.6 & 64.26 & 80.73 & 90.05 & 66.44 & 40.7 & 54.4\\ 

\hline
\multirow{3}{*}{\shortstack{Number of\\ Citations}} & $\mu$ & 243.81$^{*}$ & 412.26$^{*}$ & 301.29 & 449.46 & 230.42$^{*}$ & 552.37$^{*}$ & 191.99 & 274.36\\ 

 & $M$ & 30.97 & 70.72 & 52.54 & 145.96 & 32.45 & 135.07 & -0.25 & -8.65\\ 

 & $\sigma$ & 2017.23 & 1860.57 & 2005.05 & 1842.0 & 1577.27 & 2083.55 & 2289.88 & 1703.39\\ 

\hline
\end{tabular}
\end{adjustbox}
\end{table}

\begin{table}[htbp]
\centering
\caption{Characterizing differences in \textbf{collaborators retained} between retracted ($a$) and non-retracted ($a'$) authors stratified by reason. $p$-Value is calculated using two-sided Welch's t-test; *$p<0.05$; **$p<0.01$; ***$p<0.001$.}
\label{tab:collabAnalysisByReason_retained}
\begin{adjustbox}{max width=.95\textwidth}
\begin{tabular}{|c|c|c|c|c|c|c|c|c|c|}
\hline
\multirow{2}{*}{\textbf{Metric}} & &
\multicolumn{2}{c|}{\textbf{Overall}} & 
\multicolumn{2}{c|}{\textbf{Misconduct}} & \multicolumn{2}{c|}{\textbf{Plagiarism}} & \multicolumn{2}{c|}{\textbf{Mistake}}\\
\cline{3-10}
 & & \textbf{$a$} & \textbf{$a'$} & \textbf{$a$} & \textbf{$a'$} & \textbf{$a$} & \textbf{$a'$} & \textbf{$a$} & \textbf{$a'$} \\
\multirow{3}{*}{\shortstack{Academic\\ Age}} & $\mu$ & 14.08$^{***}$ & 15.41$^{***}$ & 13.6$^{*}$ & 15.05$^{*}$ & 12.92$^{***}$ & 15.33$^{***}$ & 15.64 & 16.01\\ 

 & $M$ & 14.1 & 14.6 & 13.76 & 14.42 & 12.45 & 14.13 & 15.4 & 15.0\\ 

 & $\sigma$ & 7.44 & 7.09 & 7.45 & 6.78 & 7.45 & 7.28 & 8.0 & 7.03\\ 

\hline
\multirow{3}{*}{\shortstack{Number of\\ Papers}} & $\mu$ & 70.78$^{*}$ & 78.15$^{*}$ & 73.71 & 82.14 & 63.62 & 69.02 & 76.44 & 81.5\\ 

 & $M$ & 52.2 & 60.5 & 56.59 & 61.3 & 44.55 & 57.84 & 61.0 & 61.89\\ 

 & $\sigma$ & 76.66 & 81.68 & 100.34 & 104.22 & 73.48 & 53.98 & 64.89 & 72.39\\ 

\hline
\multirow{3}{*}{\shortstack{Number of\\ Citations}} & $\mu$ & 1482.94 & 1566.66 & 1481.65 & 1801.16 & 1176.27 & 1043.61 & 1835.4 & 1866.87\\ 

 & $M$ & 684.75 & 776.79 & 818.35 & 902.49 & 409.83 & 577.98 & 1091.9 & 977.36\\ 

 & $\sigma$ & 3112.48 & 2614.25 & 1962.55 & 2876.94 & 4878.68 & 1316.46 & 2268.15 & 2881.3\\ 

\hline

\hline
\end{tabular}
\end{adjustbox}
\end{table}

\begin{table}[htbp]
\centering
\caption{Characterizing differences in \textbf{collaborators gained} between retracted ($a$) and non-retracted ($a'$) authors stratified by reason. $p$-Value is calculated using two-sided Welch's t-test; *$p<0.05$; **$p<0.01$; ***$p<0.001$.}
\label{tab:collabAnalysisByReason_gained}
\begin{adjustbox}{max width=.95\textwidth}
\begin{tabular}{|c|c|c|c|c|c|c|c|c|c|}
\hline
\multirow{2}{*}{\textbf{Metric}} & &
\multicolumn{2}{c|}{\textbf{Overall}} & 
\multicolumn{2}{c|}{\textbf{Misconduct}} & \multicolumn{2}{c|}{\textbf{Plagiarism}} & \multicolumn{2}{c|}{\textbf{Mistake}}\\
\cline{3-10}
 & & \textbf{$a$} & \textbf{$a'$} & \textbf{$a$} & \textbf{$a'$} & \textbf{$a$} & \textbf{$a'$} & \textbf{$a$} & \textbf{$a'$} \\
 \multirow{3}{*}{\shortstack{Academic\\ Age}} & $\mu$ & 8.0 & 8.18 & 8.35 & 8.31 & 6.79$^{**}$ & 7.98$^{**}$ & 8.88 & 8.44\\ 

 & $M$ & 7.58 & 7.8 & 7.64 & 8.02 & 6.33 & 7.65 & 8.82 & 8.0\\ 

 & $\sigma$ & 5.79 & 5.0 & 6.85 & 4.75 & 4.99 & 5.3 & 5.09 & 4.88\\ 

\hline
\multirow{3}{*}{\shortstack{Number of\\ Papers}} & $\mu$ & 38.76 & 35.73 & 45.22$^{*}$ & 35.22$^{*}$ & 32.66 & 34.38 & 41.79 & 38.51\\ 

 & $M$ & 28.0 & 28.56 & 28.36 & 31.5 & 22.56 & 28.45 & 34.56 & 27.07\\ 

 & $\sigma$ & 45.09 & 33.12 & 62.37 & 26.5 & 34.39 & 30.54 & 43.3 & 40.43\\ 

\hline
\multirow{3}{*}{\shortstack{Number of\\ Citations}} & $\mu$ & 893.43$^{*}$ & 773.14$^{*}$ & 1111.78$^{**}$ & 724.87$^{**}$ & 612.04 & 636.8 & 1087.9 & 978.41\\ 

 & $M$ & 386.88 & 386.71 & 505.7 & 478.81 & 283.61 & 301.77 & 624.41 & 433.88\\ 

 & $\sigma$ & 1509.15 & 1231.03 & 1932.21 & 881.96 & 1158.63 & 962.49 & 1597.08 & 1585.58\\ 

\hline
\end{tabular}
\end{adjustbox}
\end{table}

\begin{table}[htbp]
\centering
\caption{Characterizing differences in \textbf{collaborators retained} vs. \textbf{collaborators lost} between retracted ($a$) and non-retracted ($a'$) authors. $\mu$ represents the average of the difference between retained collaborators and lost collaborators, $M$ represents the median of the difference between retained collaborators and lost collaborators. $p$-Value is calculated using two-sided Welch's t-test; *$p<0.05$; **$p<0.01$; ***$p<0.001$.}
\label{tab:collabAnalysisByReason_diffdiff}
\begin{adjustbox}{max width=.95\textwidth}
\begin{tabular}{|c|c|c|c|c|c|c|c|c|c|}
\hline
\multirow{2}{*}{\textbf{Metric}} & &
\multicolumn{2}{c|}{\textbf{Overall}} & 
\multicolumn{2}{c|}{\textbf{Misconduct}} & \multicolumn{2}{c|}{\textbf{Plagiarism}} & \multicolumn{2}{c|}{\textbf{Mistake}}\\
\cline{3-10}
 & & \textbf{$a$} & \textbf{$a'$} & \textbf{$a$} & \textbf{$a'$} & \textbf{$a$} & \textbf{$a'$} & \textbf{$a$} & \textbf{$a'$} \\
\hline
\multirow{3}{*}{\shortstack{Academic\\ Age}} & $\mu$ & 2.01 & 2.52 & 1.69 & 2.17 & 1.63 & 2.79 & 2.38 & 2.27\\ 

 & $M$ & 1.44 & 2.05 & 0.91 & 1.59 & 1.01 & 1.86 & 1.71 & 2.07\\ 

 & $\sigma$ & 7.0 & 6.89 & 6.57 & 6.55 & 7.87 & 7.31 & 7.38 & 6.79\\ 

\hline
\multirow{3}{*}{\shortstack{Number of\\ Papers}} & $\mu$ & 21.98 & 27.5 & 25.64 & 28.08 & 20.05 & 22.99 & 18.77 & 27.28\\ 

 & $M$ & 11.13 & 15.58 & 9.21 & 16.09 & 8.81 & 14.21 & 12.86 & 14.11\\ 

 & $\sigma$ & 65.32 & 68.6 & 88.16 & 85.34 & 57.09 & 49.74 & 58.49 & 65.18\\ 

\hline
\multirow{3}{*}{\shortstack{Number of\\ Citations}} & $\mu$ & 243.81$^{*}$ & 412.26$^{*}$ & 247.83 & 542.26 & 246.52 & 162.64 & 170.8$^{*}$ & 497.87$^{*}$\\ 

 & $M$ & 30.97 & 70.72 & 32.5 & 115.06 & 1.72 & 25.98 & 56.13 & 108.79\\ 

 & $\sigma$ & 2017.23 & 1860.57 & 1551.78 & 1977.17 & 2617.51 & 1036.74 & 1811.95 & 2248.41\\ 

\hline

\end{tabular}
\end{adjustbox}
\end{table}

% stratfiied by attention
\begin{table}[htbp]
\centering
\caption{Characterizing differences in \textbf{collaborators retained} between retracted ($a$) and non-retracted ($a'$) authors stratified by attention. $p$-Value is calculated using two-sided Welch's t-test; *$p<0.05$; **$p<0.01$; ***$p<0.001$.}
\label{tab:collabAnalysisByAttention_retained}
\begin{adjustbox}{max width=.75\textwidth}
\begin{tabular}{|c|c|c|c|c|c|c|c|}
\hline
\multirow{2}{*}{\textbf{Metric}} & &
\multicolumn{2}{c|}{\textbf{Overall}} & 
\multicolumn{2}{c|}{\textbf{High attention}} & \multicolumn{2}{c|}{\textbf{Low attention}}\\
\cline{3-8}
 & & \textbf{$a$} & \textbf{$a'$} & \textbf{$a$} & \textbf{$a'$} & \textbf{$a$} & \textbf{$a'$}\\
\hline
\multirow{3}{*}{\shortstack{Academic\\ Age}} & $\mu$ & 14.08$^{***}$ & 15.41$^{***}$ & 14.17$^{**}$ & 17.02$^{**}$ & 14.07$^{***}$ & 15.25$^{***}$\\ 

 & $M$ & 14.1 & 14.6 & 14.0 & 16.0 & 14.16 & 14.5\\ 

 & $\sigma$ & 7.44 & 7.09 & 6.78 & 7.0 & 7.51 & 7.08\\ 

\hline
\multirow{3}{*}{\shortstack{Number of\\ Papers}} & $\mu$ & 70.78$^{*}$ & 78.15$^{*}$ & 76.46 & 105.41 & 70.19 & 75.33\\ 

 & $M$ & 52.2 & 60.5 & 70.57 & 76.0 & 51.27 & 57.94\\ 

 & $\sigma$ & 76.66 & 81.68 & 57.61 & 142.31 & 78.36 & 72.13\\ 

\hline
\multirow{3}{*}{\shortstack{Number of\\ Citations}} & $\mu$ & 1482.94 & 1566.66 & 1844.83$^{*}$ & 2650.77$^{*}$ & 1445.56 & 1454.69\\ 

 & $M$ & 684.75 & 776.79 & 1121.0 & 1535.08 & 607.6 & 720.02\\ 

 & $\sigma$ & 3112.48 & 2614.25 & 1893.2 & 3640.57 & 3210.45 & 2459.55\\ 
\hline
\end{tabular}
\end{adjustbox}
\end{table}

\begin{table}[htbp]
\centering
\caption{Characterizing differences in \textbf{ collaborators gained} between retracted ($a$) and non-retracted ($a'$) authors stratified by attention. $p$-Value is calculated using two-sided Welch's t-test; *$p<0.05$; **$p<0.01$; ***$p<0.001$.}
\label{tab:collabAnalysisByAttention_gained}
\begin{adjustbox}{max width=.75\textwidth}
\begin{tabular}{|c|c|c|c|c|c|c|c|}
\hline
\multirow{2}{*}{\textbf{Metric}} & &
\multicolumn{2}{c|}{\textbf{Overall}} & 
\multicolumn{2}{c|}{\textbf{High attention}} & \multicolumn{2}{c|}{\textbf{Low attention}}\\
\cline{3-8}
 & & \textbf{$a$} & \textbf{$a'$} & \textbf{$a$} & \textbf{$a'$} & \textbf{$a$} & \textbf{$a'$}\\
\hline

\multirow{3}{*}{\shortstack{Academic\\ Age}} & $\mu$ & 8.0 & 8.18 & 8.92 & 9.14 & 7.9 & 8.08\\ 

 & $M$ & 7.58 & 7.8 & 8.72 & 8.74 & 7.45 & 7.72\\ 

 & $\sigma$ & 5.79 & 5.0 & 5.79 & 5.44 & 5.78 & 4.94\\ 

\hline
\multirow{3}{*}{\shortstack{Number of\\ Papers}} & $\mu$ & 38.76 & 35.73 & 45.34 & 46.94 & 38.06$^{*}$ & 34.55$^{*}$\\ 

 & $M$ & 28.0 & 28.56 & 41.69 & 36.24 & 26.86 & 27.79\\ 

 & $\sigma$ & 45.09 & 33.12 & 36.21 & 50.95 & 45.89 & 30.45\\ 

\hline
\multirow{3}{*}{\shortstack{Number of\\ Citations}} & $\mu$ & 893.43$^{*}$ & 773.14$^{*}$ & 1301.76 & 1318.39 & 850.47$^{*}$ & 715.77$^{*}$\\ 

 & $M$ & 386.88 & 386.71 & 1049.62 & 643.87 & 349.8 & 371.01\\ 

 & $\sigma$ & 1509.15 & 1231.03 & 1286.08 & 2287.21 & 1524.91 & 1046.12\\ 

\hline

\end{tabular}
\end{adjustbox}
\end{table}

\begin{table}[htbp]
\centering
\caption{Characterizing differences in \textbf{collaborators retained} vs. \textbf{collaborators lost} between retracted ($a$) and non-retracted ($a'$) authors stratified by attention. $\mu$ represents the average of the difference between retained collaborators and lost collaborators, $M$ represents the median of the difference between retained collaborators and lost collaborators. $p$-Value is calculated using two-sided Welch's t-test; *$p<0.05$; **$p<0.01$; ***$p<0.001$.}
\label{tab:collabAnalysisByAttention_diffdiff}
\begin{adjustbox}{max width=.75\textwidth}
\begin{tabular}{|c|c|c|c|c|c|c|c|}
\hline
\multirow{2}{*}{\textbf{Metric}} & &
\multicolumn{2}{c|}{\textbf{Overall}} & 
\multicolumn{2}{c|}{\textbf{High attention}} & \multicolumn{2}{c|}{\textbf{Low attention}}\\
\cline{3-8}
 & & \textbf{$a$} & \textbf{$a'$} & \textbf{$a$} & \textbf{$a'$} & \textbf{$a$} & \textbf{$a'$}\\
\hline

\multirow{3}{*}{\shortstack{Academic\\ Age}} & $\mu$ & 2.01 & 2.52 & 0.99 & 2.65 & 2.11 & 2.51\\ 

 & $M$ & 1.44 & 2.05 & 0.82 & 2.39 & 1.5 & 2.0\\ 

 & $\sigma$ & 7.0 & 6.89 & 6.89 & 6.25 & 7.01 & 6.95\\ 

\hline
\multirow{3}{*}{\shortstack{Number of\\ Papers}} & $\mu$ & 21.98 & 27.5 & 19.8 & 38.51 & 22.2 & 26.39\\ 

 & $M$ & 11.13 & 15.58 & 13.26 & 21.66 & 10.95 & 14.79\\ 

 & $\sigma$ & 65.32 & 68.6 & 54.67 & 112.76 & 66.32 & 62.41\\ 

\hline
\multirow{3}{*}{\shortstack{Number of\\ Citations}} & $\mu$ & 243.81$^{*}$ & 412.26$^{*}$ & 258.92 & 735.57 & 242.28 & 379.61\\ 

 & $M$ & 30.97 & 70.72 & 150.81 & 270.85 & 26.6 & 57.72\\ 

 & $\sigma$ & 2017.23 & 1860.57 & 1666.29 & 2327.79 & 2050.03 & 1804.89\\ 

\hline
\end{tabular}
\end{adjustbox}
\end{table}

\clearpage

\section*{Supplementary Note 1: Data preprocessing filters}
\label{supplementarysec:filtering}
\addcontentsline{toc}{section}{Supplementary Note 1: Data preprocessing filters}
As part of our data cleaning process, we apply the following pre-processing filters on the Retraction Watch (RW) data set of 26,504 retracted papers. We highlight each pre-processing step and provide details on the number of records eliminated in each step.
\begin{itemize}

    \item \textbf{Removing bulk retractions% and duplicate titles
    :} Over 7,000 retractions in RW originated from a single publisher, the Institute of Electrical and Electronics Engineers (IEEE) \cite{McCook_2018}, most of which were retracted in bulk. Bulk retractions have a different mechanism not related to the content or science in the paper, but to the publisher or the conference. As such, we remove all retracted papers that were retracted in bulk. We do so by identifying all sets of five or more papers retracted by the same publisher on the same day. After applying this filter, we retain 15,034 out of the initial 26,504 retracted papers.
    
    \item \textbf{Removing retractions before 1990 and after 2015:} Throughout this paper we examine several outcomes that take shape post-retraction. As such, we allow for a five year window post-retraction for all authors so that we can determine if they are still pursuing careers in scientific publishing, and determine the size of their collaboration network. Naturally, this limits our ability to analyze retractions post-2015, as our data set extends only until 2020. We also exclude retractions prior to 1990 due to the scarcity of the data in prior years. After applying this filter to the remaining 15,034 papers, 6,704 remain.
    
\end{itemize}
Next, we match the titles of the 6,704 papers with those in the Microsoft Academic Graph (MAG). This matching process allowed us to identify 6,188 retracted paper titles. These papers were authored by 23,620 retracted authors. We filter these retracted authors as follows: 

\begin{itemize}
\item \textbf{Removing authors with multiple retractions:} 
Throughout the paper, we exclude authors with multiple retractions as our analysis treats a retraction as a singular event. Including scientists with multiple retractions could create challenges for our analysis, as the career impacts of multiple retractions likely differ from those of a single retraction. Additionally, establishing a clear timeline of publishing careers ``before'' and ``after'' the retraction would be challenging in these cases, as previous setbacks can influence the perceived impact of subsequent retractions. Therefore, to assess the effect of a single retraction, we focus on authors experiencing this event once in our main analyses.

%However, we find that many authors with multiple retractions have them clustered within a year (see Supplementary Figure~\ref{supplementaryfig:multiple_retractions}). Given that we examine publishing careers before and after the retraction year, as a middle ground, we also conduct a robustness analysis including only authors whose multiple retractions occur within the same year.

This resulted in 19,030 authors with single retractions and 5,275 retracted papers.

\item \textbf{Removing papers with missing fields:} Out of the 5,275 remaining retracted papers, 697 papers had missing information on at least one of the following variables: (i) affiliation at the start of the author's career; (ii) affiliation in the year of retraction; (iii) discipline; and (iv) gender. This resulted in 14,579 authors and 4,581 retracted papers.

\end{itemize}

Consequently, our final filtered sample consists of 4,581 retracted papers and 14,579  authors.

\clearpage
\newpage

\section*{Supplementary Note 2: Description of the confounders of the matching experiment}
\label{supplementarysec:matching}
\addcontentsline{toc}{section}{Supplementary Note 2: Description of the confounders of the matching experiment}

For every retracted author $a\in A$, with a paper $p$ that is retracted in the year $y_{a}$, we perform a matching process such that every retracted author $a$ is matched to a set of non-retracted authors $A'\setminus \{A \cup C_a\}$, where $C_a$ is a set of collaborators of the author $a$ at any point in time, before or after $y_{a}$.

A potential match of $a$, $a^{'} \in A^{'}\setminus \{A \cup C_a\}$, is selected when the following conditions hold:

\begin{itemize}
    \item \textbf{Gender:} $a$ and $a'$ have same gender; for details of how gender was identified, see the Materials and Methods section ``Creating author and paper level features.''
    
    \item \textbf{Scientific discipline at the time of retraction:} $a$ and $a'$ have the same scientific field at the time of retraction; for details of how the scientific field was inferred, see the Materials and Methods section ``Creating author and paper level features.''
    
    \item \textbf{Academic age:} $a$ and $a'$ have published their first paper in the same year.
    
    \item \textbf{Affiliation rank on the first paper:} The affiliations of authors are ranked based on 2019 Academic Ranking of World Universities\footnote{https://www.shanghairanking.com/rankings/arwu/2019, Accessed: December 2019.}. If $a$'s affiliation rank, $r_{a}$, 
    is between 1-100, they are matched to $a'$ such that $|{r_{a}} - {r_{a'}}| \leq 2$ where 
    $r_{a'}$ is the affiliation rank for $a'$. 
    If $r_a$ is between 101-1000, we look for matches in the following pre-defined bins by the Shanghai Ranking: $[101-150]$; $[151-200]$; $[201-300]$; $[301-400]$; $[401-500]$; $[501-600]$; $[601-700]$; $[701-800]$; $[801-900]$; $[901-1000]$. All authors with affiliations not ranked (i.e., $> 1000$) by the Academic Ranking of World Universities are matched exactly on the basis of the text field describing affiliations.
    
    \item \textbf{Affiliation rank at the time of retraction:} $a$ and $a'$ have the same affiliation in the year $y_{a}$. Matches are created based on the same process described above.

\end{itemize}

The closest matches are selected from the potential matches using a threshold of maximum distance between retracted and non-retracted authors, and calibrated distance function based on the number of papers, citations, and collaborators at the time of retraction described in Materials and Methods section ``Analytical sample for the matching experiment.''

\clearpage
\newpage

\section*{Supplementary Note 3: Description of the calculating physical distance, and divergence in field}
\label{supplementarysec:distance}
\addcontentsline{toc}{section}{Supplementary Note 3: Description of the calculating physical distance, and divergence in field}

\textbf{Physical distance}

To calculate the physical distance of a focal author and their collaborators, we take the following steps.

\begin{enumerate}
    \item Given all collaborators and collaborations for each retracted and matched author, we first identify the first collaboration for each author-collaborator pair and take that as the basis for calculations.
    \item If the (Retraction year – Collaboration year) is greater than or equal to zero, then the collaboration is ``pre'' retraction else it is ``post'' retraction.
    \item If (Retraction year – Collaboration year) is less than -5, then those collaborations are removed, i.e., we consider collaborations in the 5 years before and after retraction.
    \item For each author and collaborator, the affiliation is identified as the affiliation that is interpolated from their nearest year before or at the time of collaboration year. All affiliations in that year are retained, in case there are multiple.
    \item For each author, we compute the number of pre- and post- retraction collaborators as this criteria will be used to filter out authors who have either 0 pre- or 0 post-retraction collaborators. For these individuals no meaningful comparison can be made, in case they do not have any pre- and or post-retraction collaborators. All the corresponding matches are removed as well, in case the retracted author does not have a defined change in distance for this reason. The same process applied for matches, i.e., the retracted author is removed if none of their matches have a well-defined difference in distance. These steps ensured that both treatment and control sets are matched closed sets, i.e., each author has a match, and each match has an author.
    \item All the collaborators who have no latitude and longitude (determined from Google Maps API, or using ChatGPT and manual verification for missing data) are removed, but authors are not removed.
    \item After this pre-processing, in a loop of 100:
    \begin{enumerate} 
        \item A random match for each retracted author is chosen. 
        \item It is ensured that both treatment and control are matched closed set. Only authors who have a match and have pre- and post- collaborators and their matches have pre-post collaborators are considered (because of this we retracted authors have a distribution rather than a single point).
        \item Distance is computed using latitude and longitude in kilometers ``as the crow flies'' rather than considering a commuting route.
        \item For each author, we compute the mean distance pre- and post- (multiple affiliations are all counted).
        \item For each author, we compute the distance by considering (post - pre), i.e., positive numbers indicate collaborators further away post-retraction compared to pre-retraction.
        \item We compute the mean of this distance for authors and matches, and record it.
        \item After 100 runs, we have 100 means for authors, and 100 means for their matches which we plot. 
    \end{enumerate}
\end{enumerate}

\textbf{Field distance}

To calculate a retracted author's distance of fields pre- and post-retraction we take the following steps.

\begin{enumerate}
    \item We look at the distance at the main field level for all papers before and after retraction. If a paper is classified as a subfield, we convert it to its main field. 
    \item In the case when papers have multiple fields, we select the one that is most probable, except when it is equally distributed among multiple fields (e.g., 1/2 and 1/2 or 1/3, 1/3, 1/3 etc.), then we count that paper a portion of a paper for each field.
    \item Based on the above logic, for each paper we have a 19-long vector that shows which field(s) the paper belongs to. We aggregate these vectors giving equal weight to each paper and re-normalize the resulting vector, which will be 19-long, and its elements add to 1.
    \item Using the Jensen–Shannon divergence \cite{Osterreicher_Vajda_2003}, we calculate the distance between the fields of an author pre- and post-retraction. 
    \item We perform these calculations for both retracted authors and their matches. When multiple matches exist, we take the average. We plot the distribution of the divergence for both groups.
\end{enumerate}

The resulting two distances appear in Figure~\ref{suppfig:prepostanalysis}.

\bibliography{supplementary}
\bibliographystyle{naturemag}

\end{document}